\documentclass[12pt]{article}
\pdfoutput=1
\usepackage[totalwidth=480pt,totalheight=640pt]{geometry}
\usepackage[centertags]{amsmath}
\usepackage[square, comma, sort&compress,numbers]{natbib}
\usepackage{array,multirow}
\usepackage{amssymb}
\usepackage{graphicx}
\usepackage{color}
\usepackage{setspace}
\usepackage{comment}

\usepackage{epsfig}
\def\half{{\fr{1}{2}}}

\def\Or[#1]{{\text{O}}\left({#1}\right)}
\def\dotl[#1,#2]{\left\langle #1, #2 \right\rangle}
\def\dotlb[#1,#2]{[ #1, #2 ]}
\def\dotp[#1,#2]{(#1) \cdot (#2)}
\def\aff[#1,#2]{\hat{#1}(#2)}
\def\n4sym{{\cal N}=4 SYM}
\def\>{\rangle}
\def\<{\langle}
\def\weight[#1,#2,#3]{\{(#1),#2,#3\}}
\def\ads[#1]{$\text{AdS}_{#1}$}

\newcommand{\ba}{\begin{eqnarray}}
\newcommand{\ea}{\end{eqnarray}}
\newcommand{\be}{\begin{equation}}
\newcommand{\ee}{\end{equation}}
\newcommand{\benn}{\begin{equation*}}
\newcommand{\eenn}{\end{equation*}}

\newcommand{\Ccal}{{\cal C}}

\newcommand{\CM}{{\cal M}}
\newcommand{\Mcal}{{\cal M}}

\newcommand{\Ncal}{{\cal N}}
\newcommand{\CO}{{\cal O}}
\newcommand{\Ocal}{{\cal O}}

\newcommand{\Lcal}{{\cal L}}
\newcommand{\nn}{\nonumber}

\newcommand\oo\infty
\newcommand\s\sigma
\newcommand\de\delta
\newcommand\De\Delta
\newcommand{\p}{\partial}
\newcommand\f\phi
\newcommand\g\gamma
\newcommand\x\times

\newcommand{\ra}{\rightarrow}
\newcommand{\lra}{\leftrightarrow}
\newcommand{\fr}{\frac}
\newcommand{\tfr}{\tfrac}
\newcommand{\comm}[2]{[#1,#2]}
\newcommand{\xp}{x^+}
\newcommand{\xm}{x^-}

\newcommand\G{\Gamma}

\newcommand{\KL}{K\"{a}ll\'{e}n-Lehmann }
\newcommand{\Poincare}{Poincar\'{e} }
\newcommand{\CFT}{\textrm{CFT}}
\newcommand{\Cmax}{{\cal C}_{\max}}
\newcommand{\Dmax}{\Delta_{\max}}
\newcommand{\kmax}{k_{\max}}
\newcommand{\Lmax}{\ell_{\max}}
\newcommand{\LambdaIR}{\Lambda_{\textrm{IR}}}
\newcommand{\phivec}{\vec{\phi}^{\,2}}

\newcommand{\Lt}{\tilde{\ell}}
\newcommand{\Ft}{\widetilde{F}}
\newcommand{\norder}[1]{%
  {:\mathrel{\mspace{1mu}#1\mspace{1mu}}:}%
}

\newcommand\lrpar{\raise .8ex\hbox{$^\leftrightarrow$} \hspace{-9pt}
\partial}
\newcommand\lpar{\raise .8ex\hbox{$^\leftarrow$} \hspace{-9pt}
\partial}
\newcommand\rpar{\raise .8ex\hbox{$^\rightarrow$} \hspace{-9pt}
\partial}
\newcommand\lrd{\raise .8ex\hbox{$^\leftrightarrow$} \hspace{-9pt}
\nabla}

\renewcommand{\hat}{\widehat}

\usepackage{hyperref}

\numberwithin{equation}{section}

\begin{document}

\begin{titlepage}

\begin{center}
\vspace{1cm}

{\Large \bf A Conformal Truncation Framework \\ \vspace{.25cm} for Infinite\hspace{.04cm}-\hspace{-.1cm}Volume Dynamics}

\vspace{0.8cm}

\normalsize
\bf{Emanuel Katz, Zuhair U.\ Khandker, Matthew T.\ Walters}
\normalsize

\vspace{.5cm}

{\it Department of Physics, Boston University, Boston, MA 02215}\\

\end{center}

\vspace{1cm}

\begin{abstract}

We present a new framework for studying conformal field theories deformed by one or more relevant operators. The original CFT is described in infinite volume using a basis of states with definite momentum, $P$, and conformal Casimir, $\mathcal{C}$. The relevant deformation is then considered using lightcone quantization, with the resulting Hamiltonian expressed in terms of this CFT basis.  Truncating to states with $\mathcal{C} \leq \mathcal{C}_{\max}$, one can numerically find the resulting spectrum, as well as other dynamical quantities, such as spectral densities of operators. This method requires the introduction of an appropriate regulator, which can be chosen to preserve the conformal structure of the basis. We check this framework in three dimensions for various perturbative deformations of a free scalar CFT, and for the case of a free $O(N)$ CFT deformed by a mass term and a non-perturbative quartic interaction at large-$N$.  In all cases, the truncation scheme correctly reproduces known analytic results. We also discuss a general procedure for generating a basis of Casimir eigenstates for a free CFT in any number of dimensions.

\end{abstract}

\bigskip

\end{titlepage}

\hypersetup{pageanchor=false}

\tableofcontents

\section{Introduction and Summary}
\label{sec:Introduction}

Many of the most interesting phenomena in modern physics can be understood through the language of quantum field theory.  This includes the 
physics of quantum critical systems, aspects of statistical physics, as well as all high energy relativistic theories.  Though QFT is an old
subject, with a rich array of techniques for computation, a robust method for characterizing evolution in real time is still lacking.  Indeed,
in the non-perturbative regime, outside of certain techniques and systems in 2D, the only
commonly used approach is lattice quantization.   However, many interesting QFTs are difficult to simulate on the lattice or
lack a lattice formulation.  In addition, even for theories that have a lattice description, it is challenging to extract 
truly dynamical quantities.  These include time-dependent correlation functions, spectral densities, and properties of
the quantum wavefunction of states (such as the PDF of the proton).  It is therefore a worthwhile goal to search for 
non-perturbative methods which may also access dynamical observables.

Hamiltonian truncation has recently gained momentum as a means of studying real-time dynamics \cite{Katz:2013qua,Katz:2014uoa,Rychkov:2014eea,Rychkov:2015vap,Brooks:1983sb,Lee:2000ac,Lee:2000xna,Elias-Miro:2015bqk,Bajnok:2015bgw,Chabysheva:2013oka,Chabysheva:2014rra,Chabysheva:2015ynr,Christensen:2016naf}.  
The basic idea is to first discretize the QFT in some manner, yielding a Hilbert space consisting of an infinite tower of
discrete basis states.  The QFT Hamiltonian is then diagonalized numerically by truncating the basis to a finite
subset of the full Hilbert space.  At the heart of any truncation procedure is an interesting conceptual question - which 
choice of basis is optimal for the calculation of a desired observable?  Namely, can one choose a truncation
scheme which is efficient?  In certain regimes, for instance highly excited states in a strongly coupled
ergodic system, the expectation is that no choice of basis will be optimal due to the complexity implied
by the eigenstate thermalization hypothesis \cite{Deutsch,Srednicki}.  However, for the lowest energy excitations, one could
hope that there is a choice of basis which efficiently captures their wavefunctions.

One strategy is to view the Hamiltonian as originating from a deformed CFT and use conformal symmetry to organize the basis. The standard implementation of this strategy is called the Truncated Conformal Space Approach.  This approach, first pioneered by Yurov and Zamolodchikov \cite{Yurov:1989yu},
uses the operator-state correspondence to study CFTs perturbed by relevant operators at finite volume
on a sphere.  The truncation is to simply consider states up to a certain maximum energy, which on a sphere is related to the dimension of the corresponding operator.
This scheme has been successfully applied to various 2D systems \cite{Yurov:1991my,Lassig:1990xy,Lassig:1990wc,Klassen:1991ze,Delfino:1996xp,Feverati:1998va,Feverati:1998dt,Bajnok:2000wm,Bajnok:2000ar,Fonseca:2001dc,Bajnok:2003dk,Feverati:2006ni,Toth:2006tj,Konik:2007cb,Lepori:2008et,Lepori:2009ip,Brandino:2010sv,Watts:2011cr,Giokas:2011ix,Beria:2013hz,Lencses:2014tba,Coser:2014lla,Lencses:2015bpa}. More recently, Hogervorst et al.~\cite{Hogervorst:2014rta} have managed to extend the method to include 
free scalar CFTs in non-integer dimensions, and studied the $\phi^4$ deformation in $d=2.5$ in the 
non-perturbative regime.

In this work, we present a new conformal truncation framework, motivated by AdS/CFT, which can be directly applied in the infinite volume limit.  
To understand this framework, consider a general CFT perturbed by a relevant operator.  
The resulting RG flow could result in a mass gap or 
perhaps a new CFT fixed point.  In the holographic description, the RG flow is described by some
sort of background, where the field dual to the relevant operator is turned on, growing in the radial
direction away from the boundary.  Fields in the bulk, which correspond to the various conformal
multiplets of the UV CFT, mix in the background of the flow.  Each of these conformal multiplets is characterized
by its spin and its eigenvalue under the conformal Casimir, $\Ccal$, which determines the mass
of the corresponding bulk field.  The naive expectation is that high mass bulk fields should decouple
from the lightest energy states in the background of the RG flow.  Indeed, one can imagine integrating them out, yielding an 
effective description involving only the lightest fields \cite{Heemskerk:2009pn,Fitzpatrick:2010zm}.  The rate of decoupling of the high mass bulk
fields from low energy observables will in general depend on dynamical details (or equivalently on
the precise background flow).  The expectation is that the amplitude for creating a light
state by a primary operator $\CO$ will have the schematic behavior
\be
\langle \CO(0)|\psi_{\textrm{light}}\rangle \sim \fr{1}{(\Ccal_\CO)^n},
\ee
with the precise value for $n$ set by the dynamics of the particular theory. If the light state is further well-localized in the bulk, the decoupling can be even more rapid \cite{Fitzpatrick:2013twa}.

Motivated by this decoupling behavior, we propose the following truncation scheme.  
First, one builds a basis of states consisting of eigenstates of the conformal Casimir.  This
basis is most conveniently expressed in momentum space, which allows one
to focus on the dynamical properties of the wavefunction, trivializing the
center of mass degree of freedom.  Thus, our basis consists of states
labeled by the spatial momentum $\vec{P}$, the invariant mass $\mu^2$, and
the bulk field labels of spin and Casimir:
\be
|\Ccal, \ell;  \vec{P}, \mu \rangle \equiv \int d^dx~e^{-iP\cdot x} \CO(x)|0\rangle,
\ee
where $\mu^2 \equiv P^2$. One can think of these states as the familiar states of the \Poincare patch of AdS,
\be
\langle\phi_\ell(x,z)|\Ccal, \ell;  \vec{P}, \mu \rangle \sim z^{\fr{d}{2}-\ell} J_{\Delta-\fr{d}{2}}(\mu z) \, e^{-iP\cdot x},
\ee
where $\phi_\ell$ denotes a bulk field of spin $\ell$, and we have ignored any polarization structure.
Our truncation scheme consists of including only states with $\Ccal \leq \Cmax$.
In practice, we need to discretize the above basis, which for a given spatial momentum still
has a continuous label $\mu$.  Hence, our basis will consist of states 
\be
|\Ccal, \ell;  \vec{P}, k\rangle = \int d\mu^2 g_k(\mu)~|\Ccal, \ell;  \vec{P}, \mu \rangle,
\ee
with discrete label $k$.  We introduce a regulator, restricting $\mu^2 \leq \Lambda^2$, and
choose $g_k(\mu)$ to be polynomials.  However, it is possible that there exists a better choice of discretization for $\mu$.
Note that this regulator is Lorentz invariant and preserves the conformal structure of the basis, in that it does
not mix states with different Casimir eigenvalue. In the simple case of a scalar operator, this regulator defines the inner product
\be
\int_0^{\Lambda^2} d\mu^2 \, \rho_\Ocal (\mu) \, g_k(\mu) g_{k'}(\mu),
\ee
where $\rho_\Ocal (\mu)$ is the spectral density of the operator $\CO(x)$.  Our polynomials are orthogonal with respect to this inner product.

Thus, our final truncation scheme consists of the finite Hilbert space spanned by all conformal multiplets with $\Ccal \leq \Cmax$, with each multiplet restricted to $k \leq \kmax$.
The expectation from holography is that we should find good convergence with $\Ccal$ due to decoupling of high mass bulk fields,
while convergence in $k$ should be poorer.  Indeed, one can think of $\kmax$ heuristically as a 
parameter controlling our resolution in the bulk. The value for $\kmax$ thus sets the effective IR cutoff for our Hamiltonian eigenvalues.

Once we have our basis, the next task is to calculate the matrix elements of the Hamiltonian involving
the relevant operator which perturbs the CFT,
\be
S = S_{\CFT} - \lambda \int d^d x \,  \Ocal_R(x).
\ee
To construct these matrix elements, we choose to work in lightcone quantization, which for Hamiltonian
methods offers two important advantages \cite{Dirac:1949cp,Brodsky:1997de}. First, the lightcone momentum $P_-$ annihilates the vacuum ($P_- |0\rangle =0$) while for any non-vacuum state
$P_- > 0$ \cite{Leutwyler:1970wn,Maskawa:1975ky}.  Consequently, when perturbing the lightcone Hamiltonian, $P_+ \rightarrow P_+ + \delta P_+$,
the perturbation $\delta P_+$ cannot mix the vacuum with any other states, due to momentum conservation.  Thus, the vacuum energy is
not renormalized.  Second, for the specific case of a free CFT, this same observation implies that matrix elements which involve particle creation from nothing must vanish.  For example, adding
a mass term does not lead to particle number violating matrix elements (as it would in standard spatial
quantization).  

We therefore need to calculate matrix elements for the lightcone Hamiltonian $P_+$ arising from the relevant operator $\Ocal_R$. This Hamiltonian is defined on a spacetime slice of fixed lightcone time $x^+$, leading to the general matrix elements
\be
\lambda \int d^{d-1}\vec{x} \, \langle \Ccal, \ell;  \vec{P}, k | \CO_R(x^+=0,\vec{x})|\Ccal', \ell';  \vec{P}', k' \rangle.
\ee
In principle, these matrix elements can be related to the appropriate CFT Wightman functions $\langle \CO_\ell \CO_R \CO_{\ell'}\rangle$ in momentum space.
Hence, they are determined entirely by the OPE coefficients and CFT kinematics, which could be conveniently parameterized in terms of AdS, for instance. This framework for computing matrix elements would allow us to start from any CFT where OPE coefficients are known explicitly or
could be found through the numerical bootstrap \cite{El-Showk:2014dwa,Gliozzi:2014jsa,Kos:2016ysd} (for CFTs without a Lagrangian description).

Here we focus on deformations of free CFTs where, as a practical matter, we can instead compute all Hamiltonian matrix elements directly using standard Fock space amplitudes, $\langle p_1,\dots,p_n|\delta P_+|k_1,\dots,k_m\rangle$.
To use these matrix elements we therefore need to express the Casimir eigenstates in terms of Fock space states,
\be 
\langle p_1,\dots,p_n| \Ccal, \ell; \vec{P}, \mu \rangle \equiv F_\Ocal(p_1,\dots,p_n) \, (2\pi)^d \delta^d\Big(\sum_i p_i - P\Big).
\ee
The conformal Casimir can be expressed as a second-order differential operator in momentum space, so finding
the wavefunctions $F_\Ocal$ amounts to finding a complete set of eigenfunctions of this differential operator for each particle number sector. The resulting functions are $d$-dependent and consist of orthogonal polynomials of particle momenta. The polynomials for the case $d=2$ were used previously in \cite{Katz:2013qua,Katz:2014uoa}, though in that work they were not obtained simply as eigenfunctions of the conformal Casimir.\footnote{A similar approach, also using a basis of polynomials in 2D, was presented in \cite{Chabysheva:2013oka,Chabysheva:2014rra,Chabysheva:2015ynr}, though this work did not use conformal structure to organize the basis.} 

Once the Hamiltonian matrix elements are computed up to $\Cmax$ and $\kmax$, the last step is to diagonalize the matrix numerically\footnote{Here we consider Hamiltonian matrices up to maximum sizes of $\sim 10^4 \times 10^4$. The matrices are typically sparse and can be diagonalized in Mathematica, with computation times on the order of minutes.} and use the resulting eigenstates to compute Lorentz invariant physical observables. The first interesting observable is the spectrum itself. The eigenvalues and eigenvectors obtained via diagonalization are an approximation to the physical spectrum of the full, interacting theory. With an approximate spectrum in hand, we can move on to study dynamical correlation functions by expanding the correlators in terms of the physical states.

In this work, we specifically consider relevant deformations of free scalar CFTs in 3D. Our choice of dimension is primarily motivated by the need to extend Hamiltonian truncation methods beyond 2D. We start by constructing the basis of Casimir eigenstates and performing several consistency checks in free field theory. We then consider the addition of perturbative $\phi^3$ and $\phi^4$ interactions, using our conformal truncation framework to reproduce the one-particle mass shift. This particular observable provides a clear means of testing the effects of the truncation parameters $\Cmax$ and $\kmax$. Matching our holographic intuition, we find rapid convergence in the conformal Casimir $\Ccal$. In fact, we are able to reproduce the $\phi^3$ mass shift to within $10\%$ by using only a \emph{single} multiplet.

We then move on to the main test of our method: the strongly-coupled $O(N)$ model. By taking the large-$N$ limit, we are able to compare our results to analytic expressions in a non-perturbative setting. Specifically, we reproduce the spectral density for $\phivec$ in the presence of a mass term and quartic interaction. The spectral density is a decomposition of the dynamical correlation function $\<\phivec \phivec\>$ in terms of the physical mass eigenstates. This Lorentz invariant observable shows the full RG flow of $\phivec$ from the original free CFT in the UV to a strongly-interacting theory in the IR, where we can extract the resulting large anomalous dimension. We again find rapid convergence in $\Cmax$, even at strong coupling, reproducing the detailed form of the correlation function.

The large-$N$ limit provides us with a precise testing ground, allowing us to focus on states with low particle number. However, it is important to note that our framework proceeds no differently for finite $N$. The basis of Casimir eigenstates and the Hamiltonian matrix elements we present here are valid for any $N$, and only need to be computed for higher particle number to study the 3D Ising and $O(N)$ models, which we plan to consider in future work~\cite{FutureUs}.

Ultimately, our proposal is a Hamiltonian truncation framework that computes Lorentz invariant dynamical observables, is formulated directly in infinite volume, and can be applied in $d>2$. This framework utilizes holographic intuition to organize the Hilbert space according to the conformal Casimir, which we show to be an efficient truncation parameter. We hope this combination of features provides a new tool for studying strongly-coupled QFTs.   

The paper is organized as follows.  In section \ref{sec:Review} we review the lightcone quantization of a free scalar field in 3D and present the contributions to the lightcone Hamiltonian written in terms of Fock space modes.  We also summarize the analytic results that we later reproduce with our truncation method. In section \ref{sec:MasslessBasis} we describe the general procedure for determining a Casimir basis for free CFTs in 3D and explicitly construct the basis for the case of two and three particles. We also briefly comment on how this procedure can be generalized to higher dimensions.   In section \ref{sec:MasslessResults}
we warm up by considering observables that can be computed using a single Casimir multiplet (truncated at $\kmax$). Specifically, we reproduce the spectral densities of the operators $\phi^2$ and $\phi^3$ in the original free CFT, as well the large-$N$ $\phivec$ spectral density in the presence of a non-perturbative quartic interaction. 

In section \ref{sec:MassiveBasis} we then consider the addition of a mass term. The mass introduces an important subtlety due to the fact that its Hamiltonian matrix elements contain divergences.  Rearranging the Casimir basis slightly allows
one to avoid these divergences at the price of mixing a Casimir eigenstate with all eigenstates below it.
Though the resulting basis states are no longer strictly Casimir eigenstates, the truncation parameter $\Cmax$
still captures the maximum Casimir used to construct the basis.
Section \ref{sec:MassiveResults} contains the bulk of our numerical results with comparisons to
analytic expressions.  First, we compute the spectral densities of the operators $\phi^2$ and $\phi^3$
in the presence of a mass deformation of the CFT. Since the mass term mixes basis states, we vary the parameter $\Cmax$ to study the convergence of the truncation. Next, we consider the one-particle mass shift due to perturbative $\phi^3$ and $\phi^4$ interactions, testing the convergence in both $\Cmax$ and $\kmax$. Finally, we reproduce the RG flow of the singlet operator $\vec{\phi}^2$ resulting from
deforming the free $O(N)$ CFT by a mass term and non-perturbative quartic interaction at
large-$N$. Even for this strongly-coupled example, we see fast convergence in $\Cmax$.       
We conclude and discuss future directions in section \ref{sec:Discussion}. We also provide a self-contained set of appendices presenting the details of our calculations.


\section{Scalar Field Theory on the Lightcone}
\label{sec:Review}

The starting point for the computations in this paper is a UV CFT consisting of a free massless scalar field $\phi$, with the Lagrangian
\be
\Lcal_{\CFT} = \half \norder{\p_\mu \phi \p^\mu \phi}.
\label{eq:CFTLagrangian}
\ee
The notation $\norder{\Ocal}$ indicates that the operator is normal-ordered, but henceforth we will suppress this notation, with the understanding that $\emph{all}$ local operators are to be normal-ordered. We also consider the more general case where there are $N$ free scalar fields $\phi_i$. In this section we review, for 3D scalar field theory in lightcone quantization, the contributions to the Hamiltonian coming from different relevant deformations. We then discuss the Lorentz invariant observables we later compute using our truncation method.


\subsection{Lightcone Hamiltonian}
\label{sec:Hamiltonian}

We work in $2+1$ dimensions, using lightcone (or lightfront) coordinates, which are defined by combining a particular spatial direction $x$ with the time coordinate $t$ to form $x^\pm \equiv \fr{1}{\sqrt{2}}(t \pm x)$. The resulting Lorentzian metric is
\be
ds^2 = 2 d\xp d\xm - dx^{\perp2}.
\ee
In lightcone quantization, the new coordinate $\xp$ is treated as the ``time'' direction, while the other lightcone coordinate $\xm$ and transverse direction $x^\perp$ are the ``spatial'' directions. These coordinates have corresponding momenta $p_\mu \equiv i \p_\mu$, such that
\be
p^2 = 2 p_+ p_- - p_\perp^2.
\label{eq:psquared}
\ee

To study the IR dynamics of a given theory, we need to approximate the physical spectrum of low-mass eigenstates. In a frame with total spatial momentum $\vec{P}$, this means diagonalizing the invariant mass operator
\be
M^2 \equiv 2P_+ P_- - P_\perp^2.
\ee
As our basis states will be eigenstates of total momentum, we are free to choose any total momentum frame. Without loss of generality, we choose to work in a frame with fixed lightcone momentum $P_-$ and transverse momentum $P_\perp = 0$. Given this choice, we see that diagonalizing the operator $M^2$ is equivalent to diagonalizing the lightcone Hamiltonian $P_+$.

Since the UV theory is free, we can expand the massless scalar field $\phi$ in terms of the usual Fock space modes, 
\be
\phi(x) = \int \fr{d^2p}{(2\pi)^2 \sqrt{2p_-}} \left( e^{-ip \cdot x} a_p + e^{ip \cdot x} a^\dagger_p \right),
\label{eq:PhiDef}
\ee
where the creation and annihilation operators $a^\dagger$ and $a$ satisfy the commutation relation 
\be
\comm{a_p}{a^\dagger_q} = (2\pi)^2 \de^2(p-q).
\ee

Operators like $P_+$ can likewise be expanded in terms of these modes. As a simple example, let's first consider the unperturbed CFT Hamiltonian. This operator arises solely from the ``kinetic term'' Lagrangian in eq.~(\ref{eq:CFTLagrangian}). As shown in appendix \ref{app:Interactions}, this Lagrangian leads to the following mode expansion for $P_+$,
\be
P_+^{(\CFT)} = \int \fr{d^2p}{(2\pi)^2} \, a^\dagger_p a_p \, \fr{p_\perp^2}{2p_-}.
\label{eq:PCFT}
\ee
This expression is easily understood by noting that the Hilbert space of our UV CFT consists of states with free massless particles, which obey the equation of motion
\be
2p_+ p_- - p_\perp^2 = 0.
\ee
Solving this equation for $p_+$, we obtain precisely the function of momenta in the integrand for $P_+$. The free lightcone Hamiltonian simply corresponds to a sum over the number of particles in a given state, each weighted by their on-shell value for $p_+$.\footnote{Given that $P_+^{(\CFT)}$ is quadratic in $\phi$, one might have expected its mode expansion in eq.~(\ref{eq:PCFT}) to also contain terms proportional to $a^\dagger a^\dagger$ and $aa$. However, as detailed in appendix \ref{app:Interactions}, these terms vanish in lightcone quantization. This is a consequence of momentum conservation, together with positivity of lightcone momenta. Ultimately, any term in the full Hamiltonian containing only $a$'s or only $a^\dagger$'s vanishes in lightcone quantization.}

The simplest deformation of our UV theory is the addition of the mass term
\be
\de \Lcal = -\half m^2 \phi^2.
\ee
Including this operator adds the following term to the lightcone Hamiltonian,
\be
\de P_+^{(m)} = \int \fr{d^2p}{(2\pi)^2} \, a^\dagger_p a_p \, \fr{m^2}{2p_-}.
\ee
Like $P_+^{(\CFT)}$, this correction consists of a sum over the number of particles, weighted by a factor proportional to $m^2$. This operator corresponds to a shift in the lightcone energy of each individual particle, consistent with the massive equation of motion
\be
2p_+ p_- - p_\perp^2 = m^2.
\ee

We also consider the additional relevant operators
\be
\de \Lcal = -\fr{1}{3!} g \phi^3 - \fr{1}{4!} \lambda \phi^4.
\ee
The resulting corrections to the Hamiltonian can again be expanded in terms of Fock space modes. Starting with the cubic interaction, we obtain
\be
\de P_+^{(g)} = \fr{g}{2} \int \fr{d^2p \, d^2q}{(2\pi)^4 \sqrt{8 p_- q_-(p_- + q_-)}} \left( a^\dagger_p a^\dagger_q a_{p+q} + a^\dagger_{p+q} a_p a_q \right).
\ee
This operator clearly has different structure than the previous $P_+$ contributions. Rather than simply consisting of a weighted sum over particles, this cubic interaction mixes states whose particle numbers differ by one. The quartic interaction is somewhat similar,
\be
\de P_+^{(\lambda)} = \fr{\lambda}{24} \int \fr{d^2p \, d^2q \, d^2k}{(2\pi)^6 \sqrt{8 p_- q_- k_-}} \left( \fr{4 a^\dagger_p a^\dagger_q a^\dagger_k a_{p+q+k}}{\sqrt{2(p_- + q_- + k_-)}} + h.c. + \fr{6 a^\dagger_p a^\dagger_q a_k a_{p+q-k}}{\sqrt{2(p_- + q_- - k_-)}} \right).
\ee
As we can see, this Hamiltonian correction contains two distinct terms, one which mixes states whose particle numbers differ by two and another which preserves particle number.

We can consider various combinations of these contributions to the lightcone Hamiltonian, in order to study a range of potential IR dynamics. We specifically focus on the case where the coupling scales $g$ and $\lambda$ are perturbatively small compared to the mass scale $m$. This restriction allows us to both simplify the truncation calculations and compare the results with analytic expectations, in order to study the effectiveness of the overall method.

We also consider the case of $N$ free scalar fields, for which the Lagrangian in eq.~(\ref{eq:CFTLagrangian}) generalizes to
\be
\Lcal_{\textrm{CFT}} = \sum_{i=1}^N \half \p_\mu \phi_i \p^\mu \phi_i.
\ee
From now on, we suppress the explicit sum over flavors, with the convention that repeated indices are summed over. This generalized UV Lagrangian now has an $O(N)$ symmetry, associated with arbitrary rotations of the vector $\phi_i$.

The calculation of the associated lightcone Hamiltonian is almost identical to the case with one scalar field, leading to the similar expression
\be
P_+^{(\CFT)} = \int \fr{d^2p}{(2\pi)^2} \, a^\dagger_{p,i} a_{p,i} \, \fr{p_\perp^2}{2p_-}.
\ee
We see that the Hamiltonian has the same kinematic structure as before, with the added constraint that it preserves flavor, only linking particles with the same index.

We then consider the $O(N)$-symmetric relevant deformations
\be
\de \Lcal = -\half m^2 \phi_i^2 - \fr{1}{4} \lambda \, \phi_i^2 \phi_j^2.
\ee
Unsurprisingly, the mass correction to $P_+$ is almost identical to the single field case,
\be
\de P_+^{(m)} = \int \fr{d^2p}{(2\pi)^2} \, a_{p,i}^\dagger a_{p,i} \, \fr{m^2}{2p_-}.
\ee
The quartic correction is somewhat more complicated. As discussed in appendix \ref{app:Interactions}, this interaction leads to three distinct contributions to $P_+$. However, only one of these terms is unsuppressed in the large-$N$ limit, simplifying the resulting Hamiltonian correction to
\be
\de P^{(\lambda)}_+ = \fr{\lambda}{2} \int \fr{d^2p \, d^2q \, d^2k}{(2\pi)^6 \sqrt{8p_- q_- k_-}} \, \fr{a^\dagger_{p,i} a^\dagger_{q,i} a_{k,j} a_{p+q-k,j}}{\sqrt{2(p_- + q_- - k_-)}}.
\ee
This dominant contribution preserves particle number. In the large-$N$ limit, we can therefore treat sectors with different particle numbers as independent, with mixing between these sectors suppressed by $1/N$.


\subsection{\KL Spectral Density and IR Dynamics}

The conformal truncation recipe involves diagonalizing the invariant mass operator $M^2$ in a truncated Hilbert space of states. The resulting eigenvalues and eigenvectors are an approximation to the physical spectrum of the full, interacting theory. This computation of the spectrum in turn allows us to construct dynamical correlation functions. 

We specifically focus on two-point functions. For a given two-point function, a natural object for us to study is the \KL spectral density, which precisely encodes the decomposition of the correlator in terms of the physical mass eigenstates. In this section, we briefly review the definition of the spectral density and provide the analytic expressions that we will reproduce using conformal truncation.  

The \KL spectral density of a local operator is defined as the overlap of that operator with mass eigenstates as a function of their invariant mass $\mu^2$,
\be
\rho_\Ocal(\mu) \equiv \sum_i |\<\Ocal(0)|\mu_i\>|^2 \de(\mu^2 - \mu_i^2).
\ee
The spectral density can be used to compute the two-point function of $\Ocal(x)$ via the relation
\be
\<\Ocal(x) \Ocal(0)\> = \int d\mu^2 \rho_\Ocal(\mu) \int \fr{d^2P}{(2\pi)^2 2P_-} \,e^{-iP\cdot x} = \int d\mu^2 \rho_\Ocal(\mu) \, \fr{e^{-\mu x}}{4\pi x}.
\label{eq:SpectralRep}
\ee
This expression is known as the \KL spectral representation of $\<\Ocal(x) \Ocal(0)\>$, which corresponds to a sum over all possible intermediate mass eigenstates, weighted by the free propagator. As we can see, in $d=3$ the spectral density amounts to a Laplace transform of the original position-space two-point function. In practice, it will be simpler for us to work with the integrated spectral density,
\be
I_\Ocal(\mu) \equiv \int_{0}^{\mu^2} d\mu^{\prime \, 2} \, \rho_\Ocal(\mu') = \sum_{\mu_i \leq \mu} |\<\Ocal(0)|\mu_i\>|^2.
\ee
We therefore just need to compute the cumulative overlap of our approximate mass eigenstates with any operator $\Ocal(x)$ to calculate this integrated density and reproduce the associated correlation function.

In this work, we use our truncated basis of states to compute the spectral densities for $\phi^2$ and $\phi^3$ in free scalar field theory. One might also consider computing the spectral density for $\phi$, but this expression is actually trivial, as it only receives a contribution from the single Lorentz multiplet of the one-particle state,
\be
\rho_\phi(\mu) = \de(\mu^2 - m^2).
\label{eq:rhophi}
\ee

Our truncation results can then be compared to the exact \KL densities, which we can extract from known free field theory correlators. For example, the $\phi^2$ two-point function is
\be
\<\phi^2(x) \phi^2(0)\> = 2! \left( \fr{e^{-mx}}{4\pi x} \right)^2 = \fr{e^{-2mx}}{8\pi^2 x^2}.
\ee
By eq.\ (\ref{eq:SpectralRep}), we can take an inverse Laplace transform of this correlator to obtain the resulting spectral density
\be
\rho_{\phi^2}(\mu) = \fr{1}{4\pi\mu} \, \theta(\mu-2m),
\label{eq:PhiSquareDensity}
\ee
where $\theta(x)$ is the Heaviside step function.

Unlike the $\phi$ spectral density, we see that $\phi^2$ has contributions from a continuum of states, with invariant mass $\mu \geq 2m$. This continuum is precisely the set of two-particle states, such that we can interpret the spectral density of $\phi^2$ as the two-particle density of states. We can integrate this spectral density to obtain the theoretical prediction,
\be
I_{\phi^2}(\mu) \equiv \int_{4m^2}^{\mu^2} d\mu^{\prime \, 2} \, \rho_{\phi^2}(\mu') = \fr{1}{2\pi} (\mu-2m),
\ee
which we compare to our truncation results in section~\ref{sec:MasslessResults} for the massless case and in section~\ref{sec:MassiveResults} for the massive case.

Similarly, the \KL density for $\phi^3$ corresponds to the three-particle density of states, which we can derive from the correlation function
\be
\<\phi^3(x) \phi^3(0)\> = \fr{3 e^{-3mx}}{32\pi^3 x^3}.
\ee
The resulting integrated spectral density is
\be
I_{\phi^3}(\mu^2) = \int_{9m^2}^{\mu^2} d\mu^{\prime \, 2} \, \fr{3(\mu^\prime - 3m)}{16\pi^2\mu^\prime} = \fr{3}{16\pi^2}(\mu-3m)^2.
\ee

\begin{figure}[t!]
\begin{center}
\includegraphics[width=0.55\textwidth]{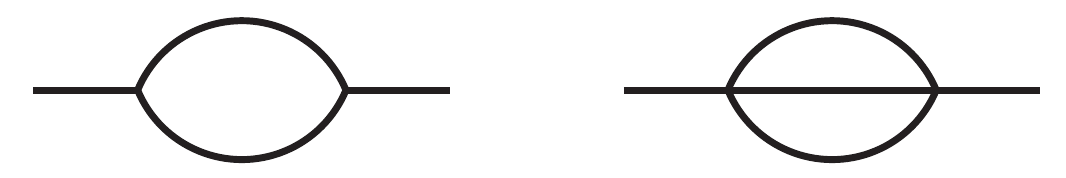}
\caption{The leading perturbative mass corrections due to $\phi^3$ (left) and $\phi^4$ (right) self-interactions. The analytic expressions can be compared with the one-particle mass eigenvalue obtained from the conformally truncated operator $M^2$.}
\label{fig:MassShifts} 
\end{center}
\end{figure}

While these spectral densities are a useful test of the completeness and convergence of our basis, we'd like to move beyond free field theory to study the effects of interactions. As mentioned earlier, we specifically consider the addition of the relevant operators
\be
\de \Lcal = - \fr{1}{3!} g \phi^3 - \fr{1}{4!} \lambda \phi^4.
\ee
The simplest Lorentz invariant observable associated with these interactions is the leading perturbative correction to the one-particle mass. These corrections can of course be calculated analytically from the Feynman diagrams in figure \ref{fig:MassShifts}.

As we can see from these diagrams, both leading mass corrections arise at second order in perturbation theory. The resulting cubic mass correction is straightforward to evaluate, leading to the constant shift
\be
\de m^2_{(g)} = - \fr{g^2\log 3}{16\pi m}.
\ee
Na\"{i}vely, one might expect the leading quartic correction to also appear at one loop, but this diagram is removed by normal-ordering the operator $\norder{\phi^4}$, which simply amounts to a redefinition of the bare mass $m$. The leading contribution, which now arises at two loops, is logarithmically divergent and therefore sensitive to our UV cutoff $\Lambda$. While this UV dependence makes the overall mass shift scheme-dependent, the divergent term is universal, leading to the theoretical prediction
\be
\de m^2_{(\lambda)} = - \fr{\lambda^2}{96\pi^2} \log \Lambda + \text{finite}.
\ee

In section \ref{sec:MassiveResults}, we compute the one-particle mass in the case where the couplings $g$ and $\lambda$ are perturbatively small compared to $m$. For each coupling, we then compare the resulting approximate mass eigenvalue to these predicted mass corrections, as a simple first test of our method for interacting systems.

\begin{figure}[t!]
\begin{center}
\includegraphics[width=0.8\textwidth]{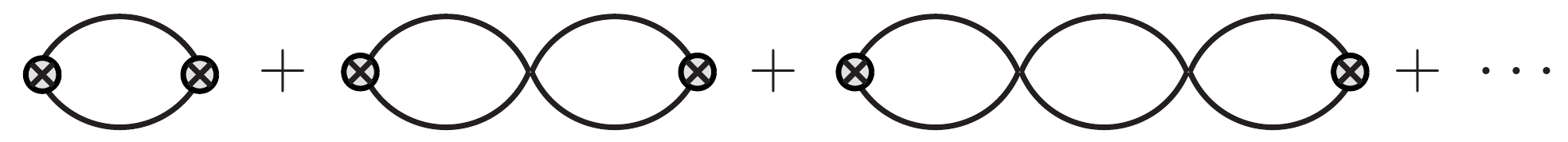}
\caption{The leading contributions to the two-point function $\<\phivec(x) \phivec(0)\>$ in the large-$N$ limit. This infinite set of diagrams can be resummed to obtain an analytic expression for the associated spectral density at finite coupling $\kappa \equiv \lambda N$.}
\label{fig:LargeNSum} 
\end{center}
\end{figure}

We also consider the generalization of our framework to the case of $N$ scalar fields $\phi_i$, with the associated Lagrangian
\be
\Lcal = \Lcal_{\CFT} + \de \Lcal = \half \p_\mu \phi_i \p^\mu \phi_i - \half m^2 \phi_i^2 - \fr{1}{4} \lambda \phi_i^2 \phi_j^2.
\ee
This system greatly simplifies in the large-$N$ limit, such that we can make analytic predictions at finite effective coupling $\kappa \equiv \lambda N$. For example, the dynamical two-point function for the operator
\be
\phivec \equiv \fr{1}{\sqrt{N}} \phi_i^2,
\label{eq:PhiVecDef}
\ee
receives its leading contributions from the sum of loop diagrams in figure \ref{fig:LargeNSum}, with all other contributions suppressed by $1/N$ \cite{Moshe:2003xn}.

Given the simple structure of these diagrams, the resulting geometric series can actually be resummed to obtain the associated spectral density
\be
\rho_{\phivec}(\mu) = \fr{\fr{1}{4\pi\mu}}{\left(1 + \fr{\kappa}{8\pi\mu}\log\Big(\fr{\mu+2m}{\mu-2m}\Big) - \fr{\kappa}{8\pi\mu}\log\Big(\fr{\Lambda+\mu}{\Lambda-\mu}\Big)\right)^2 + \left( \fr{\kappa}{8\mu}\right)^2},
\ee
which holds for any fixed coupling $\kappa$. Note that we have also included potential effects from the UV cutoff $\Lambda$ in the intermediate loops.

While this expression is rather complicated, we can understand its basic structure by considering the massless limit $m\ra0$ and $\Lambda\ra\infty$, with the simplified result
\be
\rho_{\phivec}(\mu) \approx \fr{\fr{1}{4\pi\mu}}{1 + \Big( \fr{\kappa}{8\mu}\Big)^2} \qquad (\mu \gg m).
\ee
Unsurprisingly, at high energies $\mu \gg \kappa$ we recover the free field spectral density of our UV CFT, given in eq.~(\ref{eq:PhiSquareDensity}). However, as we move to the IR, the spectral density is deformed by the presence of interactions, such that we obtain the low-energy behavior
\be
\rho_{\phivec}(\mu) \approx \fr{16\mu}{\pi\kappa^2} \qquad (\mu \ll \kappa).
\ee
We therefore see that the presence of interactions leads to an anomalous dimension for $\phivec$, shifting the scaling dimension from the free value $\De_{\phivec}=1$ in the UV to $\De_{\phivec}=2$ in the IR. This critical behavior can also be extracted from the integrated spectral density, which in the massless case takes the simple form
\be
I_{\phivec}(\mu) = \fr{\mu}{2\pi} - \fr{\kappa}{16\pi} \tan^{-1}\left(\fr{8\mu}{\kappa}\right).
\ee

We use our truncation framework to directly compute the large-$N$ spectral density for $\phivec$ at finite coupling $\kappa$, first for $m=0$ in section~\ref{sec:MasslessResults}, then for the massive case in section~\ref{sec:MassiveResults}. In both cases, we reproduce the expected RG flow and critical behavior, demonstrating the use of this method in studying strongly-coupled dynamics.


\section{Basis of Conformal Casimir Eigenstates}
\label{sec:MasslessBasis}

In order to truncate and diagonalize the Hamiltonian for interacting scalar field theories, we need to construct a complete basis of states. Motivated by AdS/CFT, our proposed basis is defined within the UV CFT of free field theory and consists of eigenstates of the conformal quadratic Casimir. These eigenstates are labeled by the associated Casimir eigenvalue $\Ccal$, Lorentz spin $\ell$, ``spatial'' momentum $\vec{P}$, and invariant mass $\mu^2 \equiv P^2$, and can be built from local operators $\Ocal(x)$ via the Fourier transform
\be
|\Ccal,\ell;\vec{P},\mu\> \equiv \int d^3x \, e^{-iP\cdot x} \Ocal(x)|0\>.
\ee
For Lorentz invariant observables, we are free to choose a particular reference frame with fixed total momentum $\vec{P}$.

Any local operator defines a continuum of Casimir eigenstates with arbitrary invariant mass $\mu$, but we can discretize this basis by introducing the \emph{weight functions} $g_k(\mu)$,
\be
|\Ccal,\ell;\vec{P},k\> \equiv \int d\mu^2 g_k(\mu) \int d^3x \, e^{-iP\cdot x} \Ocal(x)|0\>.
\ee
In order to define a discrete set of orthogonal weight functions, we impose a hard cutoff on this integration over the invariant mass, restricting to $\mu^2 \leq \Lambda^2$. Our basis is therefore organized into \emph{Casimir multiplets}, one for each operator $\Ocal$, consisting of the states $|\Ccal,\ell;\vec{P},k\>$ with $k=0,1,2,...\,$. We now have two independent parameters we can use to truncate this basis: the maximum Casimir eigenvalue $\Cmax$ and the number of weight functions $\kmax$ for each Casimir multiplet.

In this section, we present our basis of Casimir eigenstates for the case of free scalar field theory. These basis states can be written in terms of $n$-particle states using the Fock space expansion for $\phi$, which allows us to express the conformal Casimir as a differential operator acting on functions of particle momenta. Obtaining a complete basis of states is therefore equivalent to finding the eigenfunctions of this differential form for the Casimir. After discussing the general method for constructing these eigenfunctions, we present the explicit form of the basis for states with two and three particles.


\subsection{Constructing the Basis}

Our basis states are built from local operators $\Ocal(x)$, which can all be constructed using the scalar field $\phi$, with the schematic form
\be
\Ocal(x) = \sum_{\{m_n\}} C^\Ocal_{\{m_n\}} \, \p^{m_1} \phi(x) \p^{m_2} \phi(x) \cdots \p^{m_n} \phi(x).
\ee
Because particle number is conserved in the original UV theory, each operator can be written as a sum over terms with a fixed number of $\phi$ insertions. Inserting a complete set of momentum eigenstates, we can rewrite these operators as $n$-particle states weighted by powers of the individual particle momenta,
\be
\begin{split}
|\Ccal,\ell;\vec{P},\mu\> &= \int \fr{d^2p_1 \cdots d^2p_n}{(2\pi)^{2n} 2p_{1-} \cdots 2p_{n-}} \<p_1,\cdots,p_n|\Ccal,\ell;\vec{P},\mu\> \, |p_1,\cdots,p_n\> \\
&= \int \fr{d^2p_1 \cdots d^2p_n}{(2\pi)^{2n} 2p_{1-} \cdots 2p_{n-}} \, (2\pi)^3 \de^3\Big( \sum_i p_i - P\Big) \, F_\Ocal(p) |p_1,\cdots,p_n\>.
\end{split}
\ee
Each Casimir eigenstate is therefore characterized by a specific polynomial of particle momenta,
\be
F_\Ocal(p) \equiv \<\Ocal(0)|p_1,\cdots,p_n\> = \sum_{\{m_n\}} C^\Ocal_{\{m_n\}} \, p_1^{m_1} \cdots p_n^{m_n}.
\label{eq:BasisPoly}
\ee
Naively, one might expect these polynomials $F_\Ocal(p)$ to be functions of all three momentum components. However, because the scalar field $\phi$ satisfies the equation of motion
\be
2p_+ p_- - p_\perp^2 = 0,
\ee
the lightcone energy $p_+$ of each individual particle is not an independent degree of freedom. The basis functions $F_\Ocal(p)$ can therefore be written solely in terms of the spatial momenta $p_-,p_\perp$.

To determine the structure of these basis functions, we can write the conformal Casimir as a differential operator and then solve for the resulting eigenfunctions. The quadratic Casimir of the conformal group is defined in terms of the conformal generators as
\be
\Ccal \equiv -D^2 - \half (P_\mu K^\mu + K_\mu P^\mu) + \half L_{\mu\nu} L^{\mu\nu}.
\ee
As discussed in appendix \ref{app:CasimirBasis}, we can use the transformation properties of $\phi$ to derive the differential form for $\Ccal$ acting on the generic $n$-particle function in eq.~(\ref{eq:BasisPoly}), obtaining
\be
\begin{split}
\Ccal = &\sum_{i<j} \Bigg[ - 2p_{i-} p_{j-} (\p_{i-} - \p_{j-})^2 + (p_{i-} - p_{j-}) (\p_{i-} - \p_{j-}) + (p_{i\perp} - p_{j\perp}) (\p_{i\perp} - \p_{j\perp}) \\
& \, - \, 2(p_{i-} p_{j\perp} + p_{i\perp} p_{j-}) (\p_{i-} - \p_{j-}) (\p_{i\perp} - \p_{j\perp}) - \fr{(p_{i-} p_{j\perp} + p_{i\perp} p_{j-})^2}{2p_{i-} p_{j-}} (\p_{i\perp} - \p_{j\perp})^2 \Bigg] \\
& \, + \, \fr{1}{4}n(n-6),
\end{split}
\label{eq:GeneralCasimir}
\ee
where the sum is over all particle pairs. In deriving this form for $\Ccal$ we have used the fact that the corresponding local operators are built from the scalar field $\phi$, with scaling dimension $\De_\phi = \half$.

We now need to find all eigenfunctions of this operator with eigenvalues $\Ccal \leq \Cmax$. Fortunately, we can simplify this procedure by noting that these Casimir eigenfunctions can be organized into representations of the Lorentz group. Because we are using lightcone coordinates, our basis no longer has manifest Lorentz symmetry, such that different components of the same spin multiplet correspond to distinct eigenfunctions $F_\Ocal(p)$ with the same Casimir eigenvalue. However, these distinct basis functions are still related by Lorentz transformations. For each spin multiplet, we therefore only need to obtain the basis function for a single component, then act with the Lorentz generators $L_{\mu\nu}$ to obtain the remaining components.

Specifically, we need to act with a combination of \Poincare generators which preserves the total momentum $P$. For $d=3$, there is one such combination of generators, which is the Pauli-Lubanski pseudoscalar \cite{Lubanski}
\be
W \equiv \half \epsilon^{\mu\nu\rho} P_\mu L_{\nu\rho}.
\ee
This operator is the generator of the Wigner little group, as it automatically commutes with the total momentum,
\be
\comm{W}{P_\mu} = 0.
\ee
We can again use the transformation properties of $\phi$ to derive the differential form for $W$, obtaining
\be
W = P_- \sum_i \left( p_{i\perp} \p_{i-} + \fr{p_{i\perp}^2}{2p_{i-}} \p_{i\perp} \right) - P_\perp \sum_i p_{i-} \p_{i-} - \sum_i \fr{p_{i\perp}^2}{2p_{i-}} \sum_j p_{j-} \p_{j\perp},
\ee
where each sum is over particle number.

We thus have a general procedure for constructing a basis of Casimir eigenstates. For each local operator with spin, we need to find the $\Ccal$ eigenfunction for only one of the components, then act with $W$ to generate the remaining basis functions for that spin multiplet. In the following two subsections, we implement this procedure explicitly for the two- and three-particle basis states, obtaining the full set of Casimir eigenfunctions.

However, each basis function $F_\Ocal(p)$ is still associated with a continuum of Casimir eigenstates, parameterized by the invariant mass $\mu^2 \equiv P^2$. We can discretize these Casimir multiplets by introducing the weight functions $g_k(\mu)$, leading to the basis states
\be
|\Ccal,\ell;\vec{P},k\> \equiv \int d\mu^2 g_k(\mu) \int \fr{d^2p_1 \cdots d^2p_n}{(2\pi)^{2n} 2p_{1-} \cdots 2p_{n-}} \, (2\pi)^3 \de^3\Big( \sum_i p_i - P\Big) \, F_\Ocal(p) |p_1,\cdots,p_n\>.
\ee
We can construct a complete basis of weight functions $g_k(\mu)$ by considering the inner product between two such Casimir eigenstates, which leads to a natural measure of integration. The resulting basis then consists of the set of functions $g_k(\mu)$ which are orthogonal with respect to this measure.

This inner product actually diverges as the invariant mass $\mu \ra \infty$. In order to define a normalizable basis, we therefore need to impose a UV cutoff of some kind to regulate the inner product. Our proposed UV regulator is a hard cutoff $\Lambda$ on the invariant mass,
\be
\mu^2 \leq \Lambda^2.
\ee
This Lorentz invariant cutoff limits the range of integration to a finite interval, such that we can obtain a discrete basis of polynomials.

The resulting basis states are therefore characterized by the Casimir eigenvalue $\Ccal$, spin $\ell$, and the degree of the weight function $k$. We can then truncate this basis to only those states with $\Ccal \leq \Cmax$ and $k \leq \kmax$.

Though here we restrict ourselves to $d=3$, it is important to note that this general procedure can be applied in any number of dimensions. The method for obtaining the conformal Casimir differential operator presented in appendix~\ref{app:CasimirBasis} can be repeated for higher $d$ by including the additional conformal generators associated with the new transverse directions. Similarly, there will be additional Pauli-Lubanski generators needed to obtain the full spin multiplet associated with each local operator. For example, in $d=4$ there are two independent generators,
\be
W_1 \equiv P_+ L_{-\perp_1} - P_- L_{+\perp_1}, \quad W_2 \equiv P_+ L_{-\perp_2} - P_- L_{+\perp_2}.
\ee
As we can see, there is one such generator for each transverse direction. In general, one therefore needs to use the resulting Casimir differential operator to find the eigenfunction for a single component of each spin multiplet, then act with the various $W_i$ to construct the remaining basis states.


\subsection{Two-Particle States}

As an example of our general procedure, let's consider the two-particle case. These states are built from operators with two insertions of $\phi$, which take the schematic form
\be
\Ocal^{(2)}_\ell(x) \sim \phi(x) \lrpar_{\mu_1} \cdots \lrpar_{\mu_\ell} \phi(x) - \textrm{traces}.
\ee
Because $\phi$ satisfies the equation of motion,
\be
\p^2 \phi(x) = 0,
\ee
these operators correspond to higher-spin conserved currents, one for each spin $\ell$, which in $d=3$ have two independent components.

To obtain the basis functions for these operators, we only need to find the Casimir eigenfunction for one of the two spin components, then act with the Pauli-Lubanski generator to obtain the other. Without loss of generality, we can choose this first component for each current to be the ``all minus'' term,
\be
\Ocal^{(2)}_{\ell-}(x) \sim \phi(x) \lrpar_- \cdots \lrpar_- \phi(x).
\ee
The basis functions for these particular operators therefore only depend on the lightcone momenta $p_-$. After fixing the total momentum by imposing the constraints
\be
p_{1-} = p_-, \quad p_{2-} = P_- - p_-, \quad p_{1\perp} = p_{\perp}, \quad p_{2\perp} = -p_\perp,
\ee
we can then find the all minus two-particle states by solving for the eigenfunctions of the simplified differential operator,
\be
\Ccal^{(2)}_- = -2 - 2p_-(P_- - p_-) \fr{\p^2}{\p p_-^2} + (2p_- - P_-) \fr{\p}{\p p_-},
\ee
where the subscript $\Ccal_-$ indicates this is only the form of the Casimir when acting on the all minus component $\Ocal_{\ell-}$. Note that this operator has no derivatives with respect to the total momentum $P_-$, which is consistent with the fact that the Casimir commutes with all of the conformal generators.

Given this simple form for the conformal Casimir, we can easily solve for the resulting eigenfunctions, which consist of Jacobi polynomials in $p_-$,
\be
F^{(2)}_{\ell-}(p) = P^{(-\half,-\half)}_\ell\Big(\fr{2p_-}{P_-} - 1\Big).
\label{eq:CasimirState2P}
\ee
As expected, there is a single basis function for each spin $\ell$, with the associated eigenvalue
\be
\Ccal^{(2)}_\ell = (2\De_\phi+\ell)(2\De_\phi+\ell-3) + \ell(\ell+1) = 2\ell^2-2.
\ee

We can also confirm that these expressions match the precise form for the primary operators $\Ocal^{(2)}_{\ell}(x)$. For example, consider the $\ell=2$ state, which simply corresponds to the stress-energy tensor
\be
T_{\mu\nu} = \p_\mu\p_\nu\phi^2 + 2\eta_{\mu\nu} \p_\alpha\phi \p^\alpha \phi - 8\p_\mu\phi \p_\nu \phi.
\ee
Focusing on the component $T_{--}$, we can then obtain the momentum space form
\be
T_{--} = (p_{1-} + p_{2-})^2 - 8p_{1-} p_{2-} = P_-^2 - 8 p_-(P_- - p_-),
\ee
which matches the associated Jacobi polynomial,
\be
P^{(-\half,-\half)}_2\Big(\fr{2p_-}{P_-} - 1\Big) = \fr{3}{8P_-^2} \Big( P_-^2 - 8 p_-(P_- - p_-) \Big).
\ee

Just like for the conformal Casimir, we can derive a differential form for the Pauli-Lubanski generator,
\be
W = \fr{p_\perp^2 (P_- - 2p_-)}{2p_-(P_- - p_-)} \fr{\p}{\p p_\perp} + p_\perp \fr{\p}{\p p_-}.
\ee
For each conserved current, we can then act with $W$ on the basis function $F_{\ell-}$ to obtain the other independent component,
\be
F^{(2)}_{\ell\perp}(p) = \fr{p_\perp}{P_-} \, P^{(\half,\half)}_{\ell-1}\Big(\fr{2p_-}{P_-} - 1\Big).
\ee
Because $W$ commutes with $\Ccal$, these new basis functions are also eigenfunctions of the conformal Casimir, with the same $\ell$-dependent eigenvalues.

We can again check that these expressions match the expected form for each operator. For the stress-energy tensor, this new polynomial corresponds to the independent component,
\be
T_{-\perp} = (p_{1-} + p_{2-})(p_{1\perp} + p_{2\perp}) - 4 p_{1-} p_{2\perp} - 4 p_{1\perp} p_{2-} = 4 p_\perp (2p_- - P_-).
\ee
This expression again matches the corresponding Jacobi polynomial,
\be
\fr{p_\perp}{P_-} \, P^{(\half,\half)}_1\Big(\fr{2p_-}{P_-} - 1\Big) = \fr{3}{8P_-^2} \Big( 4 p_\perp (2p_- - P_-) \Big).
\ee
The remaining components of $T_{\mu\nu}$ do not correspond to independent Casimir eigenfunctions, but are related to $T_{--}$ and $T_{-\perp}$ by the equations of motion. For example, in this particular reference frame $T_{\perp\perp} \sim P^2 T_{--}$.

The basis functions $F_{\ell-}$ and $F_{\ell\perp}$ are respectively even/odd under the parity transformation
\be
p_\perp \ra -p_\perp.
\ee
In this work, we only consider interactions which preserve parity, such that we can focus solely on the even sector built from all minus states. From now on, we will therefore suppress the subscript in $F_{\ell-}$, with the understanding that we are always referring to the parity-even component.

While we now have a complete basis of Casimir eigenstates, we must impose an additional restriction on the resulting basis functions. These operators are built from a single scalar field, which means that the two particles in a given state are indistinguishable. Our basis polynomials must therefore be invariant under the exchange $p_1 \lra p_2$, or equivalently,
\be
p_- \ra P_- - p_-, \quad p_\perp \ra -p_\perp.
\ee

Requiring our states to be symmetric under this exchange reduces our basis to only operators with even spin $\ell$. More generally, the $n$-particle basis states must be invariant under the full symmetric group $S_n$, which corresponds to all permutations of particle momenta. The details of this symmetrization procedure are discussed in appendix \ref{app:Symmetrize}.

Each of these symmetric Casimir eigenfunctions is associated with an infinite number of basis states, which take the form
\be
|\ell;k\> \equiv \int d\mu^2 g^{(2)}_k(\mu) \int \fr{dp_- \, dp_\perp}{(2\pi)^2 4 p_- (P_- - p_-)} (2\pi) \, \de\bigg(\fr{p_\perp^2 P_-}{2p_-(P_- - p_-)} - \fr{\mu^2}{2P_-}\bigg) \, F^{(2)}_\ell(p) |p,P-p\>.
\ee
In order to determine the form of the weight functions $g_k(\mu)$, we need to consider the resulting inner product
\be
\<\ell;k|\ell';k'\> = 2!\int \fr{d\mu^2}{\mu} \, g^{(2)}_k(\mu) \, g^{(2)}_{k'}(\mu) \int \fr{dp_-}{2\sqrt{p_- (P_- - p_-)}} \, F^{(2)}_\ell(p) \, F^{(2)}_{\ell'}(p).
\ee
The Casimir basis functions are automatically orthogonal with respect to this Lorentz invariant measure, such that we can just focus on the integral over $\mu^2$,
\be
\int \fr{d\mu^2}{\mu} \, g^{(2)}_k(\mu) \, g^{(2)}_{k'}(\mu) = \de_{kk'}.
\label{eq:2pWeightIntegral}
\ee
After imposing the UV cutoff $\mu^2 \leq \Lambda^2$, we find that the complete set of orthogonal weight functions consists of the Legendre polynomials,
\be
g^{(2)}_k(\mu) = P_{2k}\Big(\fr{\mu}{\Lambda}\Big).
\ee

We now have a complete, discrete basis of two-particle Casimir eigenstates, parameterized by the spin $\ell$ and degree $k$. We can use this basis to construct the lightcone Hamiltonian $P_+$ for various interactions, truncate at some $\Cmax$ and $\kmax$, then diagonalize the resulting matrix to obtain an approximate IR spectrum.


\subsection{Three-Particle States}

This same approach can then be applied to states with higher particle number. In this work, we only consider basis states with up to three particles, but it is straightforward to see how the structure of the basis generalizes from there.

Because the full three-particle conformal Casimir commutes with the Casimir operator associated with a two-particle subsector, operators with three insertions of $\phi$ can be built recursively from the two-particle operators of the previous section, taking the schematic form
\be
\Ocal^{(3)}_\ell(x) \sim \phi(x) \lrpar_{\mu_1} \cdots \lrpar_{\mu_{\ell_1}} \Ocal^{(2)}_{\ell_2}(x) - \textrm{traces}.
\label{eq:3ParticleOperator}
\ee
The two-particle operator is a conserved current, such that the scaling dimension and Casimir eigenvalue of the full operator are automatically fixed by the spin $\ell = \ell_1 + \ell_2$. However, the resulting three-particle operators are \emph{not} conserved, with $2\ell+1$ independent components for each $\Ocal^{(3)}_\ell$.

Just like in the two-particle case, we only need to find the Casimir eigenfunction for one component of each operator, then act with $W$ to obtain the other spin components. Fixing the reference frame leaves only two free momenta $p_1$ and $p_2$, since
\be
p_{3-} = P_- - p_{1-} - p_{2-}, \quad p_{3\perp} = -p_{1\perp} - p_{2\perp}.
\ee
We can then find the all minus component of each multiplet by solving for the eigenfunctions of the simplified differential operator,
\be
\begin{split}
\Ccal^{(3)}_- &= -\fr{9}{4} - 2p_{1-}(P_- - p_{1-}) \fr{\p^2}{\p p_{1-}^2} - 2p_{2-}(P_- - p_{2-}) \fr{\p^2}{\p p_{2-}^2} \\
& \, + \, 4 p_{1-} p_{2-} \fr{\p}{\p p_{1-}} \fr{\p}{\p p_{2-}} + (3p_{1-} - P_-) \fr{\p}{\p p_{1-}} + (3p_{2-} - P_-) \fr{\p}{\p p_{2-}}.
\end{split}
\ee
The resulting eigenfunctions are again Jacobi polynomials,
\be
F^{(3)}_{\ell}(p) = \Big(1 - \fr{p_{1-}}{P_-}\Big)^{\ell_2} \, P^{(2\ell_2,-\half)}_{\ell_1}\Big(\fr{2p_{1-}}{P_-} - 1\Big) \, P^{(-\half,-\half)}_{\ell_2}\Big(\fr{2p_{2-}}{P_- - p_{1-}} - 1\Big).
\label{eq:CasimirState3P}
\ee
As we can see, the last term in this expression is a two-particle Casimir eigenfunction with spin $\ell_2$ and total momentum $P_- - p_{1-}$, confirming the recursive structure described in eq.~(\ref{eq:3ParticleOperator}). This behavior generalizes to the higher particle states, with an additional Jacobi polynomial for each new particle.

These three-particle eigenfunctions are parameterized by two non-negative integers $\ell_1$ and $\ell_2$, with the associated Casimir eigenvalues,
\be
\Ccal^{(3)}_{\ell} = (3\De_\phi+\ell_1+\ell_2)(3\De_\phi+\ell_1+\ell_2-3) + (\ell_1+\ell_2)(\ell_1+\ell_2+1) = 2\ell^2 + \ell - \fr{9}{4}.
\ee

Now that we have the all minus components, we can obtain the remaining eigenfunctions by acting with the Pauli-Lubanski generator,
\be
W^{(3)} = \left( \fr{p_{1\perp}^2}{2p_{1-}} - \fr{\mu^2 p_{1-}}{2P_-^2} \right) \fr{\p}{\p p_{1\perp}} + \left( \fr{p_{2\perp}^2}{2p_{2-}} - \fr{\mu^2 p_{2-}}{2P_-^2} \right) \fr{\p}{\p p_{2\perp}} + p_{1\perp} \fr{\p}{\p p_{1-}} + p_{2\perp} \fr{\p}{\p p_{2-}}.
\ee
We can parameterize these additional components by introducing the new label $m_\perp$,
\be
F^{(3)}_{\ell,m_\perp}(p) \sim W^{m_\perp} F^{(3)}_{\ell,0}(p),
\ee
where $m_\perp$ ranges from $0$ to $2\ell$. As discussed in appendix~\ref{app:CasimirBasis}, it will be simpler to express the resulting basis functions in terms of invariant masses, rather than the transverse momenta $p_{1\perp}$, $p_{2\perp}$. We can then define the new variables,
\be
\mu_1^2 \equiv \mu^2 \cos^2\theta = \mu^2 - (p_2 + p_3)^2, \quad \mu_2^2 \equiv \mu^2 \sin^2\theta = (p_2+p_3)^2.
\ee
The second variable $\mu_2$ is the invariant mass of the two-particle operator built from $p_2$ and $p_3$.

Acting with $W$ on the all minus basis functions, we find that the new $m_\perp \neq 0$ basis states take the schematic form
\be
F^{(3)}_{\ell,m_\perp}(p) = \mu^{m_\perp} \Big( f(p_-) \cos m_\perp\theta + \bar{f}(p_-) \sin m_\perp\theta \Big),
\label{eq:SchematicMPerp}
\ee
where the functions $f,\bar{f}$ generally consist of Jacobi polynomials in $p_-$. The resulting spin multiplets are therefore built from Casimir eigenfunctions with periodicity in $\theta$ set by $m_\perp$.

Following this procedure, we can build up a complete basis of Casimir eigenstates, parameterized by the three labels $\ell_1,\ell_2,m_\perp$. However, we then need to restrict this basis to states which are invariant under the exchanges,
\be
p_1 \lra p_2, \quad p_2 \lra p_3, \quad p_3 \lra p_1.
\ee
Because the Pauli-Lubanski generator is manifestly symmetric under these permutations, we only need to symmetrize the $m_\perp=0$ states, as the remaining components generated by $W$ will automatically be symmetric. A more detailed discussion of this symmetrization procedure can be found in appendix~\ref{app:Symmetrize}.

Once we have obtained the set of symmetric Casimir eigenfunctions, we then need to find the associated three-particle weight functions $g_k(\mu)$. Just like the two-particle case, we can do so by considering the inner product,
\be
\begin{split}
&\<\ell,m_\perp;k|\ell',m_\perp';k'\> \\
&\qquad = 3!\int d\mu^2 g^{(3)}_k(\mu) \, g^{(3)}_{k'}(\mu) \int \fr{dp_{1-} \, dp_{2-} \, d\theta}{32\pi^2\sqrt{p_{1-} p_{2-} P_- (P_- - p_{1-} - p_{2-})}} F^{(3)}_{\ell,m_\perp}(p) F^{(3)}_{\ell',m_\perp'}(p).
\end{split}
\ee
While the Casimir eigenfunctions are automatically orthogonal with respect to this measure, we can see from eq.~(\ref{eq:SchematicMPerp}) that each basis function comes with an overall factor of $\mu^{m_\perp}$. This factor then modifies the integration measure for $g_k(\mu)$,
\be
\int d\mu^2 \mu^{2m_\perp} g^{(3)}_k(\mu) \, g^{(3)}_{k'}(\mu) = \de_{kk'}.
\ee
After imposing the UV cutoff $\Lambda$, the resulting basis of weight functions consists of the Zernike polynomials
\be
g^{(3)}_k(\mu) = R^{m_\perp}_{2k+m_\perp}\Big(\fr{\mu}{\Lambda}\Big).
\ee

As shown in appendix~\ref{app:MassiveMatrix}, we only consider interactions which preserve periodicity in $\theta$, allowing us to focus solely on the $m_\perp=0$ sector. More generally, though, one would need to include the full set of $m_\perp$ states to obtain a complete basis.


\section{Warmup: Single Casimir Multiplet}
\label{sec:MasslessResults}

We now have a complete basis of Casimir eigenstates for scalar field theory in $2+1$ dimensions. As a simple warmup, we can first use this basis to reproduce the spectral densities of local operators in the original UV CFT. To do so, we need to evaluate the matrix elements
\be
\<\Ccal,\ell;\vec{P},k|M^2|\Ccal',\ell';\vec{P}',k'\>,
\ee
then diagonalize the resulting truncated matrix. However, in the original free CFT the invariant mass is simply given by
\be
M^2 = \mu^2.
\ee
The CFT Hamiltonian therefore only mixes the weight functions $g_k(\mu)$ within a given Casimir multiplet, but \emph{does not} mix distinct Casimir eigenfunctions $F_\Ocal(p)$. This block-diagonal structure for the Hamiltonian naturally follows from the fact that it commutes with the conformal Casimir.

We thus only need to use a \emph{single} Casimir multiplet to calculate the CFT spectral density for each operator $\Ocal(x)$. This simplification allows us to first focus solely on the effects of truncating the size of the individual multiplets, set by $\kmax$. In this section, we reproduce the spectral densities of $\phi^2$ and $\phi^3$, comparing our truncation results to the known analytic expressions. We then generalize our basis to the case of distinct fields in order to study the spectral density of $\phivec$ in the $O(N)$ model. In the limit of large-$N$, the addition of a quartic self-interaction does not mix distinct Casimir multiplets, consistent with expectations from AdS. We therefore again only need a single multiplet to reproduce the full, strongly-coupled RG flow for $\phivec$.


\subsection{CFT Spectral Densities}
\label{sec:MasslessDensities}

To calculate the spectral density for any operator $\Ocal(x)$, we need to first truncate and diagonalize the invariant mass operator
\be
M^2 \equiv 2P_+ P_- - P_\perp^2.
\ee
Because our basis is built from total momentum eigenstates, this is equivalent to diagonalizing the lightcone Hamiltonian $P_+$. We can then use the Hamiltonian matrix elements calculated in appendix~\ref{app:MasslessMatrix} to construct the matrix form of $M^2$ in our Casimir basis.

Diagonalizing this matrix gives us a spectrum of approximate mass eigenstates $|\mu_i\>$. We can then reproduce the integrated spectral density for $\Ocal$ by calculating its overlap with these approximate eigenstates,
\be
I_\Ocal(\mu) \equiv \sum_{\mu_i\leq\mu} |\<\Ocal(0)|\mu_i\>|^2.
\ee
Since the CFT Hamiltonian doesn't mix distinct operators, we only need the Casimir multiplet associated with $\Ocal$, as this operator has no overlap with any other multiplets.

For example, consider the spectral density for the operator $\phi^2$. Its corresponding overlap with our two-particle basis states is given by
\be
\<\phi^2(0)|\ell;k\> = 2! \int \fr{d\mu^2}{\mu} g_k^{(2)}(\mu) \int \fr{dp_-}{4\pi\sqrt{p_- (P_- - p_-)}} \, F^{(2)}_\ell(p) = \sqrt{\fr{\Lambda}{2\pi}} \, \de_{\ell,0} \, \de_{k,0}.
\label{eq:2pOverlap}
\ee
This inner product therefore projects onto a single state in our basis: the $k=0$ weight function in the $\ell=0$ Casimir multiplet, which precisely corresponds to the primary operator $\phi^2$.

\begin{figure}[t!]
\begin{center}
\includegraphics[width=0.8\textwidth]{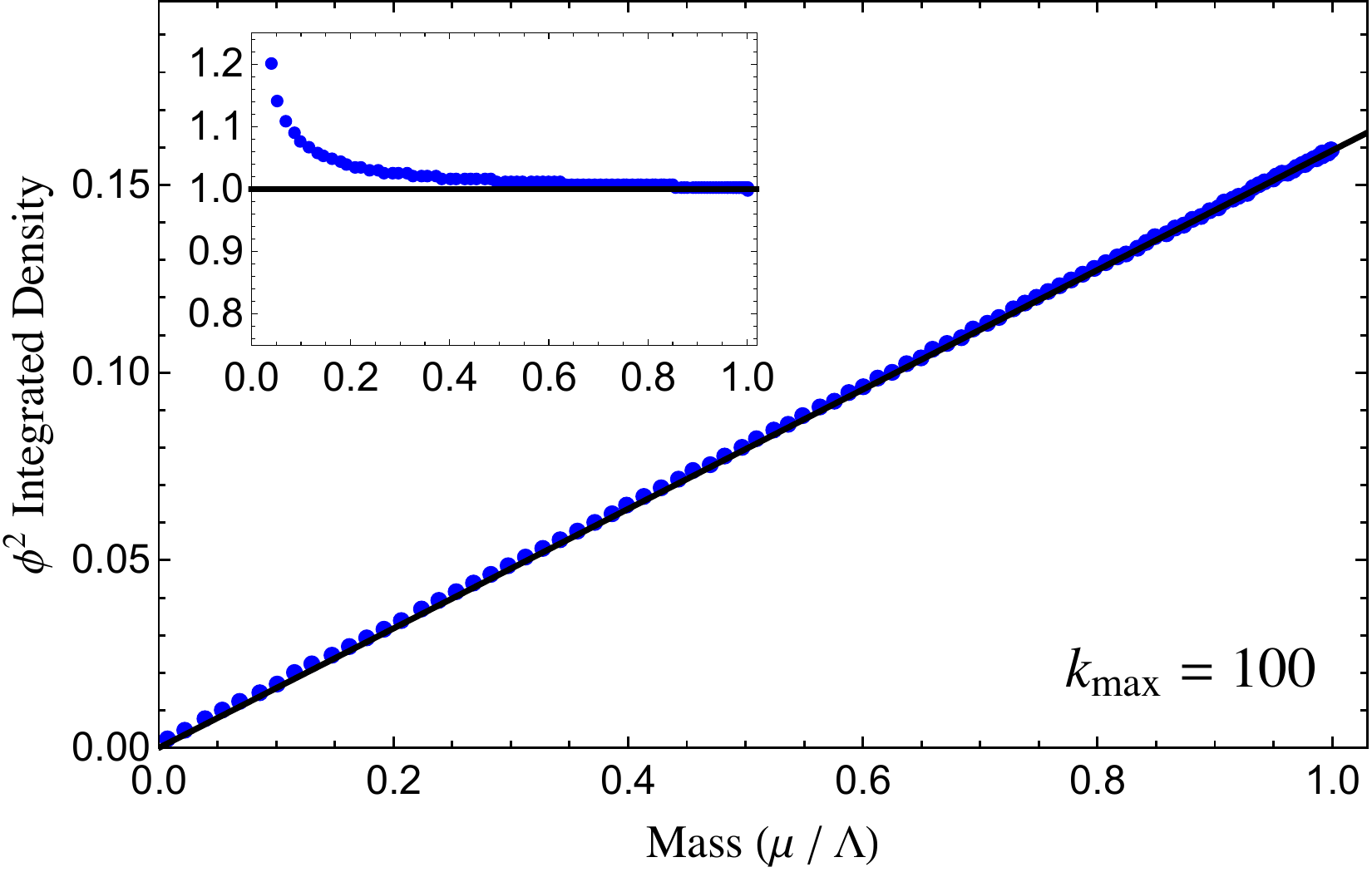}
\caption{Integrated spectral density for $\phi^2$ in massless free field theory, both the raw value (main plot) and normalized by the theoretical prediction (inset). The conformal truncation results (blue dots) are calculated from the two-particle $\ell=0$ multiplet with $\kmax=100$, and compared to the known analytic expression (black line).}
\label{fig:2pMasslessDensity} 
\end{center}
\end{figure}

We therefore only need the $\ell=0$ multiplet to compute the $\phi^2$ spectral density. However, we still need to truncate the size of the multiplet at some level $\kmax$, only keeping weight functions with $k \leq \kmax$.

The integrated spectral density obtained with $\kmax=100$ is shown in figure \ref{fig:2pMasslessDensity}, compared with the theoretical prediction
\be
I_{\phi^2}(\mu) = \int_0^{\mu^2} d\mu^{\prime \, 2} \, \rho(\mu') = \fr{\mu}{2\pi}.
\ee
While this CFT example is somewhat simple, its structure provides a useful reference point for understanding our later results, where conformal symmetry is broken.

As we can see from the main plot, the truncation results (blue dots) successfully reproduce the analytic expression (black line), confirming that only the $\ell=0$ multiplet is needed to construct the $\phi^2$ spectral density. We also see that the resulting spectrum stops precisely at $\mu = \Lambda$, due to the hard cutoff in invariant mass. However, if we normalize our results by the theoretical prediction, as shown in the subplot, we see that the truncation data begins to deviate from the analytic expression for low mass eigenvalues. This deviation arises due to the finite size of our basis, which leads to an effective IR cutoff $\LambdaIR$.

We can easily understand the appearance of this IR scale from the large $k$ behavior of the weight functions,
\be
P_{2k}\Big(\fr{\mu}{\Lambda}\Big) \approx \fr{1}{\sqrt{\pi k}} \cos \Big(\fr{2k\mu}{\Lambda}\Big) \sim \cos\Big(\fr{2\mu}{\LambdaIR}\Big) \qquad (k \ra \infty, \mu \ll \Lambda).
\ee
The parameter $\kmax$ therefore sets the intrinsic resolution of our truncation results, which manifests itself in the IR scale
\be
\LambdaIR \sim \fr{\Lambda}{\kmax}.
\ee

We can also see this emergent IR scale directly in the approximate mass eigenstates, which are simply delta functions in $\mu$. Our truncated basis states combine to reproduce these mass eigenstates as we increase $\kmax$ via the identity
\be
\sum_{k=0}^{\kmax} (2k + \tfr{1}{2}) \, P_{2k}(\mu_i) \, P_{2k}(\mu) \approx \de(\mu - \mu_i) \qquad (\kmax \gg 1).
\ee
Due to our truncation of the Hilbert space at $\kmax$, the resulting approximate mass eigenstates have a finite width, which corresponds to the cutoff $\LambdaIR$. Because of this inherent resolution, we expect to find $O(1)$ deviations in the spectral density at $\mu \sim \LambdaIR$. This matches the behavior in figure~\ref{fig:2pMasslessDensity}, where we begin to see $O(0.1)$ deviations at $\mu \sim 10\LambdaIR$. We can always improve this resolution by simply increasing $\kmax$.

This basic procedure can easily be repeated for any other operator: select the appropriate Casimir multiplet, diagonalize the truncated mass matrix, and compute the cumulative overlap with the resulting approximate mass eigenstates. For example, we can use the $\ell=2$ multiplets to construct the spectral density for each component of $T_{\mu\nu}$.

\begin{figure}[t!]
\begin{center}
\includegraphics[width=0.8\textwidth]{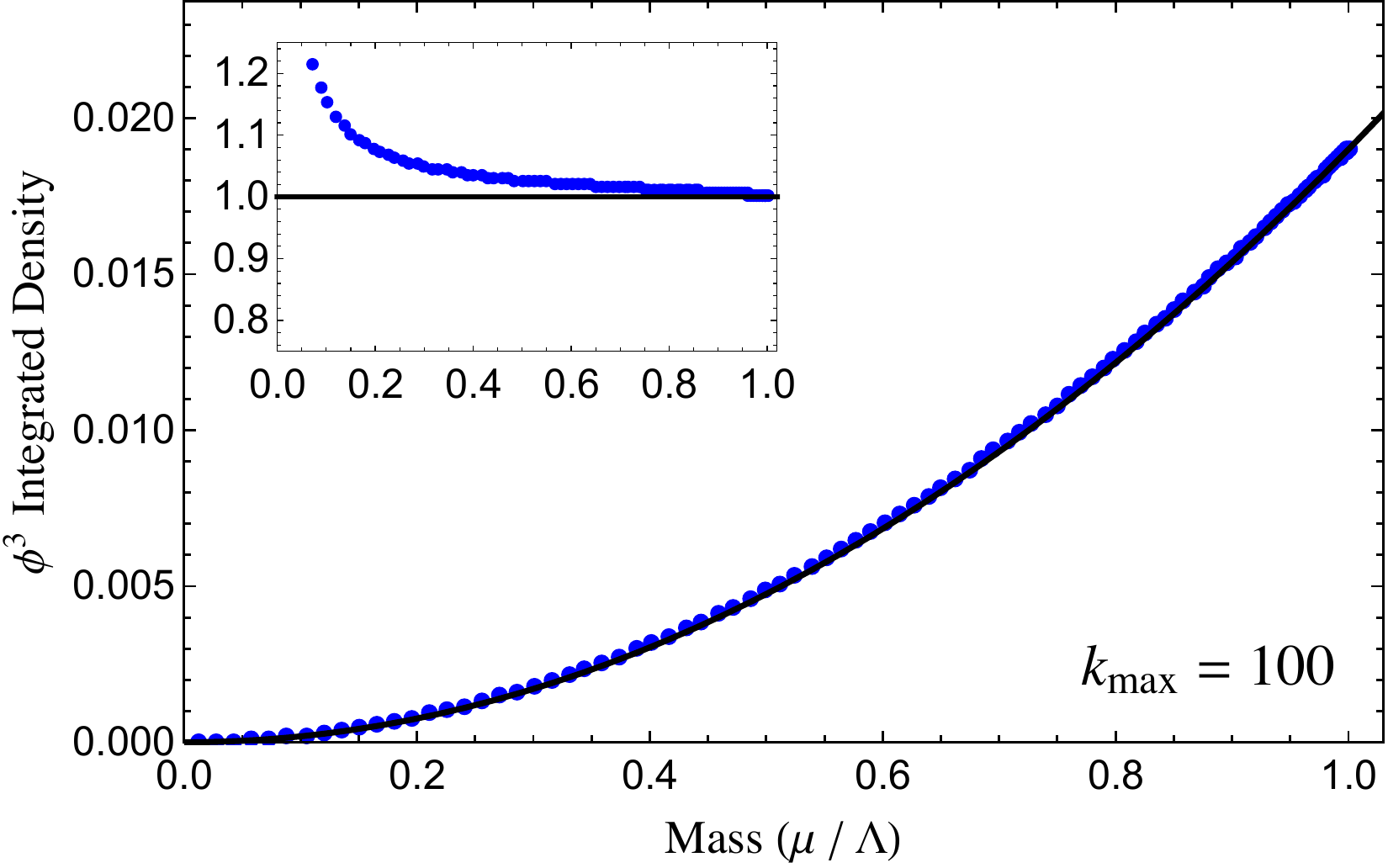}
\caption{Integrated spectral density for $\phi^3$ in massless free field theory, both the raw value (main plot) and normalized by the theoretical prediction (inset). The conformal truncation results (blue dots) are calculated from the three-particle $\vec{\ell},m_\perp=0$ multiplet with $\kmax=100$, and compared to the known analytic expression (black line).}
\label{fig:3pMasslessDensity} 
\end{center}
\end{figure}

As an example with higher particle number, let's next consider the spectral density for $\phi^3$. Its overlap with the three-particle Casimir eigenstates is given by
\be
\begin{split}
\<\phi^3(0)|\ell,m_\perp;k\> &= 3!\int d\mu^2 g^{(3)}_k(\mu) \int \fr{dp_{1-} \, dp_{2-} \, d\theta}{64\pi^3\sqrt{p_{1-} p_{2-} P_- (P_- - p_{1-} - p_{2-})}} \, F^{(3)}_{\ell,m_\perp}(p) \\
&= \sqrt{\fr{3\Lambda^2}{16\pi^2}} \, \de_{\ell_1,0 \,} \de_{\ell_2,0 \,} \de_{m_\perp,0 \,} \de_{k,0}.
\end{split}
\ee
Just like before, this inner product projects onto the lowest state in the Casimir multiplet with $\vec{\ell} = m_\perp = 0$, which corresponds to the primary operator $\phi^3$. We therefore only need to consider states from this multiplet to construct the corresponding spectral density, after truncating the multiplet at some level $\kmax$.

The integrated spectral density for $\kmax = 100$ is shown in figure~\ref{fig:3pMasslessDensity}, compared with the prediction
\be
I_{\phi^3}(\mu) = \fr{3\mu^2}{16\pi^2}.
\ee
Again, we find that the approximate conformal truncation results agree with the analytic expression up to the UV cutoff $\Lambda$, and begin to deviate as we approach $\LambdaIR \sim \Lambda/\kmax$.

We've thus demonstrated the general method for constructing spectral densities in the original UV CFT. For any operator $\Ocal(x)$, we simply need to use the weight functions with $k \leq \kmax$ to construct approximate delta functions in $\mu$, then calculate the overlap for each mass eigenstate using the associated Casimir eigenfunction.


\subsection{Large-$N$ RG Flow}
\label{sec:MasslessLargeN}

Extending our basis to the case of $N$ distinct scalar fields is rather straightforward. Each massless field $\phi_i$ can again be written in terms of Fock space modes, which leads to the same inner product and resulting momentum-dependence for the Casimir eigenstates. However, rather than restrict the basis to only permutation symmetric states, we can instead organize them into representations of the $O(N)$ flavor group. In this work, we are specifically interested in the leading large-$N$ behavior of the operator $\phivec$, such that we can restrict our basis to the two-particle $O(N)$ singlets,
\be
|\ell;k\> \equiv \int d\mu^2 \, g_k(\mu)\int \fr{d^2p_1 d^2p_2}{(2\pi)^{4} 2p_{1-} 2p_{2-}} \, (2\pi)^3 \de^3\Big( p_1 + p_2 - P\Big) \, F^{(2)}_\ell(p) \, \sum_{i=1}^N|p_{1,i},p_{2,i}\>.
\ee
Because of this simple flavor structure, these singlet states are still symmetric under the permutation $p_1 \lra p_2$. As discussed in appendix \ref{app:CasimirBasis}, the basis of two-particle $O(N)$ singlets is therefore \emph{identical} to the two-particle basis for a single scalar field, up to a slight change in the overall normalization.

We can therefore use the same basis of Casimir eigenstates to build and truncate the corresponding $N$ field Hamiltonian,
\be
P_+^{(\CFT)} = \int \fr{d^2p}{(2\pi)^2} \, a^\dagger_{p,i} a_{p,i} \, \fr{p_\perp^2}{2p_-}.
\ee
As we can see, the kinematic structure of this operator is the same as the one-particle case, which means that this Hamiltonian again doesn't mix distinct Casimir multiplets. The calculation of the integrated spectral density for $\phivec$ is therefore identical to that of the previous section, matching the $\phi^2$ results shown in figure~\ref{fig:2pMasslessDensity}.

We can then perturb the UV CFT by adding the quartic interaction,
\be
\de \Lcal = - \fr{1}{4} \lambda \phi_i^2 \phi_j^2.
\ee
This interaction clearly mixes states with different particle number. However, if we take the limit $N\ra\infty$ with the combination $\kappa \equiv \lambda N$ fixed, interactions which change particle number are suppressed by $1/N$, such that we can focus on the simpler Hamiltonian term,
\be
\de P^{(\lambda)}_+ = \fr{\lambda}{2} \int \fr{d^2p \, d^2q \, d^2k}{(2\pi)^6 \sqrt{8p_- q_- k_-}} \, \fr{a^\dagger_{p,i} a^\dagger_{q,i} a_{k,j} a_{p+q-k,j}}{\sqrt{2(p_- + q_- - k_-)}}.
\ee
Because this dominant interaction preserves particle number, we only need to consider the resulting two-particle matrix elements,
\be
\<\ell;k|\de M^2|\ell';k'\> = \fr{\lambda N}{2} \<\ell;k|\phivec(0)\> \<\phivec(0)|\ell';k'\> = \fr{\kappa\Lambda}{4\pi} \, \de_{\ell,0} \, \de_{k,0} \cdot \de_{\ell',0} \, \de_{k',0}.
\ee
As we can see, these matrix elements clearly factorize into two independent terms, each of which projects onto the basis state $\phivec$. In the large-$N$ limit, this interaction therefore \emph{only} affects the $\phivec$ Casimir multiplet!

\begin{figure}[t!]
\begin{center}
\includegraphics[width=0.8\textwidth]{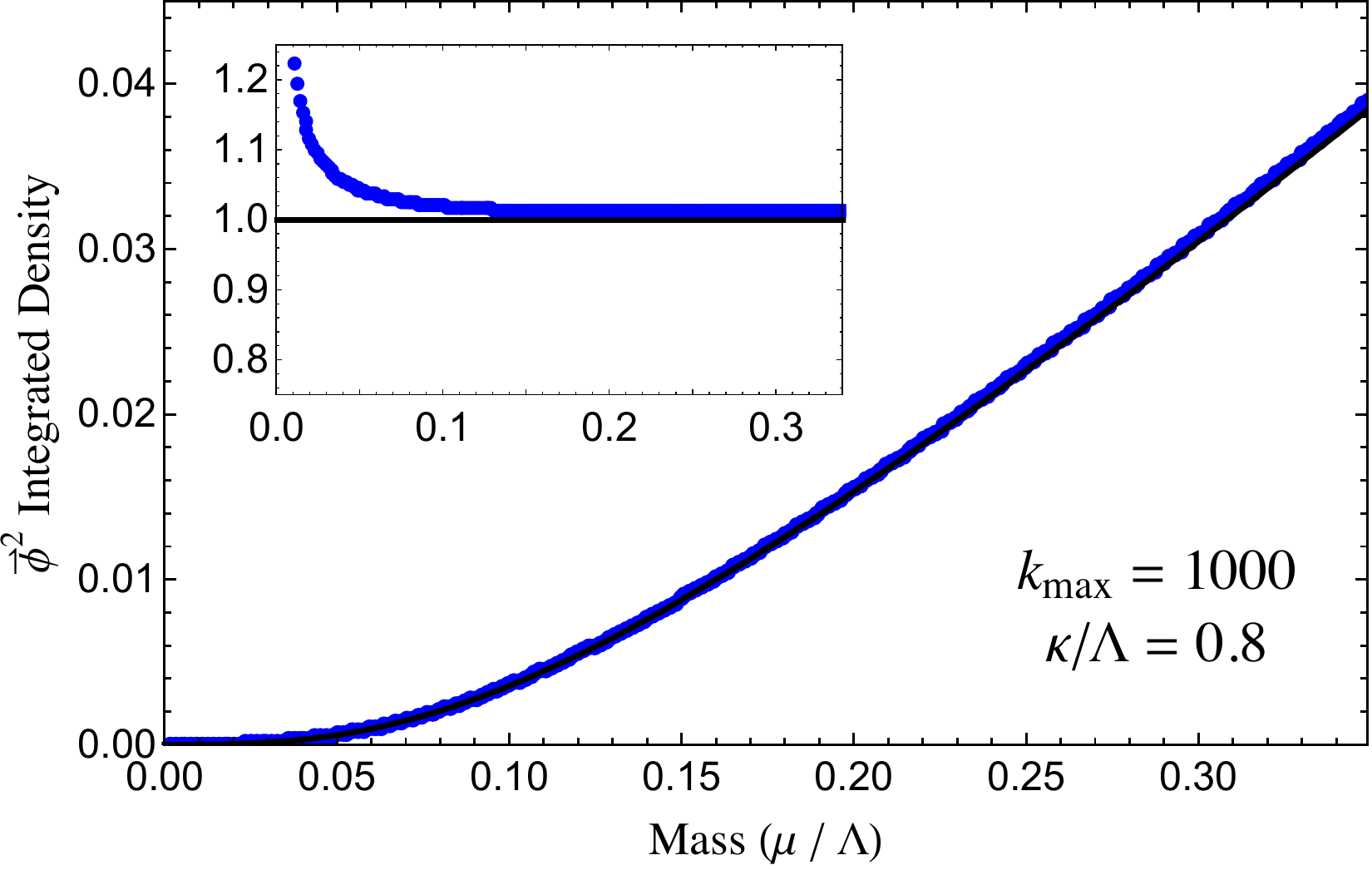}
\caption{Integrated spectral density for $\phivec$ in the large-$N$ limit with $\kappa/\Lambda=0.8$, both the raw value (main plot) and normalized by the theoretical prediction (inset). The conformal truncation results (blue dots) are calculated from the $\ell=0$ multiplet with $\kmax=1000$, and compared to the known analytic expression (black line).}
\label{fig:LargeNMasslessDensity} 
\end{center}
\end{figure}

This simple result matches our holographic intuition, as the quartic interaction just corresponds to a ``double-trace'' deformation, $(\phivec)^2$, which in the large-$N$ limit simply modifies the boundary conditions for the AdS field dual to $\phivec$ \cite{Witten:2001ua}. Our basis therefore makes the AdS perspective manifest, mixing only the components of the $\phivec$ multiplet in the resulting background flow.

We can use the $\ell=0$ multiplet, truncated at some level $\kmax$, to diagonalize this new large-$N$ Hamiltonian for any value of $\kappa$. The overlap of the approximate mass eigenstates with $\phivec$ can then be calculated using eq.~(\ref{eq:2pOverlap}), in order to obtain the associated spectral density. The resulting integrated density for $\kmax=1000$ and $\kappa/\Lambda=0.8$ is shown in figure~\ref{fig:LargeNMasslessDensity}, compared with the theoretical prediction
\be
I_{\phivec}(\mu) = \fr{\mu}{2\pi} - \fr{\kappa}{16\pi} \tan^{-1}\left(\fr{8\mu}{\kappa}\right).
\ee

\begin{figure}[t!]
\begin{center}
\includegraphics[width=0.8\textwidth]{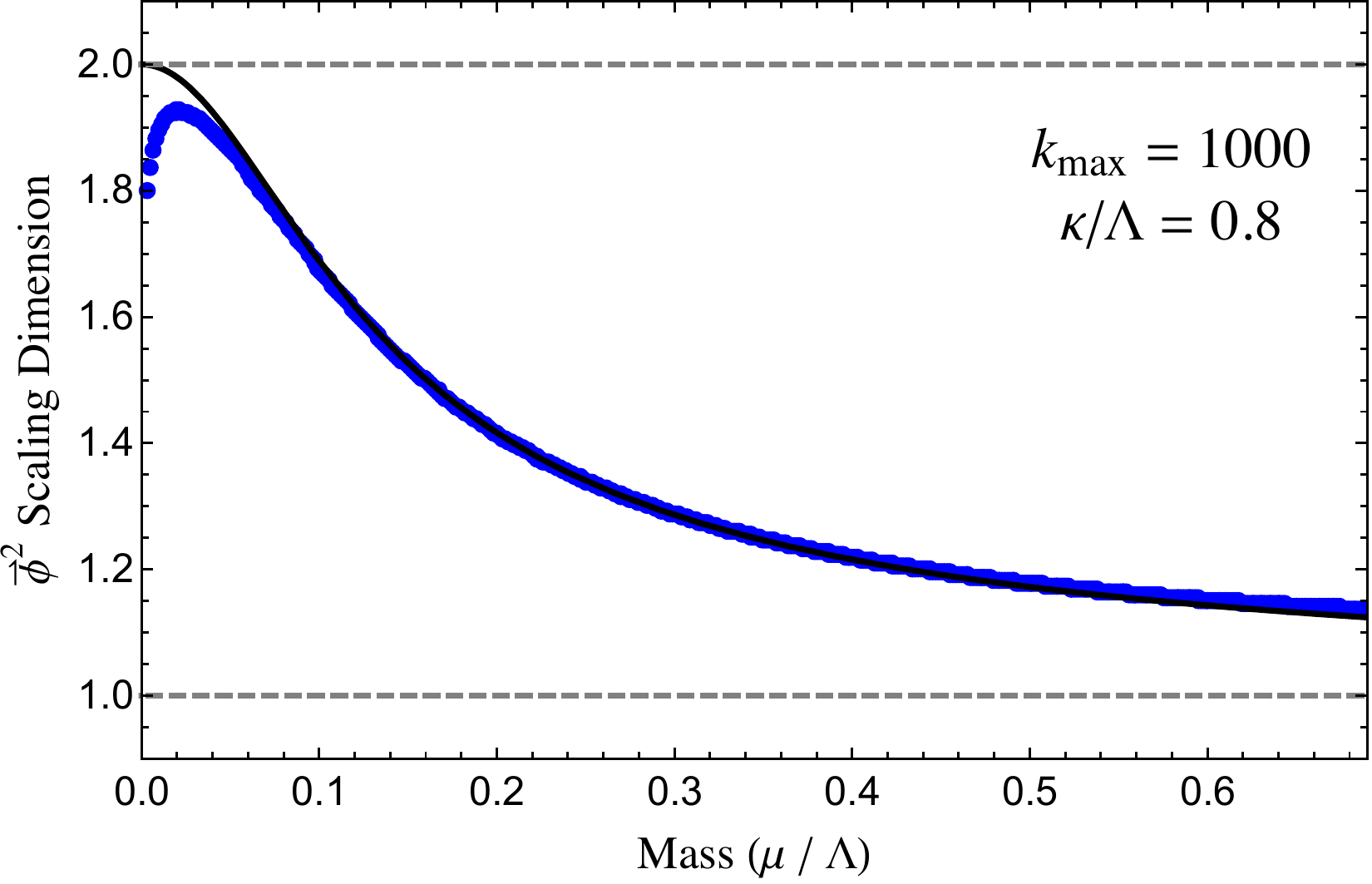}
\caption{Effective scaling dimension for $\phivec$ in the large-$N$ limit with $\kappa/\Lambda=0.8$, derived from the integrated spectral density. The conformal truncation results (blue dots) are calculated from the $\ell=0$ multiplet with $\kmax=1000$, and compared to the known analytic expression (black line).}
\label{fig:LargeNMasslessRG} 
\end{center}
\end{figure}

As we can see, the conformal truncation results successfully reproduce the analytic prediction, transitioning from the free linear behavior in the UV to the new cubic scaling in the IR. Because there is no mass gap, the resulting IR theory is clearly another CFT, but with a modified scaling dimension for $\phivec$. We can determine this new scaling dimension directly from our approximate mass eigenstates by noting that the integrated spectral density for any operator in a 3D CFT scales as
\be
I_\Ocal(\mu) \sim \mu^{2\De_\Ocal-1}.
\ee
We can therefore define an effective scaling dimension in terms of the integrated density,
\be
\De_\Ocal(\mu) \equiv \half\left(\fr{\fr{d}{d\log\mu} I_\Ocal(\mu)}{I_\Ocal(\mu)}+1\right).
\ee
The resulting effective dimension for $\phivec$ is shown in figure~\ref{fig:LargeNMasslessRG}, compared with the predicted behavior from the known spectral density.

The approximate scaling dimension asymptotes to the free value of $\De_{\phivec}=1$ for large mass eigenvalues, consistent with expectations from the original UV CFT. However, as we move to lower masses, the effective dimension increases and flows to the new value of $\De_{\phivec}=2$ in the IR, matching the AdS intuition $\De_{\phivec} \ra d-\De_{\phivec}$.

Looking at our truncation results more carefully, we see that the effective scaling dimension doesn't quite reach the IR value of $\De_{\phivec}=2$, but instead deviates from the theoretical prediction at very low masses. This deviation is a more direct manifestation of the effective IR cutoff in resolution, $\LambdaIR$. We can therefore further improve our prediction for the IR scaling dimension by simply increasing $\kmax$ and including more basis states.

Figure~\ref{fig:LargeNMasslessRG} also demonstrates the general strategy for studying an IR CFT which descends from some original UV theory. Any time the mass gap closes, we can then use the low mass behavior of the integrated densities to determine approximate values for the spectrum of scaling dimensions in the IR theory. By studying higher-point correlation functions, we can also determine the corresponding OPE coefficients, therefore extracting the IR CFT data from some known UV theory.


\section{Modified Basis for Massive Theories}
\label{sec:MassiveBasis}

So far, we have limited ourselves to the special case of massless theories. Generically, though, systems of interest will also contain the mass term
\be
\de\Lcal = -\half m^2 \phi^2.
\ee
This mass term has two important effects. First, the introduction of the mass scale $m$ breaks conformal invariance, mixing distinct Casimir eigenfunctions. We therefore need to use more than one Casimir multiplet to reconstruct the mass eigenstates and calculate the spectral densities. This result is of course unsurprising, as generically any relevant perturbation to our UV CFT will mix Casimir multiplets.

However, there is a second, more subtle consequence of adding a mass term, which is the focus of this section. To understand this effect, consider the two-particle states in a massive free theory. Acting on these states with the Hamiltonian, we can obtain the invariant mass
\be
M^2 = 2P_- \Big( P_+^{(\CFT)} + \de P_+^{(m)} \Big) = \mu^2 + \frac{m^2 P_-^2}{p_-(P_- - p_-)}.
\label{eq:MassPotential}
\ee
The $\mu^2$ term is simply the original invariant mass due to the CFT kinetic term, while we can think of the $m^2$ term as a new ``potential'' due to the mass perturbation in the Lagrangian.

As we can see, this potential has a divergence in the collinear limits $p_-\ra0, P_-$, which simply correspond to the lightcone momentum of either particle vanishing. Matrix elements for the mass term are computed by integrating this potential against the wavefunctions of two basis states. This integral diverges, because our Casimir eigenfunctions have nonzero values at the boundaries $p_- = 0, P_-$. The addition of a mass term therefore leads to \emph{divergences} in the lightcone Hamiltonian.

These divergences are a natural consequence of lightcone kinematics, as we can see from the equation of motion for a single massive particle,
\be
2 p_+ p_- - p_\perp^2 = m^2.
\ee
Due to the nonzero invariant mass, it costs an infinite amount of energy to have $p_- \ra 0$. Equivalently, the lightcone momentum for any physical state is strictly positive.

Given these divergences, there are two ways to proceed. A simple, brute force strategy is to leave our basis of Casimir eigenfunctions untouched and introduce a small collinear cutoff, $\epsilon \ll 1$, restricting the range of integration to
\be
\epsilon \leq \fr{p_-}{P_-} \leq 1 - \epsilon.
\ee
The resulting matrix elements will then depend on this cutoff $\epsilon$. In the limit $\epsilon \ra 0$, the eigenstates of $M^2$ with nonzero support on the boundary of integration will have eigenvalues that diverge as $1/\sqrt{\epsilon}$. Operationally, one can set $\epsilon$ to some small but finite value, diagonalize $M^2$, keep the eigenstates with finite eigenvalues, and disregard the eigenstates with $O(1/\sqrt{\epsilon})$ eigenvalues. This is certainly a valid approach, and one can easily confirm that the resulting spectral densities match the known analytic expressions.

There is a more transparent and efficient strategy, though, which is to note that the $m^2$ potential in eq.~(\ref{eq:MassPotential}) imposes vanishing Dirichlet boundary conditions on the resulting wavefunctions. Heuristically, this is a consequence of the fact that the potential diverges at the boundary. More concretely, the null space of the divergent contribution to $M^2$ is spanned by linear combinations of Casimir eigenstates with vanishing Dirichlet boundary conditions in $p_-$.

Based on this observation, we can therefore eliminate these divergences by imposing Dirichlet boundary conditions from the start. This more efficient strategy has a mathematically well-posed prescription: start with the basis of Casimir eigenfunctions, find linear combinations with vanishing Dirichlet boundary conditions, then re-orthonormalize with respect to the inner product, obtaining a new basis of ``Dirichlet multiplets''.

Because imposing boundary conditions mixes distinct Casimir eigenfunctions, one might worry that the resulting basis can no longer be consistently truncated at some level $\Cmax$, ruining the convergence expected from conformal truncation. As we demonstrate in this section, this is not the case. The resulting Dirichlet multiplets can be organized such that Casimir eigenstates only mix with states with lower eigenvalues, which means one can still truncate the Hilbert space to $\Ccal \leq \Cmax$.

In this section, we present the new basis obtained by imposing vanishing boundary conditions on the original Casimir eigenstates. For concreteness, we specifically focus on the case of two- and three-particle states, demonstrating that imposing these boundary conditions is \emph{equivalent} to diagonalizing the divergent $M^2$ term due to the addition of mass.


\subsection{Two-Particle States}

To understand this new Dirichlet basis more concretely, let's see how it arises for the two-particle sector. The Hamiltonian correction due to the mass term,
\be
\de P_+^{(m)} = \int \fr{d^2p}{(2\pi)^2} \, a^\dagger_p a_p \, \fr{m^2}{2p_-},
\ee
leads to the two-particle $M^2$ matrix elements,
\be
\<\ell;k|2P_- \de P_+^{(m)}|\ell';k'\> = \de_{kk'} \cdot 2! \int \fr{dp_-}{4\sqrt{p_- (P_- - p_-)}} \, F^{(2)}_\ell(p) \, F^{(2)}_{\ell'}(p) \, \fr{m^2 P_-^2}{p_-(P_- - p_-)}.
\ee
Because this operator has no dependence on $\mu$, these matrix elements are automatically diagonal in $k$. The resulting divergence therefore only affects the Casimir eigenfunctions $F_\ell(p)$, and not the weight functions $g_k(\mu)$.

To see the divergence explicitly, we can introduce a cutoff $\epsilon$ on the range of integration. Using the all minus Casimir basis functions
\be
F_{\ell-}(p) = P^{(-\half,-\half)}_\ell \Big(\fr{2p_-}{P_-} - 1\Big),
\ee
we then obtain the $\epsilon$-dependent matrix elements
\be
\<\ell|2P_- \de P_+^{(m)}|\ell'\> = \fr{2m^2}{\sqrt{(1+\de_{\ell,0})(1+\de_{\ell',0})}} \left(-4 \Lmax + \fr{4}{\pi\sqrt{\epsilon}} \right).
\ee
Each of these matrix elements therefore diverges as the cutoff $\epsilon \ra 0$.

Let's now focus solely on this divergent piece, ignoring the remaining finite contributions to $M^2$. As shown in appendix~\ref{app:MassiveBasis}, any linear combination of basis functions of the form
\be
F_\ell(p) - \sqrt{2} F_0(p),
\ee
is an eigenstate of the divergent term, with eigenvalue zero. The $\ell=0$ state is a constant, so this linear combination simply alters the constant term in $F_\ell(p)$. In fact, this particular combination perfectly \emph{cancels} the constant piece of $F_\ell(p)$, such that the resulting function is zero when $p_- \ra 0,P_-$. The divergence in $M^2$ thus just rearranges our basis into linear combinations which have vanishing boundary conditions!

However, these new states with Dirichlet boundary conditions are no longer orthogonal. Orthonormalizing the basis functions, we obtain
\be
\Ft_{\ell-}(p) \equiv p_-(P_- - p_-) \, P^{(\fr{3}{2},\fr{3}{2})}_{\ell-2}\Big(\fr{2p_-}{P_-} - 1\Big).
\label{eq:MassiveBasis2P}
\ee
As we can see, this new Dirichlet basis consists of Jacobi polynomials with a modified integration measure, and manifestly vanish when $p_- \ra 0,P_-$. These basis states span the null space of the divergent term in $M^2$, such that the matrix elements built from them are all finite in the limit $\epsilon \ra 0$.

We can understand the structure of these new Dirichlet functions by expressing them in terms of the original Casimir eigenfunctions,
\be
\Ft_{\ell-}(p) = \fr{1}{\sqrt{(\ell-1)(\ell+1)}} \sum_{\ell'=0}^{\ell-2} \sqrt{1+\de_{\ell',0}} \, F_{\ell'-}(p) - \sqrt{\fr{\ell-1}{\ell+1}} \, F_{\ell-}(p).
\ee
The new states are therefore built only from Casimir eigenstates with $\ell' \leq \ell$. Because of this, we can \emph{still} restrict our basis to $\Ccal \leq \Cmax$ by truncating in $\ell$. Imposing boundary conditions on our basis to eliminate divergences thus does not compromise our conformal truncation framework.

In practice, we don't need to use Gram-Schmidt to build an orthonormal basis. We can simply note that $\Ft_\ell(p)$ must be a polynomial with no constant term, and is thus proportional to an overall factor of $p_-$,
\be
\Ft_\ell(p) = p_- f(p).
\ee
However, because our basis states are symmetric under permutations, they must actually be proportional to an overall factor of $p_-(P_- - p_-)$,
\be
\Ft_\ell(p) = p_-(P_- - p_-) f'(p).
\ee
The Dirichlet basis therefore consists of the set of polynomials which are orthogonal with respect to the new integration measure created by this overall factor. Following this procedure for the two-particle states, we obtain the modified Jacobi polynomials in eq.~(\ref{eq:MassiveBasis2P}), matching the basis we would have found via Gram-Schmidt.

To obtain the Dirichlet multiplets for other spin components, we simply need to act with the Pauli-Lubanski generator $W$ on these new all minus basis functions. However, this action is trivial for the two-particle case, as we can see by acting on the simple linear combination
\be
W \Big(F_{\ell-}(p) - \sqrt{2} F_0(p) \Big) = F_{\ell\perp}(p).
\ee
Because $W$ annihilates the constant $\ell=0$ term, we find that $\Ft_{\ell\perp} = F_{\ell\perp}$. In other words, these spin components \emph{already} have vanishing boundary conditions,
\be
F_{\ell\perp}(p) = \fr{\mu}{P_-^2} \sqrt{p_-(P_- - p_-)} \, P^{(\half,\half)}_{\ell-1}\Big(\fr{2p_-}{P_-} - 1\Big).
\ee

Finally, we see that imposing these new boundary conditions does not ruin the permutation symmetry of our basis, as it does not mix even $\ell$ states with odd $\ell$ states. Our symmetric basis therefore still consists only of states with $\ell$ even.

We now have a complete basis of two-particle states which satisfy the Dirichlet boundary conditions required by lightcone kinematics. This basis preserves the conformal truncation structure of the original Casimir eigenstates and can be used to reproduce the IR spectrum of any 3D scalar field theory, both with and without a mass gap.


\subsection{Three-Particle States}

In the three-particle case, the ``potential'' resulting from the mass term takes the form
\be
\de M^2 = m^2 P_- \left(\fr{1}{p_{1-}} + \fr{1}{p_{2-}} + \fr{1}{P_- - p_{1-} - p_{2-}}\right).
\ee
As we can see, this potential diverges when any of the individual lightcone momenta go to zero. We therefore need to impose Dirichlet boundary conditions such that our basis functions vanish in any of these three limits.

Because these boundary conditions only affect $p_-$, they do not mix states with different values of $m_\perp$. As shown in appendix~\ref{app:MassiveBasis}, the resulting Dirichlet basis for $m_\perp=0$ is given by
\be
\Ft_{\ell,0}(p) \equiv p_{1-} p_{2-} (P_- - p_{1-} - p_{2-}) \Big(1 - \fr{p_{1-}}{P_-}\Big)^{\ell_2-2} P^{(2\ell_2,\fr{3}{2})}_{\ell_1-1}\Big(\fr{2p_{1-}}{P_-} - 1\Big) \, P^{(\fr{3}{2},\fr{3}{2})}_{\ell_2-2}\Big(\fr{2p_{2-}}{P_- - p_{1-}} - 1\Big).
\ee
The structure of these basis states is quite similar to that of the two-particle case, in that these functions have an overall factor which explicitly enforces the Dirichlet boundary conditions. These new Jacobi polynomials are again simply the orthogonal set of functions for the integration measure resulting from this overall factor.

Because these functions are polynomials, with the total degree fixed by $\vec{\ell}$, they can be expanded solely in terms of Casimir eigenfunctions with $\ell_1' \leq \ell_1, \ell_2' \leq \ell_2$. The conformal truncation structure is therefore again preserved by this change of basis, despite mixing Casimir eigenstates, such that we can still restrict the basis to $\Ccal \leq \Cmax$.

Though we do not need such states in this work, one can then obtain the remaining $m_\perp \neq 0$ components by acting with $W$ on these Dirichlet basis functions, as discussed in appendix~\ref{app:MassiveBasis}.


\section{Conformal Truncation Results}
\label{sec:MassiveResults}

We now have a full conformal truncation framework for scalar field theories in $2+1$ dimensions. In this section, we test this framework in several simple settings where we can compare with exact analytic expressions. Specifically, we consider the following scenarios:
\begin{enumerate}
\itemsep0em
\item[(1)] Free scalar field theory, 
\item[(2)] Perturbative $\phi^3$ theory,
\item[(3)] Perturbative $\phi^4$ theory,
\item[(4)] $O(N)$ model with $N\rightarrow\infty$.
\end{enumerate}

Each of these settings tests an important feature of our overall prescription. In free field theory, we reproduce the \KL spectral densities for the operators $\phi^2$ and $\phi^3$ in the presence of a mass term. These examples test the completeness of our Hilbert space of states. In general, the spectral density for any operator describes the decomposition of its two-point function in terms of physical intermediate states. If the vector space from which these states are derived is incomplete, overcomplete, or improperly normalized, the resulting spectral density will be incorrect, even at the free field level.  

Moving beyond free field theory, the $\phi^3$ and $\phi^4$ interactions test whether our truncation results agree with perturbation theory in the weak-coupling regime. In both cases, the observable we consider is the mass of the one-particle state, which is shifted by interactions. Here, the $\phi^3$ example is particularly important, because the resulting one-particle mass shift is finite and primarily sensitive to IR physics. The convergence of our calculated mass shift with respect to $\Cmax$ is thus a useful gauge of our ability to reconstruct the low-mass spectrum. 

Finally, moving beyond perturbation theory, the $O(N)$ model at large-$N$ allows us to test our prescription in a truly strongly-interacting setting. Here, the spectral density of $\vec{\phi}^2$ can be computed analytically by summing an infinite series of diagrams. In reproducing the full non-perturbative result, we confirm that our truncation procedure can be used to obtain dynamical correlation functions at strong coupling.

For each of these cases, we proceed as follows. Given a particular Hamiltonian, $P_+$, we construct the corresponding Lorentz invariant operator
\be
M^2 \equiv 2P_+ P_- - P_\perp^2.
\ee
We then choose a maximum Casimir eigenvalue, $\Cmax$, and a maximum degree for the weight functions, $\kmax$, at which to truncate the Hilbert space. We compute matrix elements of $M^2$ in this truncated Hilbert space and numerically diagonalize the resulting matrix to find its eigenvalues $\mu_i$ and eigenstates $\left|\mu_i\right\rangle$. The eigenstates are used to compute integrated spectral densities, which for a given operator $\Ocal$ are defined as
\be
I_\Ocal(\mu) \equiv \int_{0}^{\mu^2} d\mu^{\prime \, 2} \, \rho_\Ocal(\mu') = \sum_{\mu_i \leq \mu} |\<\Ocal(0)|\mu_i\>|^2.
\label{eq:I}
\ee
It is worth emphasizing that once the $\left|\mu_i\right\rangle$ are obtained numerically, the inner products on the right-hand side are calculable purely in terms of UV CFT data. This is because the operator $\Ocal$ is defined in the UV and the eigenstates $\left|\mu_i\right\rangle$ are just linear combinations of the UV Hilbert space states, whose overlap with $\Ocal$ is known.


\subsection{Massive Free Field Theory}
\label{sec:fft}

First, we consider the case of free field theory for a single massive scalar field,
\be
\Lcal = \half \p_\mu \phi \p^\mu \phi + \half m^2 \phi^2.
\ee
As discussed in section~\ref{sec:Review}, the associated lightcone Hamiltonian is
\be
P_+ = P_+^{(\CFT)} + \delta P_+^{(m)} = \int \fr{d^2p}{(2\pi)^2} \, a^\dagger_p a_p \, \left( \fr{p_\perp^2}{2p_-} +   \fr{m^2}{2p_-}\right).
\ee
This Hamiltonian conserves particle number and can thus be diagonalized independently in each $n$-particle sector. 

The matrix elements for this Hamiltonian are sensitive to two free parameters: the bare mass $m$ and our UV cutoff $\Lambda$. The mass scale sets the threshold for the resulting $M^2$ eigenvalues, while the cutoff sets the allowed range. For our free field theory results, we express all mass scales in units of $\Lambda$, with $m/\Lambda = 0.1$.

The one-particle sector in free field theory is trivial, consisting of a single state of mass $m$. The two-particle sector, however, has a continuum of states with masses starting at $2m$. After diagonalizing the two-particle Hamiltonian, we compute the integrated spectral density for $\phi^2$, which can be compared to the analytic free field theory result,
\be    
I_{\phi^2}(\mu) = \fr{1}{2\pi} (\mu-2m).
\label{I2}
\ee

\begin{figure}[t!]
\begin{center}
\includegraphics[width=0.8\textwidth]{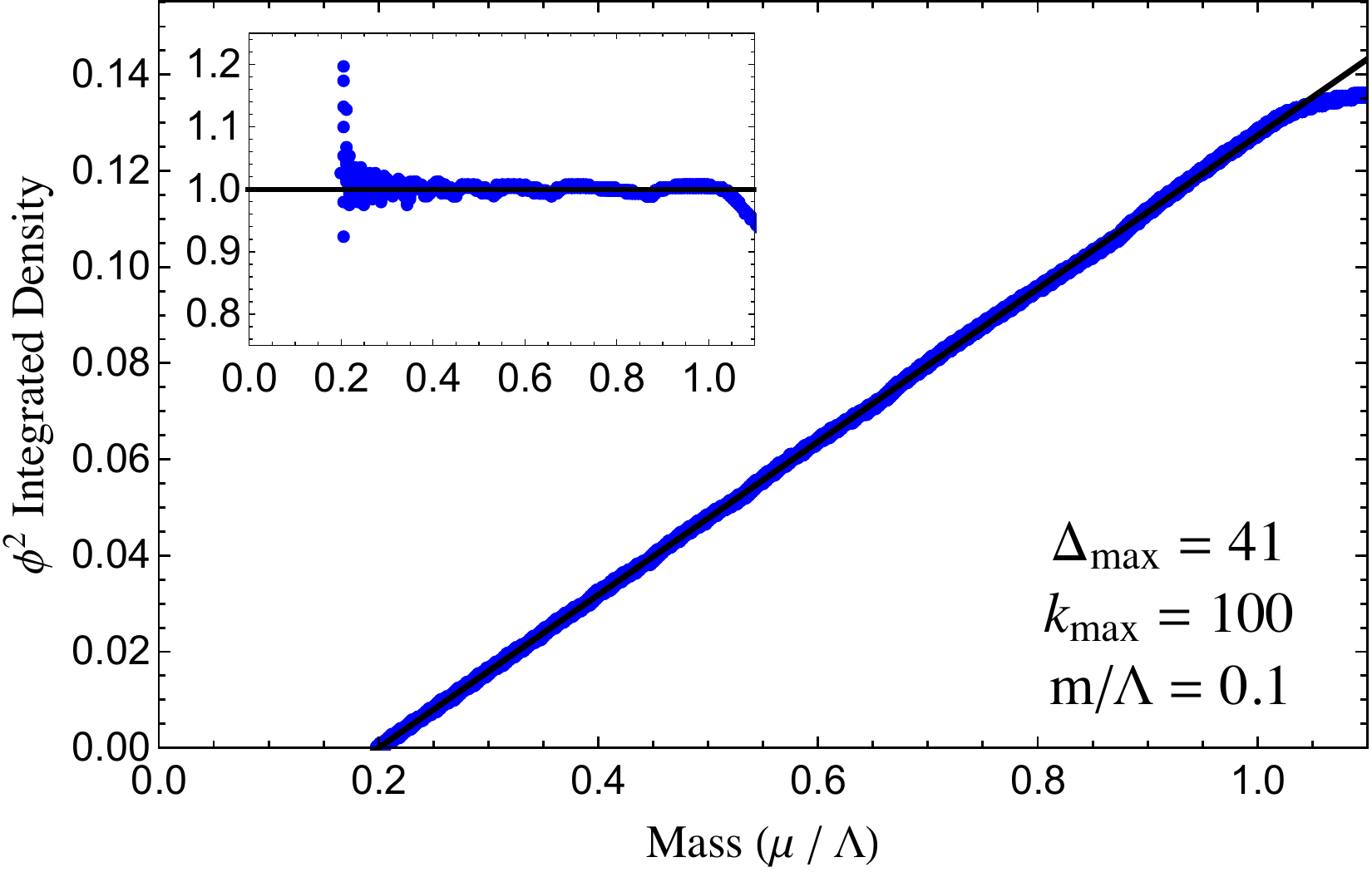}
\caption{Integrated spectral density for $\phi^2$ in massive free field theory with $m/\Lambda=0.1$, both the raw value (main plot) and normalized by the theoretical prediction (inset). The conformal truncation results (blue dots) are calculated with $\Dmax=41$ (or $\Lmax=40$) and $\kmax=100$, and compared to the known analytic expression (black line).}
\label{2pSD01} 
\end{center}
\end{figure}

The size of our truncated two-particle Hilbert space is controlled by two parameters: $\Cmax$, which controls the number of Dirichlet multiplets, and $\kmax$, which controls the size of each multiplet. For conceptual simplicity, we actually choose to report our results in terms of the maximum scaling dimension $\Dmax$, which scales linearly with the degree of our basis functions, rather than $\Cmax$, which scales quadratically. For two- and three-particle states, the scaling dimension $\Dmax$ uniquely determines the maximum Casimir eigenvalue $\Cmax$ via the relation
\be
\Cmax = \Dmax(\Dmax-3) + (\Dmax-\tfr{n}{2})(\Dmax-\tfr{n}{2}+1) \sim 2\Dmax^2 \quad (n=2,3).
\ee
In both cases, this scaling dimension is related to the maximum spin by $\Dmax = \Lmax + \fr{n}{2}$.

Our truncation results for $I_{\phi^2}$ are shown in figure~\ref{2pSD01}. The primary plot in this figure is the raw data for $I_{\phi^2}$, obtained with $\Dmax=41$ and $\kmax=100$, which corresponds to a total of 2,020 states. The solid black line is the analytic expression, eq.~(\ref{I2}). The truncation results (blue dots) successfully reproduce this spectral density from the mass threshold $2m$ up to the UV cutoff $\Lambda$.

To study this agreement in more detail, the subplot in this figure shows the ratio of our data to the analytic expression. The truncation data lies within a few percent of the theoretical prediction over much of the allowed range. The spreading of the data for $\mu$ near the threshold $2m$ indicates the presence of the IR cutoff, $\LambdaIR$, similar to the massless case discussed in section~\ref{sec:MasslessResults}. Overall, though, the agreement in these plots indicates that our truncation method correctly reproduces the $\phi^2$ two-point function in the limit of large $\Dmax$.

\begin{figure}[t!]
\begin{center}
\includegraphics[width=0.9\textwidth]{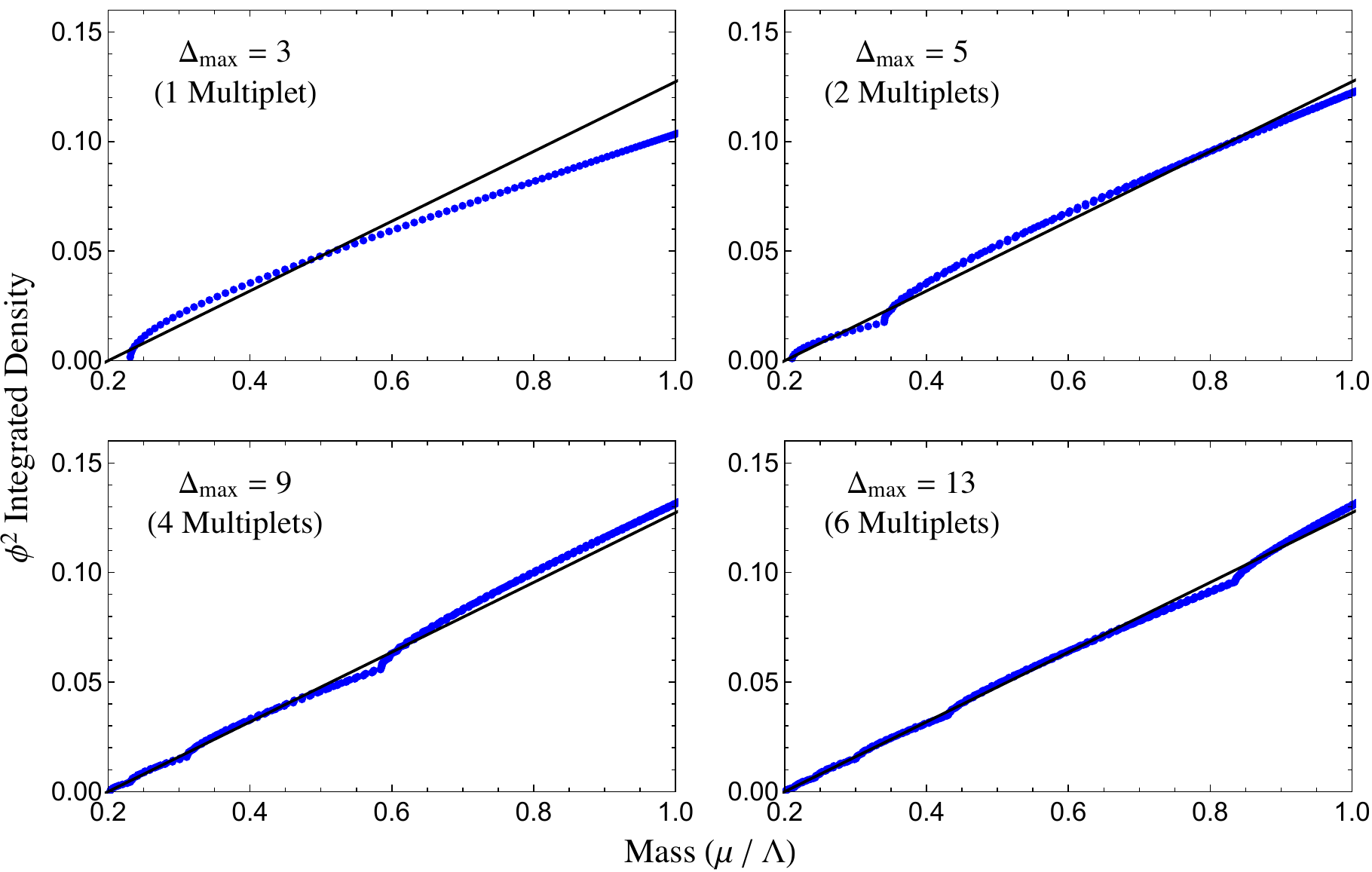}
\caption{Integrated spectral density for $\phi^2$ in massive free field theory with $m/\Lambda=0.1$. The conformal truncation results (blue dots) are calculated with $\kmax=100$ and different values of $\Dmax$, and compared to the known analytic expression (black line).}
\label{2pSD02} 
\end{center}
\end{figure}

An important question is how rapidly these truncation results converge for lower values of $\Dmax$. In figure~\ref{2pSD02}, we again show results for $I_{\phi^2}$ with $\kmax=100$, but now for $\Dmax = 3, 5, 9, 13$. These four values for $\Dmax$ respectively correspond to including 1, 2, 4, and 6 Dirichlet multiplets, or equivalently 101, 202, 404, and 606 total states. Clearly, the results with larger $\Dmax$ have better agreement with the analytic expression (black line). However, one obtains a reasonable approximation even with just a \emph{single} Dirichlet multiplet, and, at least qualitatively, there appears to be rapid convergence with increasing $\Dmax$, especially at low masses.

This convergence can be understood by studying the $M^2$ matrix elements in appendix~\ref{app:MassiveMatrix}. Specifically, if we consider the matrix elements due to the mass term in eq.~(\ref{eq:MassTermMixing}), we find the asymptotic behavior
\be
\<\Lt| \de M^2 | \Lt\> \sim m^2 \ell, \qquad \<\Lt| \de M^2 | \Lt'\> \sim \fr{m^2}{\ell} \qquad (\ell \gg \ell').
\ee
The Dirichlet multiplets with large $\ell$, or equivalently large $\De$, therefore lead to high mass states and decouple from the low $\ell$ multiplets, consistent with our AdS intuition.

Turning to the three-particle sector, we then compute the integrated spectral density for $\phi^3$, which can be compared to the analytic result, 
\be
I_{\phi^3}(\mu) = \fr{3}{16\pi^2}(\mu-3m)^2.
\label{I3}
\ee

\begin{figure}[t!]
\begin{center}
\includegraphics[width=0.8\textwidth]{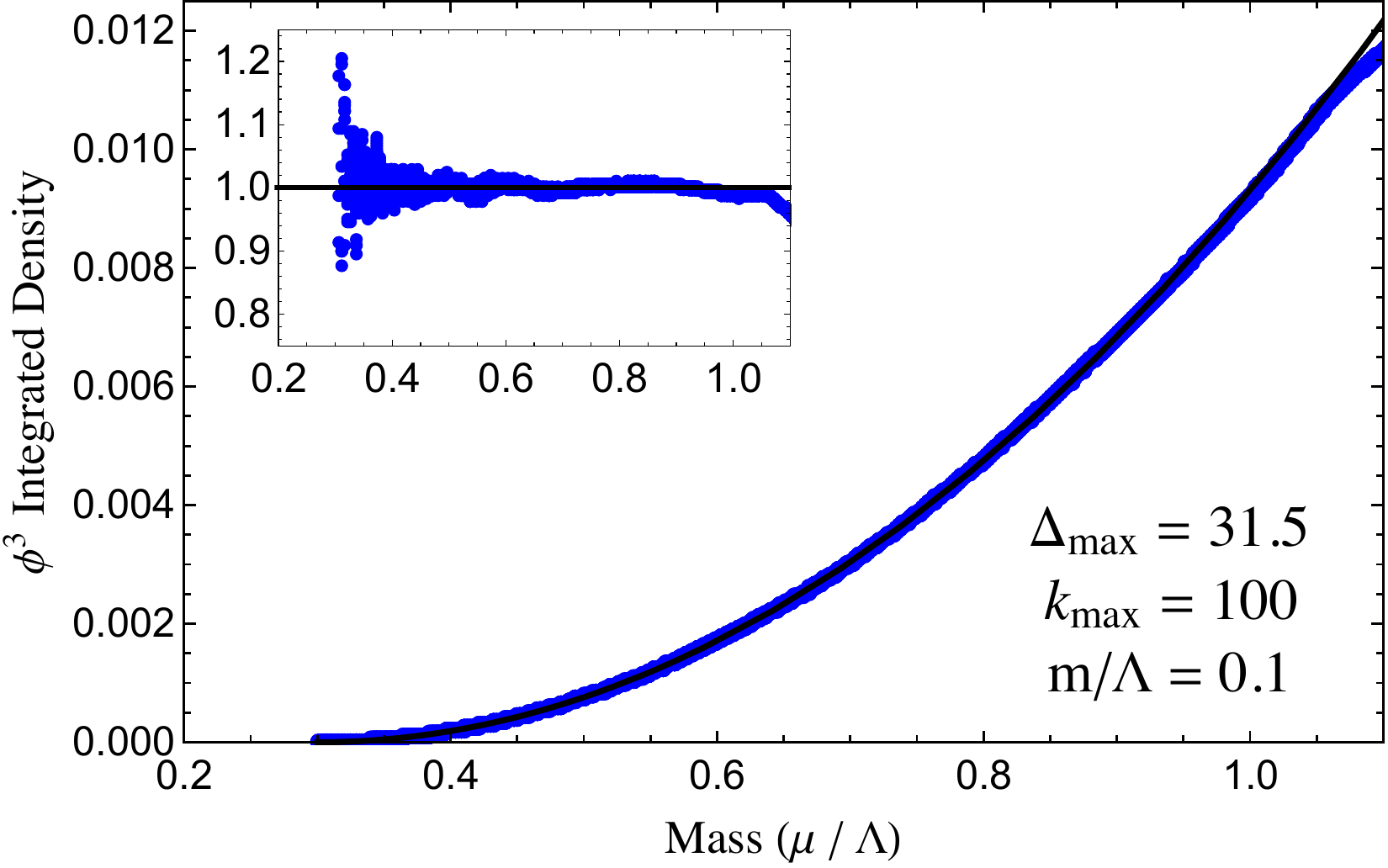}
\caption{Integrated spectral density for $\phi^3$ in massive free field theory with $m/\Lambda=0.1$, both the raw value (main plot) and normalized by the theoretical prediction (inset). The conformal truncation results (blue dots) are calculated with $\Dmax=\fr{63}{2}$ (or $\Lmax = 30$) and $\kmax=100$, and compared to the known analytic expression (black line).}
\label{3pSD01} 
\end{center}
\end{figure}

Figure~\ref{3pSD01} shows the truncation results for $\Dmax=\fr{63}{2}$ and $\kmax=100$, corresponding to a Hilbert space of 7,575 states. The main plot again shows the raw data (blue dots), which correctly reproduces the analytic expression (black line) from the mass threshold $3m$ to the UV cutoff $\Lambda$. As shown in the subplot, the conformal truncation data again lies within a few percent of the theoretical prediction until spreading out near the IR cutoff.

\begin{figure}[t!]
\begin{center}
\includegraphics[width=0.9\textwidth]{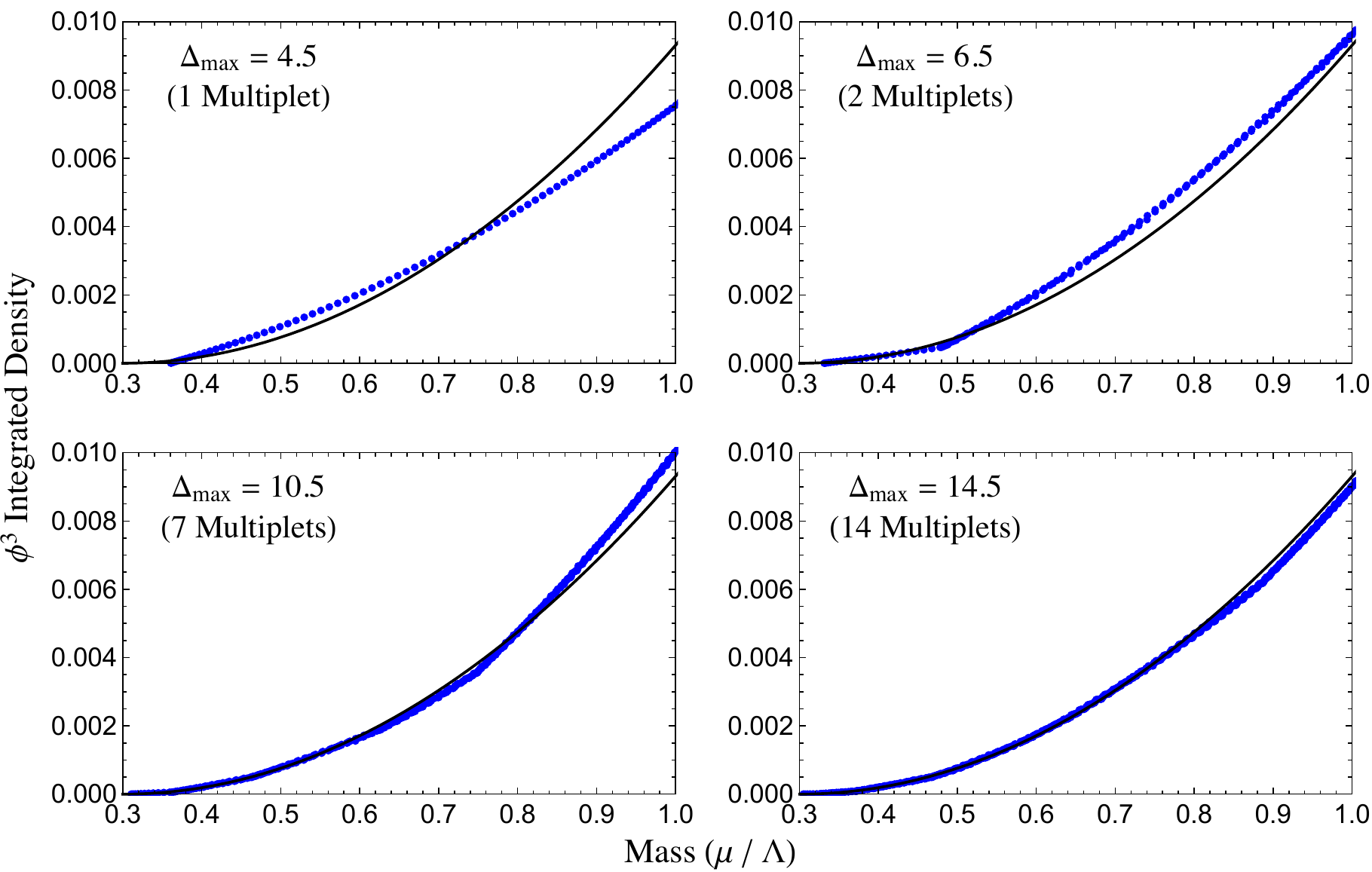}
\caption{Integrated spectral density for $\phi^3$ in massive free field theory with $m/\Lambda=0.1$. The conformal truncation results (blue dots) are calculated with $\kmax=100$ and different values of $\Dmax$, and compared to the known analytic expression (black line).}
\label{3pSD02} 
\end{center}
\end{figure}

To study the convergence of the three-particle basis, figure~\ref{3pSD02} shows the spectral density for $\Dmax=\fr{9}{2},\fr{13}{2},\fr{21}{2},\fr{29}{2}$ with $\kmax=100$. These values correspond to 1, 2, 7, and 14 Dirichlet multiplets, or 101, 202, 707, and 1414 states, respectively. As in the two-particle case, we find that the conformal truncation data rapidly converges to the analytic result as we increase $\Dmax$.

Based on these free field theory results, we therefore verify that the new Dirichlet basis for two- and three-particle states discussed in section~\ref{sec:MassiveBasis} is complete and correctly reproduces correlation functions in a massive theory. In general, this simple check can be repeated for each $n$-particle sector by reproducing the spectral density for $\phi^n$.     


\subsection{Perturbative $\phi^3$ Theory}
\label{sec:phi3}

In the presence of interactions, sectors with different particle number are generically no longer independent. For example, in perturbation theory the physical mass of the single-particle state is shifted due to mixing with higher-particle states. As a simple test of our truncation scheme for interacting theories, we first consider the addition of the $\phi^3$ interaction
\be
\de \Lcal = - \frac{1}{3!}g\phi^3,
\ee 
to the massive free theory studied in the previous subsection. Specifically, we consider this theory in the perturbative limit $g/m^{3/2} \ll 1$ and verify that our truncation results correctly reproduce the leading mass shift for the one-particle state,
\be
\de m^2 = - \fr{g^2\log 3}{16\pi m}.
\ee
We have chosen to start with $\phi^3$ theory because the above mass shift is finite and insensitive to the UV cutoff $\Lambda$. In the next subsection, we consider perturbative $\phi^4$ theory, where the mass shift is logarithmically UV divergent.

Operationally, the $\phi^3$ interaction introduces nonzero matrix elements in the Hamiltonian between states with $n$ and $n\pm1$ particles. At leading order in perturbation theory, the one-particle mass shift is therefore due to mixing with the two-particle sector. So long as we restrict ourselves to perturbatively small values for $g$, it thus suffices to consider the subspace consisting of one- and two-particle states only. For a given coupling, we can compute the Hamiltonian matrix elements in this subspace, truncate the two-particle states with the two parameters $\Dmax$ and $\kmax$, then numerically diagonalize the truncated matrix. The lowest eigenvalue, $\mu_{\min}^2$, corresponds to the physical mass of the one-particle state. The mass shift obtained via conformal truncation is then defined as
\be
\de m^2 \equiv \mu_{\min}^2 - m^2,
\ee
where $m$ is the original bare mass.

In this subsection, we will express all mass scales in terms of $m$. We set the interaction coupling to $g/m^{3/2}=0.01$, which is well within the regime of perturbation theory. As for the UV cutoff, we will actually vary the ratio $\Lambda/m$ to study its effect on $\de m^2$.

\begin{figure}[t!]
\begin{center}
\includegraphics[width=0.75\textwidth]{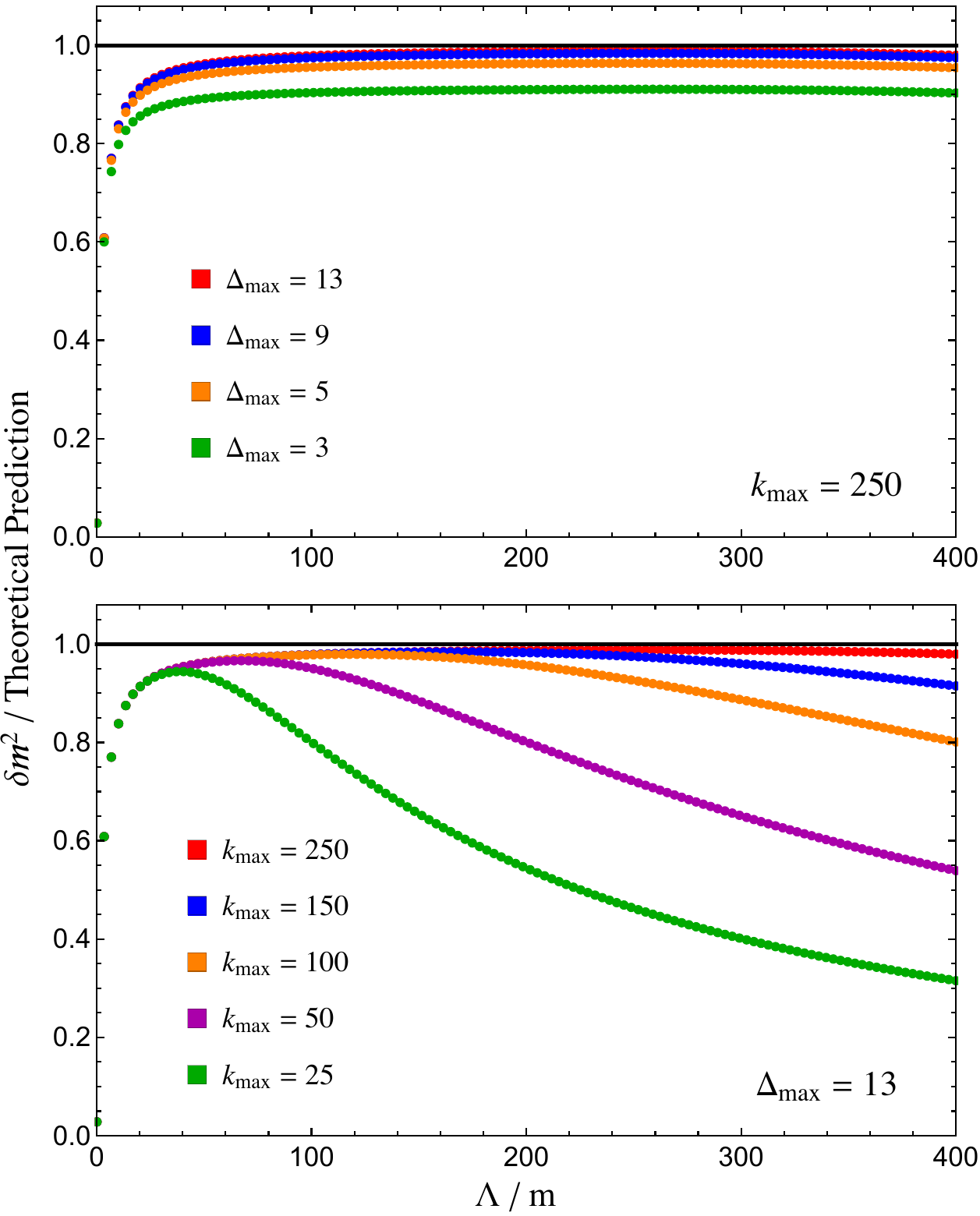}
\caption{Leading contribution to the one-particle mass due to the $\phi^3$ interaction with $g/m^{3/2}=0.01$, normalized by the theoretical prediction. The resulting mass shift is shown as a function of $\Lambda/m$ for $\kmax=250$ and several values of $\Dmax$ (top) and for $\Dmax=13$ and several values of $\kmax$ (bottom).}
\label{fig:2pShift} 
\end{center}
\end{figure}

Figure~\ref{fig:2pShift} shows our numerically-obtained mass shift $\de m^2$, normalized by the theoretical prediction, as a function of $\Lambda/m$. In the top plot, we fix $\kmax=250$ and vary $\Dmax$, while in the bottom plot we fix $\Dmax=13$ and vary $\kmax$.

Let's first consider the top plot. The values of $\Dmax=3,5,9,13$ correspond to 1, 2, 4, and 6 two-particle Dirichlet multiplets, or equivalently 252, 503, 1005, and 1507 total states (including the one-particle state). In each case, the resulting mass shift approaches the theoretical value (solid black line) from below as we increase $\Lambda/m$. Because the mass shift is negative, this means that the eigenvalue $\mu_{\min}$ approaches the true physical mass from above. In approximating this lowest eigenvalue, we can think of conformal truncation as simply a variational method, with the trial wavefunction set by the truncation parameters $\Dmax$ and $\kmax$, as well as the cutoff $\Lambda$. The eigenvalue $\mu_{\min}$ will therefore \emph{always} be greater than the true physical mass, or equivalently, these truncation results set a lower bound on the mass shift $\de m^2$.

The fact that each approximate mass shift asymptotes to a constant value indicates that the mass shift is independent of the cutoff in the limit $\Lambda \gg m$. This asymptotic value rapidly converges to the theoretical prediction as we increase $\Dmax$, indicating that our truncation results successfully reproduce perturbation theory. This behavior is expected given the convergence of the free field theory spectral densities in the previous subsection, as the perturbative mass shift is due to the exchange of these two-particle mass eigenstates. The fact that the $\Dmax=3$ results are within $10\%$ of the theoretical prediction in figure~\ref{fig:2pShift} is therefore simply a manifestation of the fact that a single Dirichlet multiplet provides a reasonable approximation to the $\phi^2$ spectral density in figure~\ref{2pSD02}.

Next, let's focus on the second plot. We've now fixed $\Dmax=13$ (6 Dirichlet multiplets) with $\kmax=25,50,100,150,250$, which correspond to 157, 307, 607, 907, and 1507 total states, respectively. Our truncation results again approach the theoretical value (black line) from below. For each $\kmax$, however, the approximate mass shift eventually reaches a peak value and then begins to fall with increasing $\Lambda/m$. This turnover arises because of the effective IR cutoff $\LambdaIR$. Increasing the UV cutoff with fixed $\kmax$ is equivalent to increasing $\LambdaIR$. For $\LambdaIR \ll m$, its effect on the mass shift is negligible. Once $\LambdaIR \sim m$, however, the finite resolution of our basis leads to deviations away from the theoretical value.

As one increases $\kmax$, though, two things happen. First, the peak value asymptotes to the theoretical prediction, and second, the peak flattens out and persists for a wider and wider range of $\Lambda/m$. The IR cutoff therefore decreases as $\kmax$ increases, and our truncation results are valid up to $\Lambda \gg m$. This rising and falling pattern is in fact also present in the top plot, though the eventual turnover occurs outside the region shown.

\begin{figure}[t!]
\begin{center}
\includegraphics[width=0.8\textwidth]{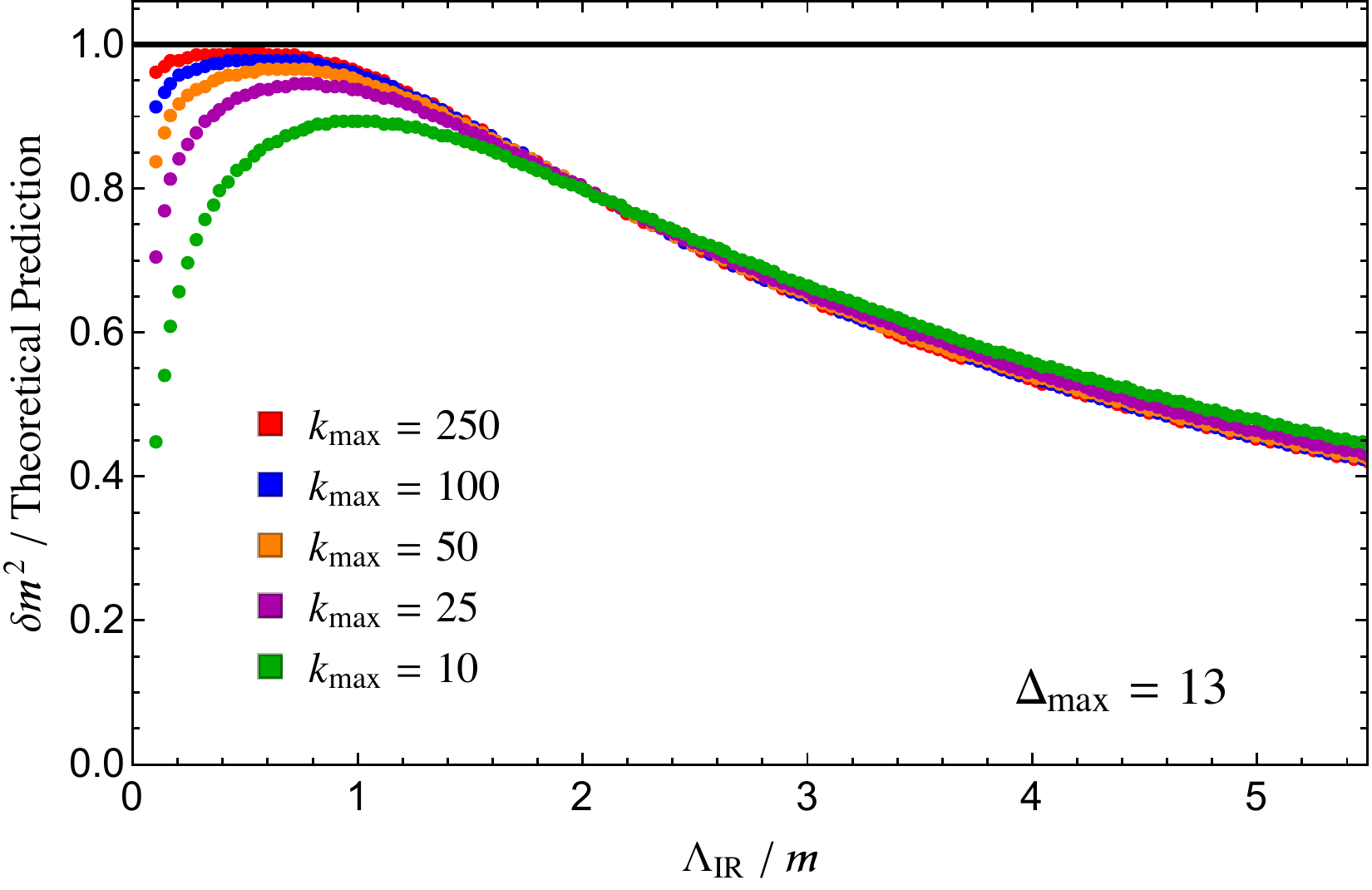}
\caption{Leading correction to the one-particle mass due to the $\phi^3$ interaction as a function of $\LambdaIR \equiv \Lambda/\kmax$, for $\Dmax=13$ (or $\Lmax=12$) and multiple values of $\kmax$. For $\LambdaIR \gtrsim m$, the mass corrections collapse to a single curve, indicating that the low mass eigenstates only depend on $\Lambda$ and $\kmax$ in this fixed ratio. For low $\LambdaIR$, the curves separate due to effects from the bare mass $m$, with the peak approaching the known theoretical value for increasing $\kmax$.}
\label{fig:2pScaleCollapse} 
\end{center}
\end{figure}

We can confirm this IR cutoff structure by instead plotting the one-particle mass shift as a function of the ratio
\be
\LambdaIR \equiv \fr{\Lambda}{\kmax},
\ee
which is shown in figure~\ref{fig:2pScaleCollapse} for multiple values of $\kmax$ with $\Dmax=13$. As we can see, for large $\LambdaIR$ the plots all collapse to a single curve. This simple behavior suggests that the approximate low-mass eigenstates depend only on this effective IR scale, as discussed in section~\ref{sec:MasslessResults}. For $\LambdaIR \lesssim m$, however, the mass shifts for distinct $\kmax$ separate, with the peak value increasing with $\kmax$.

In practice, one can place bounds on the lowest mass eigenvalue for a given $\Dmax$ and $\kmax$ by varying the UV cutoff $\Lambda$ (or equivalently $\LambdaIR$) and selecting the peak, extremum value. This approach converges rapidly in $\Dmax$, which suggests that one potentially needs few Dirichlet mutliplets, so long as $\kmax$ is sufficiently large.


\subsection{Perturbative $\phi^4$ Theory}
\label{sec:phi4}

Similarly, we can consider perturbing massive free field theory by the quartic interaction
\be
\de \Lcal = - \frac{1}{4!}\lambda\phi^4,
\ee
and again calculate the one-particle mass shift in the perturbative limit $\lambda/m \ll 1$.

\begin{figure}[t!]
\begin{center}
\includegraphics[width=0.75\textwidth]{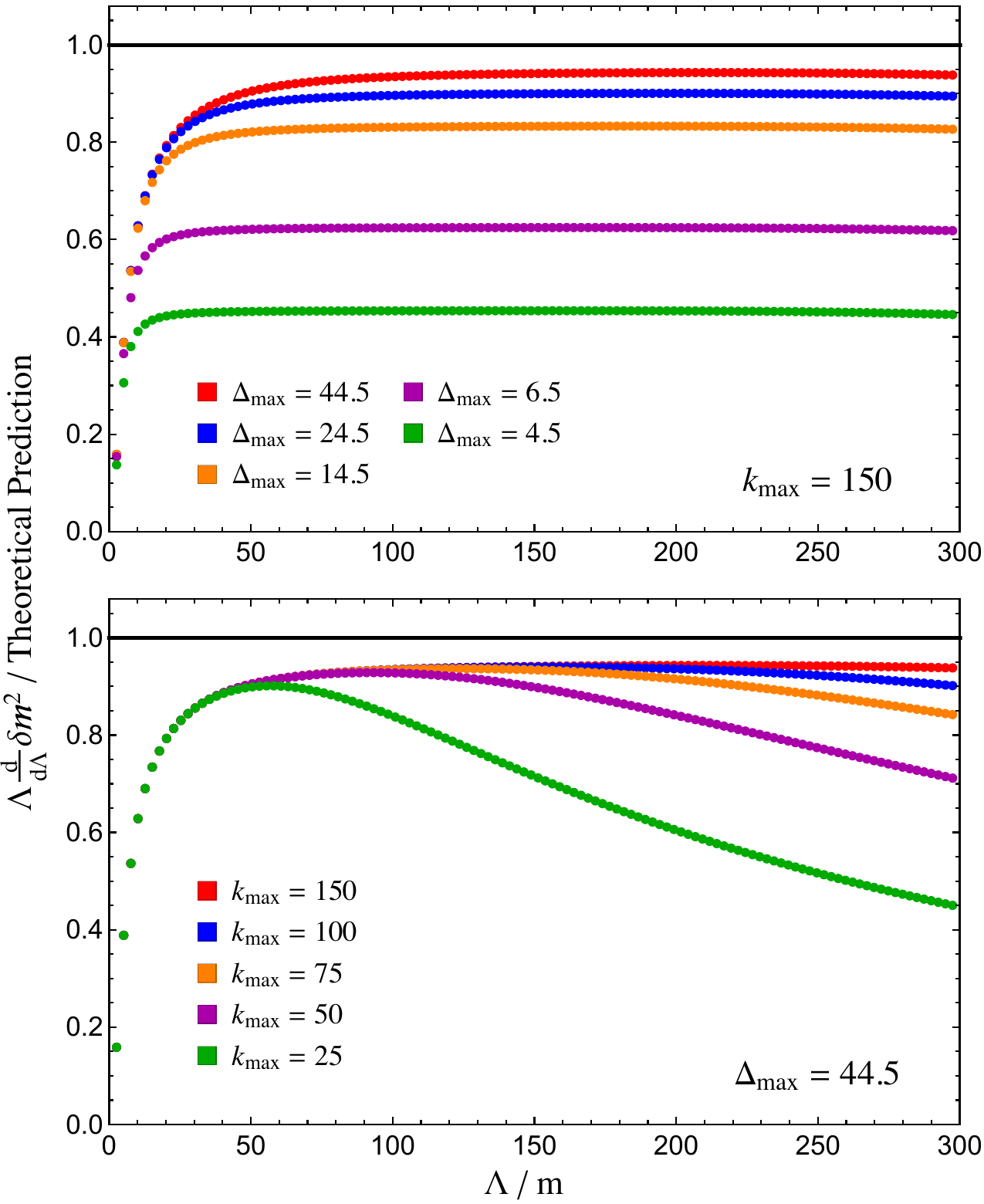}
\caption{Coefficient of log-divergent leading contribution to the one-particle mass due to the $\phi^4$ interaction with $\lambda/m=0.01$, normalized by the theoretical prediction. The resulting log-divergence is shown as a function of $\Lambda/m$ for $\kmax=150$ and several values of $\Dmax$ (top) and for $\Dmax=\fr{89}{2}$ and several values of $\kmax$ (bottom).}
\label{fig:3pShift} 
\end{center}
\end{figure}

In contrast to the $\phi^3$ interaction considered in the previous example, the resulting mass shift is logarithmically sensitive to the UV cutoff,
\be
\de m^2 = - \fr{\lambda^2}{96\pi^2}\log\Lambda + \mathrm{finite}.
\ee
While the finite term is scheme-dependent, the overall coefficient for the logarithmic divergence is universal and should be reproduced by conformal truncation. This is a nontrivial check that our UV cutoff in invariant mass is well-behaved in perturbation theory.

The $\phi^4$ interaction introduces nonzero Hamiltonian matrix elements between states with $n$ and $n\pm2$ particles. The leading correction to the one-particle mass in perturbation theory therefore arises from mixing with three-particle states. For perturbative values of $\lambda$, we can thus restrict our basis to the subspace of one- and three-particle states only. Similar to the previous subsection, we can compute the Hamiltonian matrix elements in this subspace, truncate the three-particle states in $\Dmax$ and $\kmax$, then diagonalize the resulting matrix. To isolate the logarithmic divergence in the mass shift, we compute the resulting spectrum for multiple values of $\Lambda/m$, then calculate
\be
\frac{\mathrm{d}}{\mathrm{d}\log\Lambda} \de m^2 \equiv \frac{\mathrm{d}}{\mathrm{d}\log\Lambda} \Big( \mu_{\min}^2 - m^2 \Big),
\ee  
which we compare to the theoretical value $-\fr{\lambda^2}{96\pi^2}$.

\begin{figure}[t!]
\begin{center}
\includegraphics[width=0.8\textwidth]{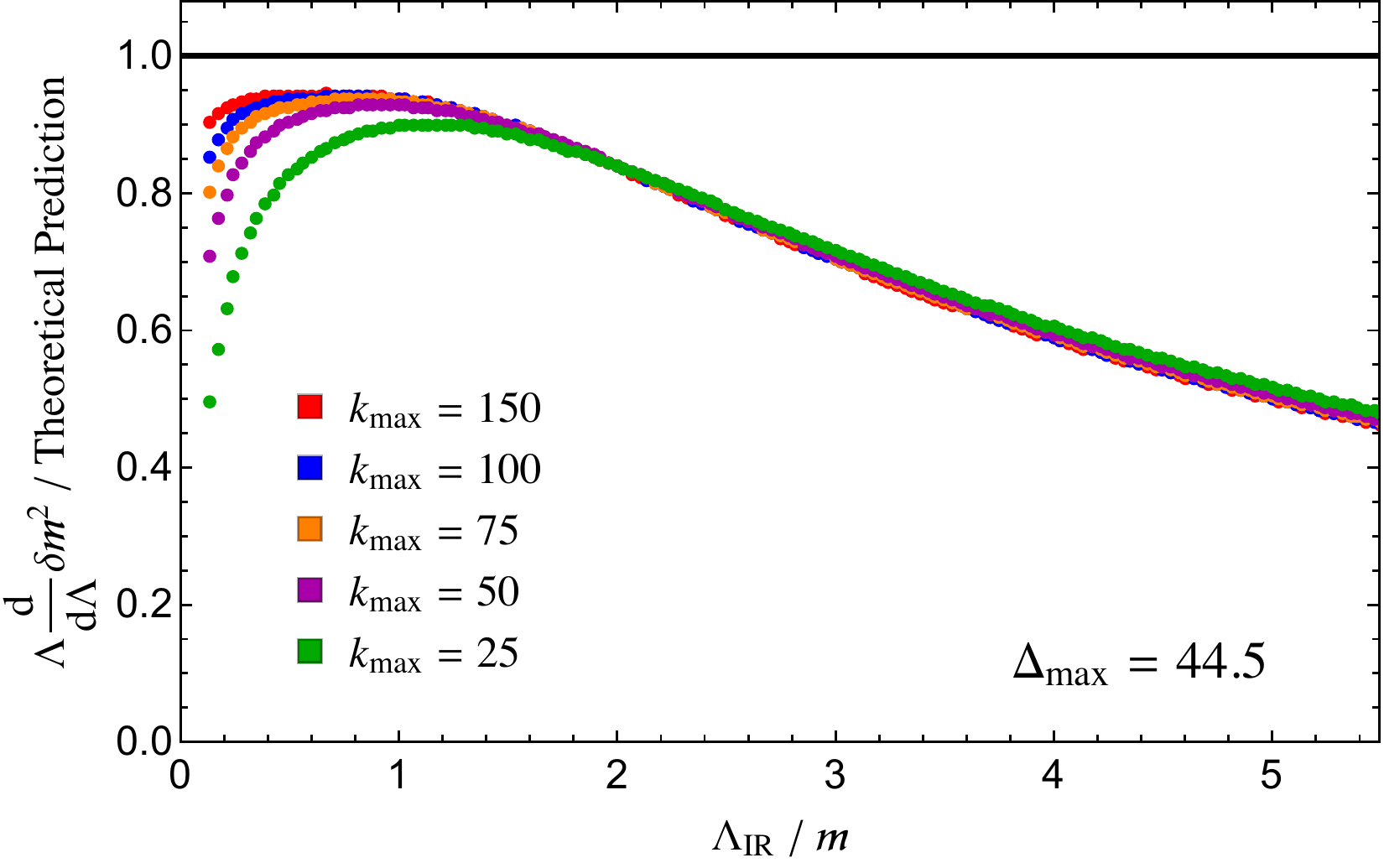}
\caption{Coefficient of log-divergent leading contribution to the one-particle mass due to the $\phi^4$ interaction as a function of $\LambdaIR \equiv \Lambda/\kmax$, for $\Dmax=\fr{89}{2}$ (or $\Lmax=43$) and multiple values of $\kmax$. For $\LambdaIR \gtrsim m$, the mass corrections collapse to a single curve, indicating that the low mass eigenstates only depend on $\Lambda$ and $\kmax$ in this fixed ratio. For low $\LambdaIR$, the curves separate due to effects from the bare mass $m$, with the peak approaching the known theoretical value for increasing $\kmax$.}
\label{fig:3pScaleCollapse} 
\end{center}
\end{figure}

Figure~\ref{fig:3pShift} shows the resulting logarithmic coefficient, normalized by the theoretical prediction, as a function of $\Lambda/m$. We have set $\lambda/m=0.01$, which is well within the regime of perturbation theory. The top plot has fixed $\kmax=150$, with $\Dmax=\fr{9}{2},\fr{13}{2},\fr{29}{2},\fr{49}{2},\fr{89}{2}$, which correspond to including 1, 2, 14, 44, and 154 Dirichlet multiplets or equivalently, a Hilbert space of 152, 303, 2115, 6645, and 23255 total states. In the bottom plot, we fix $\Dmax=\fr{89}{2}$ and vary $\kmax=25,50,75,100,150$, corresponding to 4005, 7855, 11705, 15555, and 23255 total states, respectively.

Qualitatively, we see that these results are analogous to the $\phi^3$ mass shift behavior in figure~\ref{fig:2pShift}. In particular, as both $\Dmax$ and $\kmax$ increase, the results become insensitive to $\Lambda/m$ and asymptote towards the correct value. Unlike the $\phi^3$ case, however, these plots are calculated from a derivative with respect to $\Lambda$. The fact that our results approach a constant therefore indicates that we have correctly reproduced a logarithmic UV divergence, matching our expectation from perturbation theory.

Figure~\ref{fig:3pScaleCollapse} again shows the logarithmic coefficient, but as a function of $\LambdaIR \equiv \Lambda/\kmax$. Just like the $\phi^3$ case, we see that the results for distinct values of $\kmax$ collapse to a single curve, indicating that the low-mass eigenstates only depend on the UV cutoff and $\kmax$ through this emergent IR scale. 

Finally, we can note that, compared to the $\phi^3$ results, one needs a larger value for $\Dmax$ (i.e.\ more Dirichlet multiplets) to achieve equivalent accuracy for $\phi^4$. This slower convergence occurs because we are reproducing a UV divergent observable, which is more sensitive to high-mass eigenstates than the constant mass shift in the previous subsection. In general, we expect rapid convergence in $\Dmax$ for observables which are predominantly sensitive to the low-mass spectrum and less efficient convergence as we begin to probe higher mass eigenstates.


\subsection{$O(N)$ Model at Large-$N$}
\label{sec:LargeN}

Now that we've confirmed that our truncation method correctly reproduces physical spectra in both free and weakly-coupled examples, we turn to a truly non-perturbative example in the $O(N)$ model, with the corresponding Lagrangian
\be
\Lcal = \half \p_\mu \phi_i \p^\mu \phi_i - \half m^2 \phi_i^2 - \fr{1}{4} \lambda \phi_i^2 \phi_j^2.
\ee
We consider this model in the limit $N\rightarrow\infty$, where we can compare with analytic expressions. Specifically, we use conformal truncation to compute the large-$N$ spectral density of the operator 
\be
\phivec \equiv \fr{\phi_i^2}{\sqrt{N}}.
\ee
As was discussed in section~\ref{sec:Review}, to leading order in $1/N$ with fixed $\kappa \equiv \lambda N$, this spectral density is given by
\be
\rho_{\phivec}(\mu) = \fr{\fr{1}{4\pi\mu}}{\left(1 + \fr{\kappa}{8\pi\mu}\log\Big(\fr{\mu+2m}{\mu-2m}\Big) - \fr{\kappa}{8\pi\mu}\log\Big(\fr{\Lambda+\mu}{\Lambda-\mu}\Big) \right)^2 + \left( \fr{\kappa}{8\mu}\right)^2}.
\label{LNSD}
\ee

\begin{figure}[t!]
\begin{center}
\includegraphics[width=0.8\textwidth]{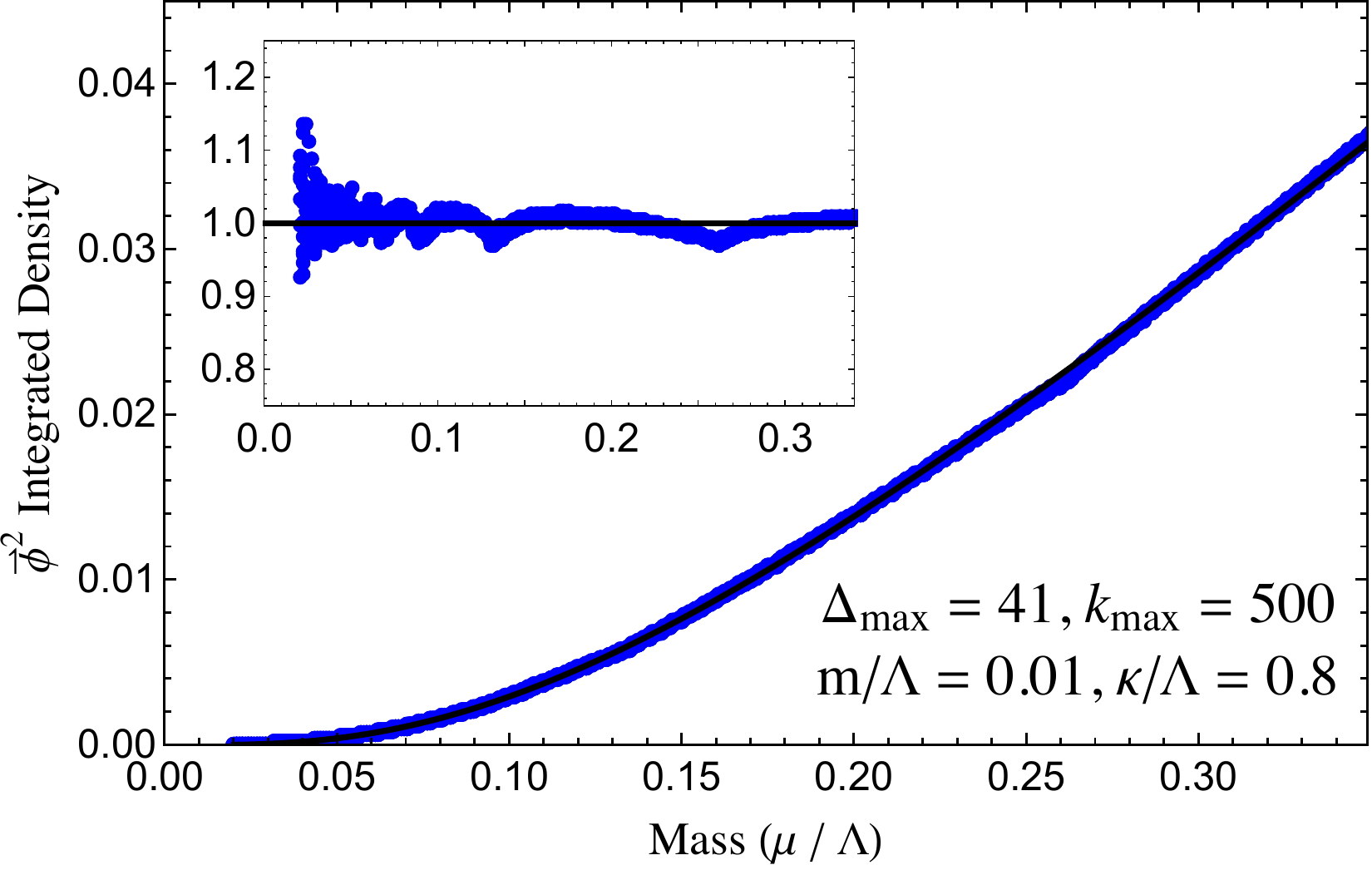}
\caption{Integrated spectral density for $\phivec$ in the large-$N$ limit with $m/\Lambda=0.01$ and $\kappa/\Lambda=0.8$, both the raw value (main plot) and normalized by the theoretical prediction (inset). The conformal truncation results (blue dots) are calculated with $\Dmax=41$ (or $\Lmax=40$) and $\kmax=500$, and compared to the known analytic expression (black line).}
\label{LN01} 
\end{center}
\end{figure}

We considered this observable in the massless case in section~\ref{sec:MasslessResults}, where the spectral density was only affected by the $\phivec$ Casimir multiplet. However, we now generalize that analysis to $m\neq0$ to demonstrate the convergence of this truncation method in $\Cmax$ for strongly-coupled theories. 

To proceed, we need a complete basis of states for the $O(N)$ theory. The original description of our basis in section~\ref{sec:MasslessBasis} was specific to $N=1$. However, for $O(N)$ singlet operators like $\phivec$, this basis requires no modification for $N>1$. This is simply a reflection of the fact that in acting with creation operators on the vacuum, there is a unique contraction of flavor indices to form a singlet. Indeed, the parameter $N$ only appears in overall normalization factors, making it easy to verify that the orthogonal polynomials forming a complete basis for the $N=1$ theory are also a complete basis for the $O(N)$ singlet sector for general $N$.

Given this basis, we need to compute the $M^2$ matrix elements using the corresponding lightcone Hamiltonian,
\be
P_+ = P_+^{(\CFT)} + \delta P_+^{(m)} + \delta P_+^{(\lambda)},
\ee 
whose precise form was discussed in section~\ref{sec:Review}. The first two terms in $P_+$ preserve particle number, and in fact the resulting $O(N)$ singlet matrix elements are completely independent of $N$. We can therefore reuse the matrix elements computed in the free field setting of subsection~\ref{sec:fft}.

We then only need to calculate the matrix elements for the interaction term $\de P_+^{(\lambda)}$. In the large-$N$ limit, any matrix elements which change particle number are suppressed by $1/N$. Thus, to leading order in $1/N$, we only need to include two-particle states to reproduce the spectral density of $\phivec$. 

After computing the large-$N$ matrix elements for the two-particle sector, we truncate the Hilbert space at a given $\Dmax$ and $\kmax$ and numerically diagonalize the truncated Hamiltonian to find the resulting approximate mass eigenstates. We then use these eigenstates to compute the integrated spectral density
\be
I_{\phivec}(\mu) \equiv \sum_{\mu_i \leq \mu} |\<\phivec(0)|\mu_i\>|^2,
\ee
which we compare to the integrated form of eq.~(\ref{LNSD}).

\begin{figure}[t!]
\begin{center}
\includegraphics[width=0.8\textwidth]{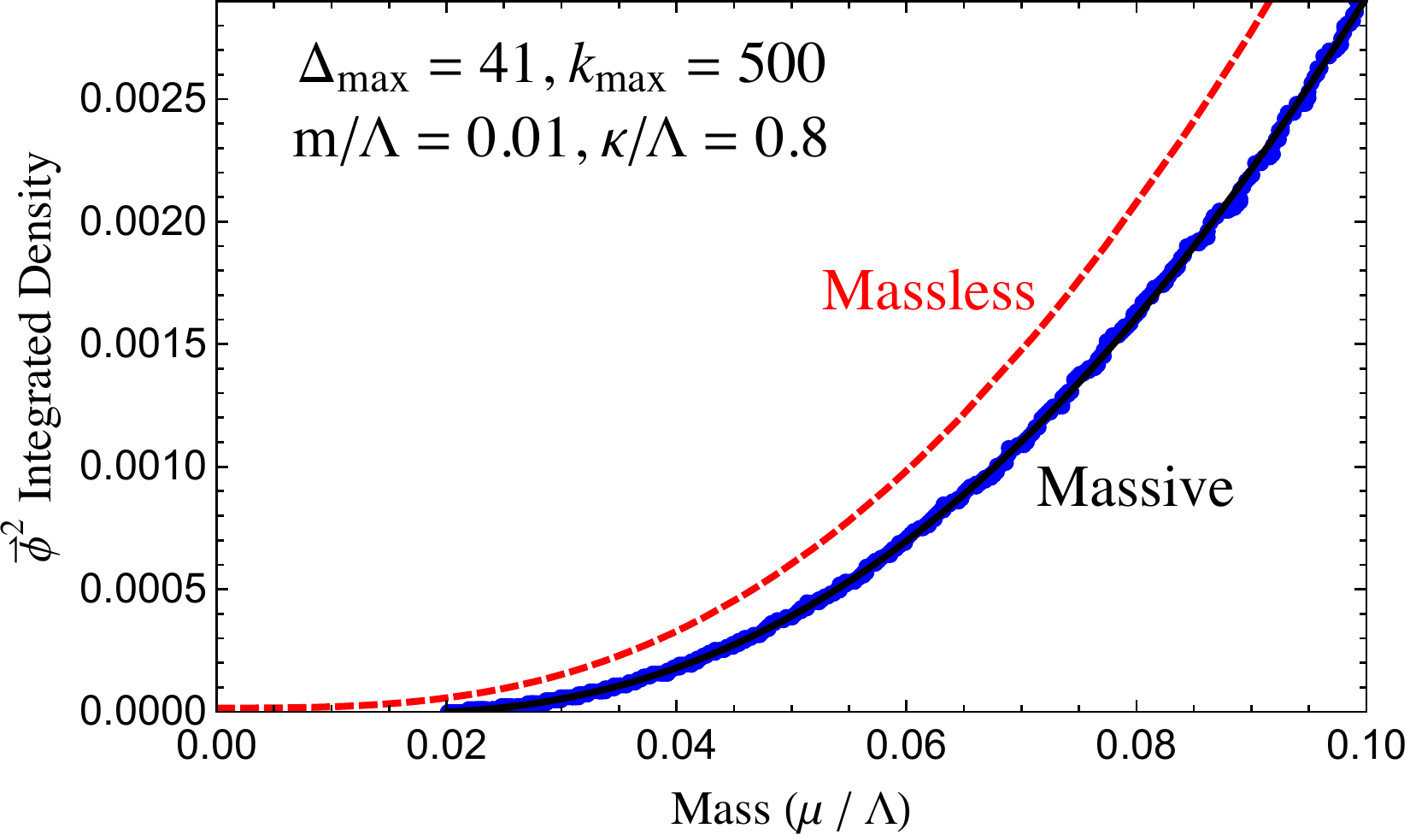}
\caption{A zoom-in of figure~\ref{LN01} into the IR. The conformal truncation results (blue dots) are compared to the theoretical prediction (solid black line), which is clearly distinguishable from the massless case (dashed red line).}
\label{LN02} 
\end{center}
\end{figure}

We can fix any mass scale in terms of the UV cutoff $\Lambda$. As the resulting spectral density has more complicated structure, we choose to set $m/\Lambda = 0.01$ to consider a wider range of invariant mass eigenvalues. We choose the effective coupling $\kappa/\Lambda = 0.8$, such that we are well within the non-perturbative regime $\kappa \gg m$.

Figure~\ref{LN01} shows the resulting integrated spectral density for $\phivec$ with $\Dmax=41$ and $\kmax=500$, corresponding to 20 Dirichlet multiplets and 10020 total states. The main plot shows the raw data (blue dots), compared to the analytic expression (solid black curve), while the subplot shows the ratio of the conformal truncation results to the theoretical prediction.

From the analytic expression in eq.~(\ref{LNSD}), we see that there are three distinct regimes for the $\phivec$ spectral density. For $\mu \gg \kappa/8$, the spectral density asymptotes to the free field theory expression reproduced in subsection~\ref{sec:fft}. As $\mu$ approaches the scale $\kappa/8$, the spectral density then transitions to a new IR theory, with a modified scaling dimension for $\phivec$. This behavior agrees with the massless case considered in section~\ref{sec:MasslessResults}.

\begin{figure}[t!]
\begin{center}
\includegraphics[width=0.9\textwidth]{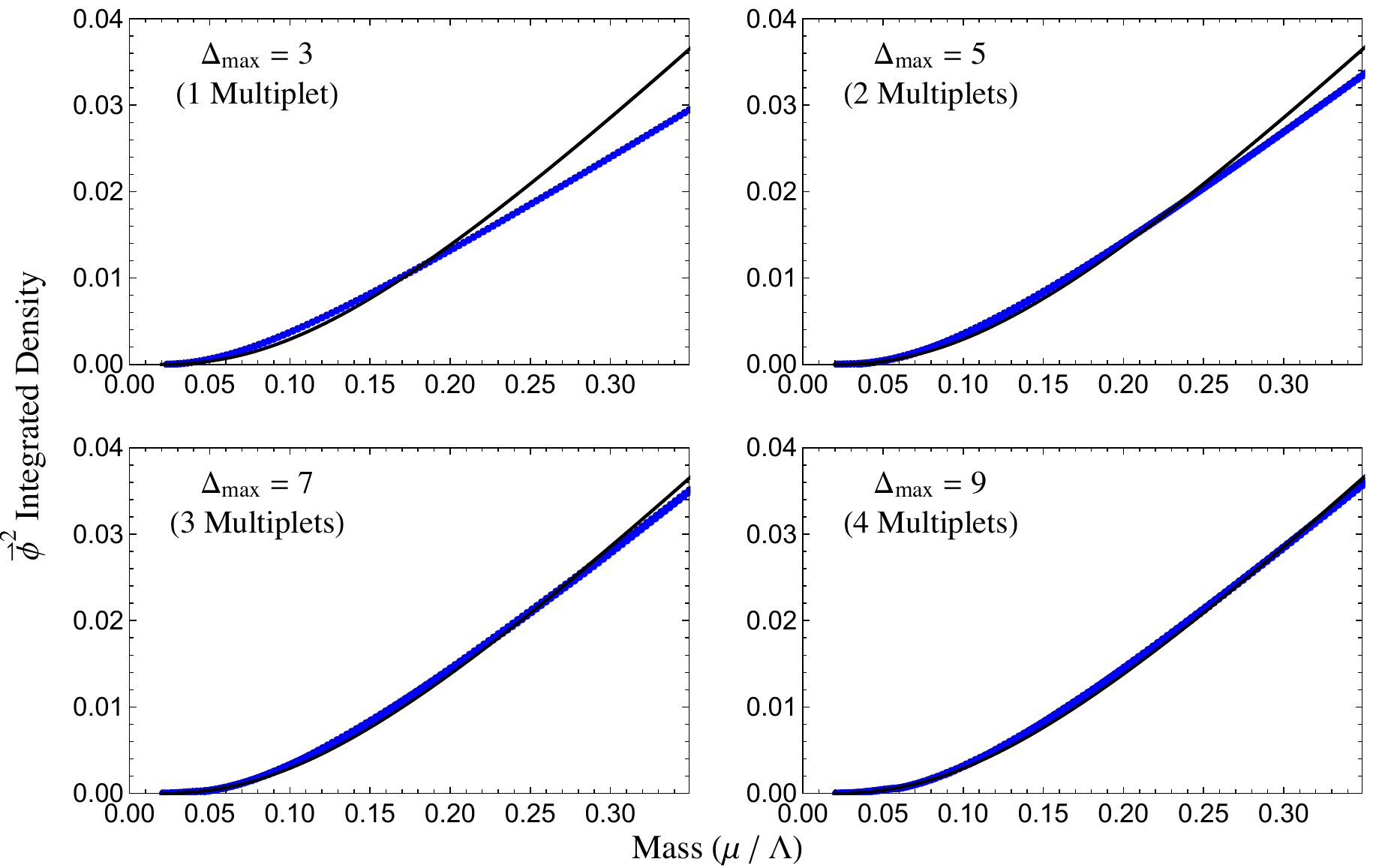}
\caption{Integrated spectral density for $\phivec$ in the large-$N$ limit with $m/\Lambda=0.01$ and $\kappa/\Lambda = 0.8$. The conformal truncation results (blue dots) are calculated with $\kmax=500$ and different values of $\Dmax$, and compared to the known analytic expression (black line).}
\label{fig:LargeNVaryDmax} 
\end{center}
\end{figure}

However, as we continue farther into the IR, to $\mu \sim m$, we find that the spectral density deviates from the massless case due to the presence of a mass gap. Figure~\ref{LN02} focuses on this IR region, comparing the truncation results to both the massless (dashed red) and massive (solid black) theoretical predictions. As we can see, the numerical results correctly reproduce the massive spectral density up to the mass threshold of $2m$.   
Our truncation method is therefore able to reproduce the entire RG flow encoded in the spectral density of $\phivec$. To study the convergence of these non-perturbative results, figure~\ref{fig:LargeNVaryDmax} shows the same integrated spectral density for $\kmax=500$ and $\Dmax = 3, 5, 7, 9$, which corresponds to including 1, 2, 3, and 4 Dirichlet multiplets, or equivalently 501, 1002, 1503, and 2004 total states. As we can see, even at strong coupling this truncation scheme needs few Dirichlet multiplets to reproduce the analytic expression (black line).

\section{Discussion and Future Directions}
\label{sec:Discussion}

The modern picture of QFT suggests that all information about a particular RG flow is contained within the original UV fixed point. Based on recent progress in understanding the structure of CFTs, it's worth revisiting the question of whether one can extract this information to access real-time dynamics in non-perturbative settings. The intuition from AdS/CFT is that dynamical observables for the lowest energy excitations are predominantly encoded by the lowest eigenstates of the conformal quadratic Casimir, $\Ccal$. Motivated by this perspective, in this paper we have proposed a new Hamiltonian truncation scheme for systems at infinite volume. We tested this framework for deformations of free scalar CFTs in 3D by diagonalizing the Hamiltonian in a finite Hilbert space defined by truncating in both $\Ccal$ and a discretization parameter $k$. Comparing our truncation scheme to known analytic results, we found rapid convergence in $\Ccal$ and slower convergence in $k$.

It is worth emphasizing that these two parameters are quite different in nature. The conformal Casimir $\Ccal$ is intimately linked to the complexity of
the basis, and by dialing it we are effectively increasing (in holographic terms) the number of bulk fields needed to describe a state. The discretization variable
$k$, on the other hand, simply controls our resolution of the continuous parameter $\mu$ by expanding the wavefunctions of Hamiltonian eigenstates in a basis of polynomials,
\be
\<\Ccal,\ell;\mu|\psi\> \equiv \rho_\Ocal(\mu) \, \psi_\Ocal(\mu) = \rho_\Ocal(\mu) \sum_k \psi_{\Ocal,k} \, g_k(\mu).
\ee
It would be interesting to explore whether there is a better discretization choice, or perhaps even an approach which avoids discretizing $\mu$ altogether, as the original AdS wavefunctions are known exactly, resulting in a fully infinite-volume computation. For instance, one could imagine inserting a complete set of states in the eigenvalue equation $M^2 |\psi\> = \mu_\psi^2 |\psi\>$ to obtain the infinite set of coupled integral equations
\be
\mu^2 \psi_\CO(\mu) +   \lambda \sum_{\CO'} \int _0^{\Lambda^2} d\mu^{\prime \, 2} \CM^{(\Ocal_R)}_{\Ccal\ell,\Ccal'\ell'} (\mu,\mu') \, \psi_{\CO'}(\mu') = \mu_\psi^2 \psi_\CO(\mu),
\ee
where the matrix elements $\CM^{(\Ocal_R)}_{\Ccal\ell,\Ccal'\ell'}$ can be obtained from the three-point functions $\langle \CO\CO_R\CO'\rangle$, as discussed in appendix~\ref{app:GeneralCFT}.  

The above system of equations is also suggestive of a holographic interpretation.   Namely, since the function $\CM(\mu,\mu')$ is
entirely determined in terms of CFT kinematics, up to the overall OPE coefficient, we could replace it with an appropriate integral over an AdS bulk point weighted by Bessel functions (i.e.\ as
dictated by the AdS representation of the Fourier transform of the three-point Wightman function). Thus, an appropriate
Bessel function transform of the the above system of integral equations may yield a set of coupled approximate differential equations in the AdS radial coordinate $z$
(at least when $\Lambda$ is taken to infinity, and in the absence of divergences).  It would be interesting to explore this picture
further and study the connection to holographic RG flows, as well as the effect of keeping $\Lambda$ finite, from the bulk perspective.
One could also study the structure of the above equations in the large-$N$ limit, where they might simplify (as indeed they did
in the case of the large-$N$ flow we considered in this paper).

From this perspective, the only consequence of working in lightcone quantization is a set of selection rules restricting the allowed matrix elements. It would be interesting to understand the emergence of these rules in a general CFT. In fact, one might wonder whether the framework of lightcone quantization is even necessary, or if the entire truncation scheme can instead be defined solely in terms of CFT Wightman functions, evaluated using AdS kinematics.

While the focus in this work has been on testing our framework in controllable examples, we now have all the tools necessary to study strongly-coupled dynamics in theories like the 3D Ising and $O(N)$ models. At a practical level, the main computational challenge is constructing a basis which is fully symmetric under particle exchanges. Here we proceeded with a brute force method of generating all Casimir eigenfunctions, then numerically solving for symmetric linear combinations. However, this approach is inefficient for states with higher particle number. In future work, we plan to instead use a more effective strategy of constructing the basis directly in terms of manifestly symmetric polynomials \cite{FutureUs}.

Because we specifically considered deformations of a free CFT, we were able to utilize the Fock space representation of operators to compute matrix elements, and indirectly therefore the OPE coefficients. However, it would be interesting to test the general framework in a context where the OPE coefficients are already known. One potential setting would be to study a relevant deformation of a minimal model in 2D.  For instance, one could directly consider the $\epsilon$ and $\sigma$ perturbations of the 2D Ising model and compare them to equivalent RG flows starting from the free 2D scalar CFT with a mass and a quartic interaction tuned to criticality \cite{FutureUs}.  This would allow for two independent means of computing the same spectral densities or other dynamical quantities.  The $Z_2$ broken phase of this model would also provide an interesting setting to test how the lightcone framework behaves under spontaneous symmetry breaking.

While we chose to focus on the test case of 3D scalar CFTs, our approach should apply equally well in any number of dimensions, so one obvious direction is to repeat this analysis in $d > 3$. It would also be useful to study the generalization of this framework to theories with fermions or gauge fields, such as scalar QED or pure Yang-Mills.

Looking ahead, the overall framework of conformal truncation motivates the development or improvement of methods for determining the OPE coefficients for CFTs. So far, numerical bootstrap results have primarily been focused on scaling dimensions, but it would be interesting to obtain additional values for OPE coefficients in theories such as the 3D Ising model \cite{Kos:2016ysd}. Hamiltonian truncation methods are a promising complement to the conformal bootstrap program, and we encourage further work on this technique.


\section*{Acknowledgments}    

We would like to thank Nikhil Anand, Rich Brower, Chris Brust, Andy Cohen, Liam Fitzpatrick, Vincent Genest, Matthijs Hogervorst, Jared Kaplan, Gustavo Marques Tavares, Jo\~{a}o Penedones, Slava Rychkov, Balt van Rees, and Yiming Xu for valuable discussions. We are especially grateful to Liam for helpful comments on the draft. This work was supported by DOE grant DE-SC0010025.  We would also like to thank the Weizmann Institute and the Simons Center at Stony Brook for hospitality while this work was completed.  This work was performed in part at the Aspen Center for Physics, which is supported by National Science Foundation grant PHY-1066293.


\appendix

\section{Overview of Interactions}
\label{app:Interactions}

In this appendix, we derive the explicit form for all contributions to the lightcone Hamiltonian, which includes both the original unperturbed operators associated with the UV theory, as well as any corrections due to the various relevant operators added to the Lagrangian. Because our original UV CFT corresponds to free field theory, we can use the Fock space decomposition of the scalar field $\phi$ to construct these Hamiltonian terms, which greatly simplifies the calculation of the resulting matrix elements.


\subsection{Conventions}

We consider scalar field theory in three spacetime dimensions, with the associated Lorentzian metric
\be
ds^2 = dt^2 - dx^2 - dy^2.
\ee
Instead of using these standard coordinates, however, we define the ``lightcone'' coordinates $x^\pm \equiv \fr{1}{\sqrt{2}}(t \pm x)$, $x^\perp \equiv y$ such that the metric then takes the form
\be
ds^2 = 2 d\xp d\xm - dx^{\perp2}.
\ee
We use the framework of lightcone quantization, where the new coordinate $\xp$ is treated as the ``time'' direction and $\xm$, $x^\perp$ are the ``spatial'' directions. These coordinates also have the associated momenta $p_\mu \equiv i \p_\mu$, such that
\be
p^2 = 2 p_+ p_- - p_\perp^2.
\ee

Our basis of states is defined within the trivial UV fixed point of free field theory containing one or more massless scalar fields. In lightcone quantization, the Hilbert space corresponds to a complete set of states defined on a spacetime slice of constant ``time'' $\xp$. The real scalar field $\phi(x)$ acting on this timeslice can then be expanded in terms of creation and annihilation operators 
\be
\phi(x) = \int \fr{dp_- dp_\perp}{(2\pi)^2\sqrt{2p_-}} \left( e^{-ip \cdot x} a_p + e^{ip \cdot x} a^\dagger_p \right),
\label{eq:PhiDefApp}
\ee
where these raising/lowering operators satisfy the canonical commutation relations
\be
\comm{a_p}{a^\dagger_q} = (2\pi)^2 \de^2(p-q).
\ee
The scalar field $\phi$ is therefore normalized such that the equal-time commutator is \cite{Dalley:1992yy}
\be
\comm{\phi(x)}{\pi(y)} \equiv \comm{\phi(x)}{\p_- \phi(y)} = \fr{i}{2} \de^2(x-y).
\ee

The basis for this theory can then be written in terms of momentum eigenstates, which are created by acting with raising operators on the vacuum,
\be
|p\> \equiv \sqrt{2p_-} \, a^\dagger_p |0\>,
\ee
such that the resulting states have the normalization
\be
\<p|q\> = 2p_- (2\pi)^2 \de^2(p-q).
\ee
Because this inner product is Lorentz invariant, we can choose to work in a particular reference frame, using a single representative of the full one-particle Lorentz multiplet.

We can construct the rest of our basis as linear combinations of the multi-particle states $|p_1,\cdots,p_n\>$. These basis states $|\Ccal,\ell;\vec{P},k\>$ can then be chosen to be total momentum eigenstates, with the universal normalization
\be
\<\Ccal,\ell;\vec{P},k|\Ccal',\ell';\vec{P}',k'\> = 2 P_- (2\pi)^2 \de^2(P-P') \, \de_{\Ccal\Ccal'} \, \de_{\ell\ell'} \, \de_{kk'}.
\label{eq:NormConvention}
\ee
Without loss of generality, we will specifically work in the reference frame with total momentum
\be
\vec{P} \equiv (P_-,P_\perp) = (P_-,0).
\ee

In the case where the UV theory contains $N$ scalar fields, each individual field has its own mode expansion
\be
\phi_i(x) = \int \fr{d^2p}{(2\pi)^2\sqrt{2p_-}} \left( e^{-ip \cdot x} a_{p,i} + e^{ip \cdot x} a^\dagger_{p,i} \right).
\ee
These raising/lowering operators satisfy the same commutation relations, with the added constraint that modes from distinct fields always commute,
\be
\comm{a_{p,i}}{a^\dagger_{q,j}} = (2\pi)^2 \de^2(p-q) \, \de_{ij}.
\ee
The basis states $|\Ccal,\ell;\vec{P},k\>$ can then be written in terms of multi-particle states built from these modes organized according their $O(N)$ flavor structure.


\subsection{Contributions to Invariant Mass Operator}

We are specifically interested in studying the low-mass eigenstates, which means we need to diagonalize the invariant mass operator $M^2$. The invariant mass can be expressed in terms of translation generators as
\be
M^2 \equiv 2 P_+ P_- - P_\perp^2.
\ee
Each of these momentum operators can then be derived from the stress-energy tensor,
\be
T_{\mu\nu} \equiv \fr{\p \Lcal}{\p (\p^\mu \phi)} \p_\nu \phi - \eta_{\mu\nu} \Lcal.
\ee
We can therefore directly relate terms in the Lagrangian to contributions to the invariant mass operator. Starting with the original UV Lagrangian,
\be
\Lcal_{\CFT} = \half \p_\mu \phi \p^\mu \phi,
\ee
we can then obtain the resulting stress-energy tensor,
\be
T_{\mu\nu} = \p_\mu \phi \p_\nu \phi - \half \eta_{\mu \nu} \p_\sigma \phi \p^\sigma \phi.
\ee
This expression for $T_{\mu\nu}$ in turn leads to the translation generators
\be
\begin{split}
P_+ &\equiv \int d^2x \, T_{-+} = \half \int d^2x \, (\p_\perp \phi)^2, \\
P_- &\equiv \int d^2x \, T_{--} = \int d^2x \, (\p_- \phi)^2, \\
P_\perp &\equiv \int d^2x \, T_{-\perp} = \int d^2x \, (\p_- \phi)(\p_\perp \phi).
\end{split}
\ee

Using our definition of $\phi(x)$ in eq.\ (\ref{eq:PhiDefApp}), we can expand these generators in terms of creation and annihilation operators. Note that all such contributions to the momentum generators are to be normal-ordered. A simple first example is $P_\perp$, which can be written as
\begin{align}
P_\perp &= \int \fr{d^2x \, d^2p \, d^2q}{(2\pi)^4\sqrt{4p_- q_-}} \, \norder{\p_-\Big( e^{ip \cdot x} a^\dagger_p + e^{-ip \cdot x} a_p \Big) \p_\perp \Big( e^{iq \cdot x} a^\dagger_q + e^{-iq \cdot x} a_q \Big)} \\
&= \int \fr{d^2x \, d^2p \, d^2q}{(2\pi)^4\sqrt{4p_- q_-}} p_- q_\perp \Big(  e^{i(p-q)\cdot x} a^\dagger_p a_q + e^{-i(p-q) \cdot x} a^\dagger_q a_p - e^{i(p+q) \cdot x} a^\dagger_p a^\dagger_q - e^{-i(p+q)\cdot x} a_p a_q \Big). \nn
\end{align}
Evaluating the spatial integral simply enforces conservation of momentum, fixing $q$ in terms of $p$. Positivity of lightcone momenta further simplifies the expression, leading to
\be
P_\perp = \int \fr{d^2p}{(2\pi)^2} \, a^\dagger_p a_p \, p_\perp.
\ee
This final result is unsurprising, as the operator $P_\perp$ should simply correspond to a sum over the number of particles, each weighted by their transverse momentum.

Looking at the original expression for $P_-$, we see that it takes an almost identical form, but with $p_\perp \ra p_-$,
\be
P_- = \int \fr{d^2p}{(2\pi)^2} \, a^\dagger_p a_p \,  p_-.
\ee
Finally, we can obtain the original lightcone Hamiltonian,
\be
P_+^{(\CFT)} = \int \fr{d^2p}{(2\pi)^2} \, a^\dagger_p a_p \, \fr{p_\perp^2}{2p_-}.
\ee
We can easily understand this expression by realizing that, for massless particles satisfying the equation of motion, the lightcone energy $p_+$ can be expressed in terms of the ``spatial'' momenta as
\be
p_+ = \fr{p_\perp^2}{2p_-}.
\ee
This integral expression for $P_+$ is therefore similar to the others, with the sum over particles weighted by their on-shell lightcone energy.

While we are using a basis of states associated with the UV fixed point of free field theory, we wish to study the resulting IR spectrum after including (some combination of) the following relevant interactions,
\be
\de \Lcal = -\half m^2 \phi^2 - \fr{1}{3!} g \phi^3 - \fr{1}{4!} \lambda \phi^4.
\ee
These contributions to the Lagrangian all lead to a shift in the stress-energy tensor of the form
\be
\de T_{\mu \nu} = \eta_{\mu \nu} \left( \half m^2 \phi^2 + \fr{1}{3!} g \phi^3 + \fr{1}{4!} \lambda \phi^4 \right).
\ee
Because this correction to $T_{\mu\nu}$ is proportional to the metric, we can easily see that it only leads to a shift in $P_+$, with no effect on the other two generators.

Let's consider each of these corrections separately, starting with the mass term. The evaluation of this `interaction' is quite similar to that of the kinetic term, leading to
\be
\begin{split}
\de P^{(m)}_+ &= \fr{m^2}{2} \int \fr{d^2x \, d^2p \, d^2q}{(2\pi)^4\sqrt{4p_- q_-}} \norder{\Big( e^{ip \cdot x} a^\dagger_p + e^{-ip \cdot x} a_p \Big) \Big( e^{iq \cdot x} a^\dagger_q + e^{-iq \cdot x} a_q \Big)} \\
&= \int \fr{d^2p}{(2\pi)^2} \, a^\dagger_p a_p \, \fr{m^2}{2p_-}.
\end{split}
\ee
The lightcone energy of each particle is therefore shifted by a contribution proportional to $m^2$, which is consistent with the on-shell expression
\be
p_+ = \fr{p_\perp^2 + m^2}{2p_-}.
\ee 

Next, we can turn to the cubic interaction, which is slightly more complicated than the previous operators. Using the same basic analysis as before, we have
\begin{align}
\de P^{(g)}_+ &= \fr{g}{3!} \int \fr{d^2x \, d^2p \, d^2q \, d^2k}{(2\pi)^6\sqrt{8p_- q_- k_-}} \norder{\Big( e^{ip \cdot x} a^\dagger_p + e^{-ip \cdot x} a_p \Big) \Big( e^{iq \cdot x} a^\dagger_q + e^{-iq \cdot x} a_q \Big) \Big( e^{ik \cdot x} a^\dagger_k + e^{-ik \cdot x} a_k \Big)} \nonumber \\
&= \fr{g}{2} \int \fr{d^2p \, d^2q}{(2\pi)^4 \sqrt{8 p_- q_- (p_- + q_-)}} \Big( a^\dagger_p a^\dagger_q a_{p+q} + a^\dagger_{p+q} a_p a_q \Big).
\end{align}
Unlike the previous generators, which simply consist of a weighted sum over all particles, this generator clearly mixes states with distinct particle numbers. The quartic interaction leads to a similar contribution,
\be
\de P^{(\lambda)}_+ = \fr{\lambda}{24} \int \fr{d^2p \, d^2q \, d^2k}{(2\pi)^6\sqrt{8p_- q_- k_-}} \left( \fr{4 a^\dagger_p a^\dagger_q a^\dagger_k a_{p+q+k}}{\sqrt{2(p_- + q_- + k_-)}} + h.c. + \fr{6 a^\dagger_p a^\dagger_q a_k a_{p+q-k}}{\sqrt{2(p_- + q_- - k_-)}} \right).
\ee

Let's now briefly turn to the case of $N$ scalar fields. Limiting ourselves to operators which preserve the $O(N)$ flavor symmetry, we obtain the Lagrangian contributions
\be
\Lcal = \half \p_\mu \phi_i \p^\mu \phi_i - \half m^2 \phi_i^2 - \fr{1}{4} \lambda \phi_i^2 \phi_j^2.
\ee
We can then repeat the same analysis as above, using the mode expansion of $\phi_i$ to obtain the corresponding lightcone Hamiltonian. The resulting kinetic and mass terms are almost identical to the single field case,
\be
\begin{split}
P_+^{(\CFT)} &= \half \int d^2x \, (\p_\perp \phi_i) (\p_\perp \phi_i) = \int \fr{d^2p}{(2\pi)^2} \, a^\dagger_{p,i} a_{p,i} \, \fr{p_\perp^2}{2p_-}, \\
\de P_+^{(m)} &= \half \int d^2x \, m^2 \phi_i^2 = \int \fr{d^2p}{(2\pi)^2} \, a^\dagger_{p,i} a_{p,i} \, \fr{m^2}{2p_-}.
\end{split}
\ee
However, the quartic interaction has more complicated flavor structure, leading to the contributions
\be
\begin{split}
\de P^{(\lambda)}_+ = \fr{\lambda}{2} \int \fr{d^2p \, d^2q \, d^2k}{(2\pi)^6 \sqrt{8p_- q_- k_-}} &\bigg( \fr{a^\dagger_{p,i} a^\dagger_{q,i} a_{k,j} a_{p+q-k,j} + 2a^\dagger_{p,i} a^\dagger_{k,j} a_{q,i} a_{p+q-k,j}}{\sqrt{2(p_- + q_- - k_-)}} \\
& \, + \, \fr{2 a^\dagger_{p,i} a^\dagger_{q,i} a^\dagger_{k,j} a_{p+q+k,j} + 2 a^\dagger_{p+q+k,j} a_{p,i} a_{q,i} a_{k,j}}{\sqrt{2(p_- + q_- + k_-)}} \bigg).
\end{split}
\ee
As we demonstrate in appendix \ref{app:MasslessMatrix}, the very first term in this Hamiltonian provides the dominant contribution in the large-$N$ limit, such that we can safely ignore the other terms in our analysis.


\section{Derivation of Conformal Casimir Eigenstates}
\label{app:CasimirBasis}

The Hilbert space of our three-dimensional UV CFT is naturally described in terms of local operators acting on the vacuum,
\be
|\Ccal,\ell;\vec{P},k\> \equiv \int d\mu^2 g_k(\mu) \int d^3x \, e^{-iP\cdot x} \Ocal(x)|0\>.
\ee
Each operator $\Ocal(x)$ therefore defines an infinite number of states, parameterized by the weight functions $g_k(\mu)$, which together form an irreducible representation of the conformal group $SO(2,2)$.

In general, the $d$-dimensional conformal group is generated by translations $P_\mu$, Lorentz transformations $L_{\mu\nu}$, dilatations $D$, and special conformal transformations $K_\mu$, which satisfy the relevant commutation relations (see \cite{Aharony:1999ti} for a more thorough discussion, including the rest of the conformal algebra)
\be
\comm{D}{P_\mu} = -i P_\mu, \quad \comm{K_\mu}{P_\nu} = 2i L_{\mu\nu} + 2i\eta_{\mu\nu} D, \quad \comm{L_{\mu\nu}}{P_\rho} = -i(\eta_{\mu\rho} P_\nu - \eta_{\nu\rho} P_\mu).
\label{eq:ConformalAlgebra}
\ee
The irreducible representations of the conformal group then are characterized by their eigenvalue under the conformal quadratic Casimir
\be
\Ccal \equiv -D^2 - \half (P_\mu K^\mu + K_\mu P^\mu) + \half L_{\mu\nu} L^{\mu\nu}.
\ee
The resulting Casimir eigenvalue for each representation is determined by the scaling dimension $\De$ and spin $\ell$ of the associated operator $\Ocal(x)$,
\be
\Ccal = \De(\De-d) + \ell(\ell+d-2).
\ee

As our particular UV CFT is free field theory containing a single massless scalar field $\phi$, the space of local operators is built from combinations of $P_\mu$ and $\phi$ of the schematic form
\be
\Ocal(x) = \sum_{\{m_n\}} C^\Ocal_{\{m_n\}} P^{m_1} \phi(x) P^{m_2} \phi(x) \cdots P^{m_n} \phi(x). 
\ee
We can then use the Fock space mode expansion of $\phi(x)$ to rewrite our basis states as
\be
|\Ccal,\ell;\vec{P},k\> = \int d\mu^2 g_k(\mu) \int \fr{d^2p_1 \cdots d^2p_n}{(2\pi)^{2n} 2p_{1-} \cdots 2p_{n-}} (2\pi)^3 \de^3\Big(\sum_i p_i - P\Big) F_\Ocal(p)|p_1,\cdots,p_n\>.
\ee 
We can thus map each local operator to a corresponding function of particle momenta,
\be
\Ocal(x) \ra F_\Ocal(p) \equiv \sum_{\{m_n\}} C^\Ocal_{\{m_n\}} \, p_1^{m_1} p_2^{m_2} \cdots p_n^{m_n}.
\ee
In order to find a complete basis for the CFT Hilbert space, we need to first derive the differential form for the conformal Casimir in momentum space, then calculate the corresponding eigenfunctions $F_\Ocal(p)$. Finally, we can use the resulting inner product to obtain an orthogonal set of weight functions $g_k(\mu)$.

We can simplify this calculation by noting that the scalar field $\phi$ satisfies the equation of motion
\be
P^2 \phi(x) \equiv (2P_+ P_- - P_\perp^2)\phi(x) = 0,
\ee
which means that we can write the conformal Casimir, and the resulting basis functions, solely in terms of the two momentum components $p_-, p_\perp$. We therefore need to first determine the action of the conformal generators on ``building blocks'' of the form
\benn
P_-^a P_\perp^k \phi(x) \ra p_-^a p_\perp^k.
\eenn

Because the Casimir commutes with translations, we only need to consider its action on operators located at the origin in order to determine its form in momentum space. Our approach will therefore be to derive the action of the individual conformal generators on building blocks located at the origin, then obtain momentum space differential operators which replicate these conformal transformations. We can then combine these differential operators together to obtain the momentum space version of the conformal Casimir. While we specifically consider the case of scalars in $d=3$, this general procedure is equally applicable in any number of dimensions and to free fields with spin.

First, we need to determine the behavior of the scalar operator $\phi$ under conformal transformations. By keeping $\phi$ at the origin, we greatly simplify the action of the conformal generators,
\be
D\phi(0) = -i\De_{\phi} \, \phi(0), \quad P_\mu \phi(0) = i\p_\mu \phi(0), \quad L_{\mu\nu} \phi(0) = K_\mu \phi(0) = 0,
\label{eq:ActionPhi}
\ee
where $\De_\phi = \half$ in three dimensions. We can then combine these $\phi(0)$ conformal transformations with the commutation relations in eq.~(\ref{eq:ConformalAlgebra}) to derive the corresponding transformations of our building blocks.

As a simple example, let's first consider the action of the dilatation operator $D$. Using its commutation relations with $P_\mu$, we can obtain the general expression
\be
D P_{\nu_1} \cdots P_{\nu_n} \phi(0) = -i(\De_\phi+n) P_{\nu_1} \cdots P_{\nu_n} \phi(0),
\ee
which we can then use to derive the action on the building block
\be
D P_-^a P_\perp^k \phi(0) = -i(\De_\phi + a + k) P_-^a P_\perp^k \phi(0).
\ee
Unsurprisingly, the dilatation operator simply counts the total number of insertions of $P$. We can then convert this conformal transformation into the momentum space differential operator
\be
D p_-^a p_\perp^k = -i \left( \De_\phi + p_{-} \fr{\p}{\p p_{-}} + p_{\perp} \fr{\p}{\p p_{\perp}} \right) p_-^a p_\perp^k = -i(\De_\phi + a + k) p_-^a p_\perp^k.
\ee
Generalizing this differential operator to act on an arbitrary function of multiple momenta $F_\Ocal(p)$, we then obtain
\be
D = -i \sum_i \Big( \De_\phi + p_{i-} \p_{i-} + p_{i\perp} \p_{i\perp} \Big),
\ee
where the sum is over particle number.

It is important to note that this expression is \emph{not} the momentum space form for dilatations. Instead, this is a differential operator which replicates the action of $D$ on operators $\Ocal$ located at the origin, and is merely an intermediate step in deriving the momentum space form for the conformal Casimir.

We can then repeat this process for other conformal generators, first deriving their action at the origin and then converting that expression into a momentum space differential operator. For Lorentz transformations, we obtain the building block transformation
\be
L_{\mu_1\mu_2} P_{\nu_1} \cdots P_{\nu_n} \phi(0) = -i \sum_i (\eta_{\mu_1\nu_i} P_{\mu_2} - \eta_{\mu_2 \nu_i} P_{\mu_1}) P_{\nu_1} \cdots \hat{P}_{\nu_i} \cdots P_{\nu_n} \phi(0),
\ee
where the notation $\hat{P}$ indicates that the operator is absent. Finally, we can consider the special conformal transformations,
\benn
\begin{split}
&K_\mu P_{\nu_1} \cdots P_{\nu_n} \phi(0) \\
&= 2(\De_\phi+n-1) \sum_i \eta_{\mu\nu_i} P_{\nu_1} \cdots \hat{P}_{\nu_i} \cdots P_{\nu_n} \phi(0) - 2 \sum_{i<j} \eta_{\nu_i\nu_j} P_\mu P_{\nu_1} \cdots \hat{P}_{\nu_i} \cdots \hat{P}_{\nu_j} \cdots P_{\nu_n} \phi(0).
\end{split}
\eenn

We can then use the resulting differential operators to derive the momentum space version of the conformal Casimir,
\be
\begin{split}
\Ccal = \sum_{i<j} &\Bigg[ 2 \De_\phi^2 - 2p_{i-} p_{j-} (\p_{i-} - \p_{j-})^2 + 2 \De_\phi (p_{i-} - p_{j-}) (\p_{i-} - \p_{j-}) \\
& \, - \, 2(p_{i-} p_{j\perp} + p_{i\perp} p_{j-}) (\p_{i-} - \p_{j-}) (\p_{i\perp} - \p_{j\perp}) + 2\De_\phi (p_{i\perp} - p_{j\perp}) (\p_{i\perp} - \p_{j\perp}) \\
& \, - \, \fr{(p_{i-} p_{j\perp} + p_{i\perp} p_{j-})^2}{2p_{i-} p_{j-}} (\p_{i\perp} - \p_{j\perp})^2 \Bigg] + \sum_i \De_\phi(\De_\phi-3).
\end{split}
\ee
We can now construct a complete basis for the UV CFT by finding the eigenfunctions of this differential operator.

However, so far we have ignored the polarization structure of operators with spin. Because we are working in lightcone quantization, our basis functions do not have manifest Lorentz symmetry, such that different polarization components of the same Lorentz representation correspond to distinct eigenfunctions $F_\Ocal(p)$, though with the same Casimir eigenvalue.

However, these distinct basis functions are still related by Lorentz transformations. For each spin multiplet, we therefore only need to obtain the basis function for a single component, then act with a combination of Lorentz generators known as the Pauli-Lubanski pseudoscalar,
\be
W \equiv \half \epsilon^{\mu\nu\rho} P_\mu L_{\nu\rho} = P_+ L_{-\perp} - P_- L_{+\perp} + P_\perp L_{+-},
\ee
to generate the remaining components. The advantage of using this particular operator is that it preserves the total momentum,
\be
\comm{W}{P_\mu} = 0.
\ee
We can therefore repeat the procedure we used for the conformal Casimir to derive the differential form for $W$, obtaining
\be
W = P_- \sum_i \left( p_{i\perp} \p_{i-} + \fr{p_{i\perp}^2}{2p_{i-}} \p_{i\perp} \right) - P_\perp \sum_i p_{i-} \p_{i-} - \sum_i \fr{p_{i\perp}^2}{2p_{i-}} \sum_j p_{j-} \p_{j\perp}.
\ee

For each operator $\Ocal$, we just need to find the $\Ccal$ eigenfunction for only one of the polarization components, then act with $W$ to generate the basis functions for the remaining components. Because the Casimir doesn't mix particle number, we can consider each $n$-particle sector independently.

Once we have a complete basis of Casimir eigenfunctions, we can then define the associated weight functions $g_k(\mu)$ as the complete basis of polynomials which are orthogonal with respect to the resulting integration measure, normalized such that
\be
\<\Ccal,\ell;\vec{P},k|\Ccal',\ell';\vec{P}',k'\> = 2P_- (2\pi)^2 \de^2(P-P') \, \de_{\Ccal\Ccal'} \, \de_{\ell\ell'} \, \de_{kk'}.
\ee

As a simple example, consider the one-particle sector. Unsurprisingly, this sector is relatively trivial, as it consists of only one operator, $\phi(x)$, with the associated one-particle state
\be
|\De_\phi;\vec{P}\> \equiv |P\>.
\ee
This basis state is automatically an eigenstate of the conformal Casimir, with eigenvalue
\be
\Ccal^{(1)} = \De_\phi(\De_\phi-3).
\ee
Unlike the higher particle case, this operator also has a unique weight function,
\be
g^{(1)}(\mu) = \de(\mu^2).
\ee


\subsection{Two-Particle States}

We can then turn to the less trivial two-particle case, with the associated differential operator
\be
\begin{split}
\Ccal^{(2)} &= 2 \De_\phi(2\De_\phi-3) - 2p_{1-} p_{2-} (\p_{1-} - \p_{2-})^2 + 2 \De_\phi (p_{1-} - p_{2-}) (\p_{1-} - \p_{2-}) \\
& \, - \, 2(p_{1-} p_{2\perp} + p_{1\perp} p_{2-}) (\p_{1-} - \p_{2-}) (\p_{1\perp} - \p_{2\perp}) - \fr{(p_{1-} p_{2\perp} + p_{1\perp} p_{2-})^2}{2p_{1-} p_{2-}} (\p_{1\perp} - \p_{2\perp})^2 \\
& \, + \, 2\De_\phi (p_{1\perp} - p_{2\perp}) (\p_{1\perp} - \p_{2\perp}).
\end{split}
\label{eq:2pCasimir}
\ee
The corresponding eigenstates are built from operators with two insertions of $\phi$, which can be written as
\be
\Ocal^{(2)}_{\mu_1 \cdots \mu_\ell}(x) \sim \phi(x) \lrpar_{\mu_1} \cdots \lrpar_{\mu_\ell} \phi(x) - \textrm{traces},
\ee
with the associated eigenvalues
\be
\Ccal^{(2)}_\ell = (2\De_\phi + \ell)(2\De_\phi + \ell - 3) + \ell(\ell+1) = 2\ell^2 - 2.
\ee

Because of the equations of motion, these operators all correspond to conserved higher-spin currents, with one such current for each spin $\ell$. These conserved currents each have only two independent components, which means there are only two Casimir eigenfunctions per $\ell$. We can choose one of these two components to correspond to the ``all minus'' operator
\be
\Ocal^{(2)}_{\ell-}(x) \sim \phi(x) \lrpar_{-} \cdots \lrpar_{-} \phi(x).
\ee

The advantage of choosing this component is that the corresponding basis function $F_{\ell-}(p)$ only depends on $p_-$. We therefore only need to find eigenfunctions of the much simpler operator
\be
\Ccal^{(2)}_- = 2 \De_\phi(2\De_\phi-3) - 2p_{1-} p_{2-} (\p_{1-} - \p_{2-})^2 + 2 \De_\phi (p_{1-} - p_{2-}) (\p_{1-} - \p_{2-}).
\ee
Because the Casimir commutes with Lorentz transformations, we can choose a particular reference frame for our eigenstates. We can therefore fix the total momentum $\vec{P}$ by imposing the constraint
\be
\begin{split}
p_{1-} &= p_-, \quad p_{2-} = P_- - p_-, \\
p_{1\perp} &= p_\perp, \quad  p_{2\perp} = -p_\perp,
\end{split}
\ee
which reduces the all minus Casimir to the simpler form
\be
\Ccal^{(2)}_- = 2 \De_\phi(2\De_\phi-3) - 2p_-(P_- - p_-) \p_-^2 + 2 \De_\phi (2p_- - P_-) \p_-.
\ee

We can then easily solve for the associated eigenfunctions by setting $\De_\phi = \half$ and introducing the dimensionless variable
\be
z \equiv \fr{p_-}{P_-},
\ee
resulting in the differential equation
\be
\Big(-2 - 2z(1-z) \p_z^2 + (2z-1) \p_z \Big) F^{(2)}_{\ell-}(z) = \Ccal^{(2)}_\ell F^{(2)}_{\ell-}(z).
\label{eq:2pCas}
\ee
The solutions to this differential equation consist of the Jacobi polynomials
\be
F^{(2)}_{\ell-}(z) = P^{(-\half,-\half)}_\ell(2z-1).
\ee

To obtain the other independent component for each conserved current, we need to act with the Pauli-Lubanski generator, which after fixing the total momentum takes the form
\be
W^{(2)} = \fr{p_\perp^2(P_- - 2p_-)}{2p_-(P_- - p_-)} \p_\perp + p_\perp \p_-.
\ee
Acting on the all minus operators, we obtain the new basis functions,
\be
F^{(2)}_{\ell\perp}(p) \equiv W^{(2)} F^{(2)}_{\ell-}(p) = \fr{p_\perp}{P_-} P^{(\half,\half)}_{\ell-1}(2z-1).
\ee
Schematically, the generator $W$ simply removes a factor of $p_-$ and replaces it with $p_\perp$. These new functions then correspond to operators with a single transverse component,
\be
\Ocal^{(2)}_{\ell\perp}(x) \sim \phi(x) \lrpar_\perp \lrpar_{-} \cdots \lrpar_{-} \phi(x).
\ee
These new polynomials are also eigenfunctions of the full two-particle Casimir in eq.~(\ref{eq:2pCasimir}), with the same eigenvalues as the all minus functions. The new basis functions $F_{\ell\perp}$ are also odd under the parity transformation,
\be
p_\perp \ra -p_\perp,
\ee
while the all minus functions $F_{\ell-}$ are manifestly even. Because we only consider interactions which preserve parity, these two sectors are completely independent. We can therefore safely ignore the parity-odd functions, and focus solely on the parity-even states,
\benn
\begin{split}
|\ell;\vec{P},k\> &\equiv \int d\mu^2 g^{(2)}_k(\mu) \int \fr{d^2p_1 d^2p_2}{(2\pi)^4 2p_{1-}2p_{2-}} (2\pi)^3 \de^3\Big(\sum_i p_i - P\Big) F^{(2)}_{\ell-}(p)|p_1,p_2\> \\
&= \int d\mu^2 g^{(2)}_k(\mu) \int \fr{dp_- \, dp_\perp}{(2\pi)^2 4p_-(P_- - p_-)} (2\pi) \de\bigg(\fr{p_\perp^2 P_-}{2p_-(P_- - p_-)} - \fr{\mu^2}{2P_-}\bigg) F^{(2)}_{\ell-}(p)|p,P-p\>.
\end{split}
\eenn
From now on, we will suppress the index in $F_{\ell-}$, with the understanding that we are always referring to the parity-even sector.

In order to construct an orthogonal basis of weight functions $g_k(\mu)$, we need to consider the inner product
\be
\begin{split}
&\<\ell;\vec{P},k|\ell';\vec{P}',k'\> \\
& \, = 2P_-(2\pi)^2 \de^2(P-P') \cdot 2!\int \fr{d\mu^2}{\mu} g^{(2)}_k(\mu) g^{(2)}_{k'}(\mu) \int \fr{dp_-}{2\sqrt{p_-(P_- - p_-)}} F^{(2)}_{\ell}(p) F^{(2)}_{\ell'}(p),
\end{split}
\ee
where the factor of $2!$ arises from the number of possible Wick contractions between two-particle states. Suppressing the overall momentum-conserving delta function, the inner product therefore factorizes into two independent pieces,
\be
\<\ell;k|\ell';k'\> = \int \fr{d\mu^2}{\mu} g^{(2)}_k(\mu) g^{(2)}_{k'}(\mu) \cdot \int \fr{dp_-}{\sqrt{p_-(P_- - p_-)}} F^{(2)}_{\ell}(p) F^{(2)}_{\ell'}(p) = \de_{kk'} \, \de_{\ell\ell'}.
\ee
Our Casimir eigenfunctions are automatically orthogonal with respect to this measure, so we simply need to properly normalize them,
\be
F^{(2)}_\ell(z) = \fr{1}{\sqrt{\Ncal_\ell}} \, P^{(-\half,-\half)}_\ell(2z-1),
\ee
with the normalization coefficient
\be
\Ncal_\ell = \fr{\G^2(\ell+\half)(1+\de_{\ell,0})}{2\G^2(\ell+1)}.
\ee

The weight functions then correspond to the complete set of polynomials which are orthogonal with respect to the integration measure
\be
\int \fr{d\mu^2}{\mu} g^{(2)}_k(\mu) g^{(2)}_{k'}(\mu) = \de_{kk'}.
\ee
To obtain a normalizable basis, we need to impose the UV cutoff $\mu^2 \leq \Lambda^2$. We can then define the new dimensionless variable
\be
r^2 \equiv \fr{\mu^2}{\Lambda^2},
\ee
to obtain the resulting weight functions,
\be
g^{(2)}_k(r) = \fr{1}{\sqrt{\Ncal_k}} \, P_{2k}(r).
\ee
These functions are Legendre polynomials, parameterized by the non-negative integer $k$, with the overall coefficient
\be
\Ncal_k = \fr{\Lambda}{2k+\half}.
\ee

We now have a complete, properly normalized two-particle basis. The Casimir eigenfunctions $F_\ell(p)$ indicate the particular Casimir multiplet associated with a given operator $\Ocal_\ell$, while the weight functions $g_k(\mu)$ indicate the particular combination of primary operator and descendants within a given multiplet.

In subsection \ref{sec:MasslessLargeN}, we use the generalization of this basis to states built from $N$ scalar fields, in order to study the large-$N$ spectral density of $\phivec$. In this limit, interactions which change particle number are suppressed by $1/N$, such that even at strong coupling we only need to consider two-particle states to compute the leading \KL density. Because our Hamiltonian preserves the $O(N)$ flavor symmetry, we can further limit ourselves to states which are $O(N)$ singlets, with the schematic structure
\benn
\begin{split}
|\ell;\vec{P},k\> &= \int d\mu^2 g^{(2)}_k(\mu) \int \fr{d^2p_1 d^2p_2}{(2\pi)^4 \sqrt{2p_{1-}2p_{2-}}} (2\pi)^3 \de^3\Big(\sum_i p_i - P\Big) F^{(2)}_{\ell}(p) \sum_{i=1}^N a^\dagger_{p_1,i} a^\dagger_{p_2,i}|0\> \\
& = \int d\mu^2 g^{(2)}_k(\mu) \int \fr{dp_- \, dp_\perp}{(2\pi)^2 4p_-(P_- - p_-)} (2\pi) \de\bigg(P_+ - \fr{\mu^2}{2P_-}\bigg) F^{(2)}_{\ell}(p) \sum_{i=1}^N|p,i;P-p,i\>.
\end{split}
\eenn
The kinematic structure of these states is clearly the same as in the single field case, as is the resulting integration measure. These singlet states are also symmetric under exchange of the two momenta, such that we can apply the same analysis as in appendix \ref{app:Symmetrize}. We therefore see that the resulting basis of states is \emph{identical} to the single field basis, with only a slight change to the overall normalization to compensate for the number of fields,
\be
F^{(2)}_{\ell}(p) \ra \fr{1}{\sqrt{N}} \, F^{(2)}_{\ell}(p).
\ee


\subsection{Three-Particle States}

Next, we can consider the three-particle states. The corresponding operators can be built recursively from the two-particle sector to obtain the general form
\be
\Ocal^{(3)}_{\mu_1 \cdots \mu_{\ell_1} \nu_1 \cdots \nu_{\ell_2}}(x) \sim \phi(x) \lrpar_{\mu_1} \cdots \lrpar_{\mu_{\ell_1}} \Big( \phi(x) \lrpar_{\nu_1} \cdots \lrpar_{\nu_{\ell_2}} \phi(x) \Big) - \textrm{traces},
\ee
with the Casimir eigenvalues
\be
\Ccal^{(3)}_\ell = (3\De_\phi + \ell)(3\De_\phi + \ell - 3) + \ell(\ell+1) = 2\ell^2 + \ell - \fr{9}{4},
\ee
where $\ell \equiv \ell_1+\ell_2$. Unlike the two-particle case, these three-particle operators are \emph{not} conserved, such that each has $2\ell+1$ independent components. There are also multiple distinct operators with a given $\ell$.

We can again construct the full basis by first solving for the all minus component for each operator, then acting with the Pauli-Lubanski operator to generate the remaining states. After imposing the constraint
\be
p_{3-} = P_- - p_{1-} - p_{2-}, \quad p_{3\perp} = -p_{1\perp} - p_{2\perp},
\ee
we can then obtain the all minus states by solving for the eigenfunctions of the simplified Casimir
\be
\begin{split}
\Ccal^{(3)}_- &= 3 \De_\phi(3\De_\phi-3) - 2p_{1-} p_{2-} (\p_{1-} - \p_{2-})^2 - 2(P_- - p_{1-} - p_{2-})(p_{1-} \p_{1-}^2 + p_{2-}\p_{2-}^2) \\
& + \, 2 \De_\phi (p_{1-} - p_{2-}) (\p_{1-} - \p_{2-}) + 2\De_\phi(2p_{1-} + p_{2-} - P_-) \p_{1-} + 2\De_\phi(p_{1-} + 2p_{2-} - P_-) \p_{2-}.
\end{split}
\ee
We can further simplify this expression by setting $\De_\phi = \half$ and introducing the variables
\be
z_1 \equiv \fr{p_{1-}}{P_-}, \quad z_2 \equiv \fr{p_{2-}}{P_- - p_{1-}},
\ee
which results in the new differential operator
\be
\Ccal^{(3)}_- = -\fr{9}{4} - 2z_1(1-z_1) \p_{z_1}^2 + (3z_1 - 1) \p_{z_1} - \fr{2z_2(1-z_2)}{1-z_1} \p_{z_2}^2 + \fr{2z_2-1}{1-z_1} \p_{z_2}.
\ee
Focusing on the $z_2$-dependence, we see that the corresponding differential operator precisely matches the two-particle Casimir in eq.~(\ref{eq:2pCas}). The eigenfunctions of this operator are therefore a product of the two-particle basis functions (for $z_2$) and a new Jacobi polynomial (for $z_1$),
\be
F^{(3)}_\ell(z) = (1-z_1)^{\ell_2} \, P^{(2\ell_2,-\half)}_{\ell_1}(2z_1-1) \, P^{(-\half,-\half)}_{\ell_2}(2z_2-1).
\ee
This recursive structure generalizes to higher particle number, such that the $n$-particle basis states are also eigenfunctions of the $(n-1)$-particle conformal Casimir.

To understand the form of these basis states, let's consider two simple examples, both of which have spin $\ell=1$. The first corresponds to $\ell_1=1,\ell_2=0$, which has the momentum space expression,
\be
\phi \, \lrpar_- \phi^2 \ra 2p_{1-} - (p_{2-} + p_{3-}) = P_- (3z_1 - 1).
\ee
As we can see, this perfectly matches the form of the Casimir eigenfunction
\be
F^{(3)}_{10}(z) = P^{(0,-\half)}_1(2z_1-1) \, P^{(-\half,-\half)}_0(2z_2-1) = \fr{1}{2P_-} \Big( P_-(3z_1-1) \Big).
\ee
Next, we can consider the $\ell_1=0,\ell_2=1$ case, which corresponds to the operator,
\be
\phi \Big( \phi \, \lrpar_- \phi \Big) \ra p_{2-} - p_{3-} = P_- (1-z_1) (2z_2 - 1).
\ee
This expression then matches the other $\ell=1$ basis function,
\be
F^{(3)}_{01}(z) = (1-z_1) \, P^{(2,-\half)}_0(2z_1-1) \, P^{(-\half,-\half)}_1(2z_2-1) = \fr{1}{2P_-} \Big( P_-(1-z_1)(2z_2-1) \Big).
\ee

To generate the remaining components for each operator, we need the differential form of the Pauli-Lubanski pseudoscalar. After fixing the total momentum, we obtain
\be
W^{(3)} = \left( \fr{p_{1\perp}^2}{2p_{1-}} - \fr{\mu^2 p_{1-}}{2P_-^2} \right) \fr{\p}{\p p_{1\perp}} + \left( \fr{p_{2\perp}^2}{2p_{2-}} - \fr{\mu^2 p_{2-}}{2P_-^2} \right) \fr{\p}{\p p_{2\perp}} + p_{1\perp} \fr{\p}{\p p_{1-}} + p_{2\perp} \fr{\p}{\p p_{2-}}.
\ee
Much like the two-particle case, this operator replaces factors of $p_-$ with $p_\perp$. However, because we have imposed a UV cutoff on the invariant mass $\mu$, rather than directly on the transverse momenta $p_\perp$, it is more straightforward to express the resulting basis functions in terms of the new variables,
\be
r^2 \cos^2\theta \equiv \fr{\mu_1^2}{\Lambda^2} = \fr{1}{\Lambda^2} \Big(\mu^2 - (p_2 + p_3)^2\Big), \quad r^2 \sin^2\theta \equiv \fr{\mu_2^2}{\Lambda^2} = \fr{1}{\Lambda^2} (p_2+p_3)^2.
\ee
These polar coordinates are defined such that $r$ is the invariant mass of the full three-particle system, in units of the cutoff, while $r\sin\theta$ is the invariant mass of the two-particle system built from $p_2$ and $p_3$.

Using these coordinates, we can then derive the new expression for $W$,
\be
W^{(3)} = r \left( 2\cos\theta \sqrt{z_1(1-z_1)} \, \p_{z_1} + 2\sin\theta \sqrt{\fr{z_2(1-z_2)}{1-z_1}} \, \p_{z_2} + \sin\theta \sqrt{\fr{z_1}{1-z_1}} \, \p_\theta \right).
\ee
As mentioned earlier, each operator with spin $\ell$ contains $2\ell+1$ independent components. We can parameterize these additional components by introducing the new label $m_\perp$,
\be
F^{(3)}_{\ell,m_\perp}(z,r,\theta) \sim W^{m_\perp} F^{(3)}_{\ell,0}(z,r,\theta),
\ee
where $m_\perp$ ranges from $0$ to $2\ell$.

The general structure of the components with $m_\perp \neq 0$ is somewhat complicated, but we can gain some intuition by again considering the operators with $\ell=1$. First, we can look at $\phi \, \lrpar_\mu \phi^2$. While we've already discussed the minus component for this operator, we can now consider the ``transverse'' component,
\be
\phi \, \lrpar_\perp \phi^2 \ra 2p_{1\perp} - (p_{2\perp} + p_{3\perp}) = 3 r\Lambda\cos\theta \sqrt{z_1(1-z_1)},
\ee
which perfectly matches the state created by acting with the Pauli-Lubanski generator on the $m_\perp=0$ state,
\be
F^{(3)}_{10,m_\perp=1}(z,r,\theta) \equiv W^{(3)} F^{(3)}_{10,m_\perp=0}(z,r,\theta) = \fr{2}{\Lambda} \Big( 3 r\Lambda\cos\theta \sqrt{z_1(1-z_1)} \Big).
\ee
Similarly, we can consider the other $\ell=1$ operator, $\phi (\phi \, \lrpar_\mu \phi)$, with the transverse component
\be
\phi \Big( \phi \, \lrpar_\perp \phi\Big) \ra p_{2\perp} - p_{3\perp} = 2r \Lambda \sin\theta \sqrt{(1-z_1)z_2(1-z_2)} - r\Lambda\cos\theta \sqrt{z_1(1-z_1)}(2z_2-1),
\ee
which agrees with the $m_\perp=1$ state
\be
\begin{split}
&F^{(3)}_{01,m_\perp=1}(z,r,\theta) \equiv W^{(3)} F^{(3)}_{01,m_\perp=0}(z,r,\theta) \\
&\qquad = \fr{2}{\Lambda} \Big( 2r \Lambda \sin\theta \sqrt{(1-z_1)z_2(1-z_2)} - r\Lambda\cos\theta \sqrt{z_1(1-z_1)}(2z_2-1) \Big).
\end{split}
\ee
We can then verify that these new $m_\perp=1$ basis functions are also eigenfunctions of the full conformal Casimir,
\be
\begin{split}
\Ccal^{(3)} &= -\fr{9}{4} - 2z_1(1-z_1) \p_{z_1}^2 + (3z_1 - 1) \p_{z_1} - \fr{2z_2(1-z_2)}{1-z_1} \p_{z_2}^2 \\
& \, + \, \fr{2z_2-1}{1-z_1} \p_{z_2} - \fr{2\sqrt{z_1 z_2(1-z_2)}}{1-z_1} \p_{z_2} \p_\theta - \fr{1}{2(1-z_1)} \p_\theta^2,
\end{split}
\ee
with the same eigenvalues as the $m_\perp=0$ components.

More generally, the full basis of three-particle Casimir eigenfunctions takes the schematic form
\be
F^{(3)}_{\ell,m_\perp}(z,r,\theta) \sim r^{m_\perp} \Big( f(z) \cos m_\perp\theta + \bar{f}(z) \sin m_\perp\theta \Big),
\label{eq:Cas3PScheme}
\ee
where the functions $f,\bar{f}$ are built from Jacobi polynomials in $z_1,z_2$. The index $m_\perp$ therefore parameterizes the periodicity in $\theta$, as well as the scaling with $r$. As we demonstrate in appendix~\ref{app:MasslessMatrix}, the interactions we consider in this work preserve this periodicity, such that the different $m_\perp$ sectors are independent. For simplicity, we therefore focus solely on the $m_\perp=0$ states, though understanding the precise basis structure at $m_\perp \neq 0$ is an important direction for future work.

Given these basis functions, our full three-particle states can then be written in the general form,
\be
|\ell,m_\perp;\vec{P},k\> \equiv \int d\mu^2 g_k^{(3)}(\mu) \int \fr{d^2p_1 \, d^2p_2 \, d^2p_3}{(2\pi)^6 2p_{1-} 2p_{2-} 2p_{3-}} (2\pi)^3 \de^3\Big( \sum_i p_i - P \Big) F^{(3)}_{\ell,m_\perp}(p) |p_1,p_2,p_3\>.
\ee
In order to normalize our basis functions and obtain the weight functions $g_k(\mu)$, we can then consider the inner product,
\be
\begin{split}
&\<\ell,m_\perp;k|\ell',m_\perp';k'\> \\
& \qquad = 3! \int d\mu^2 g_k^{(3)}(\mu) g_{k'}^{(3)}(\mu) \int \fr{d^2p_1 \, d^2p_2 \, d^2p_3}{32\pi^2 p_{1-} p_{2-} p_{3-}} \de^3\Big( \sum_i p_i - P \Big) F^{(3)}_{\ell,m_\perp}(p) F^{(3)}_{\ell',m_\perp'}(p).
\end{split}
\ee
Using the dimensionless variables introduced earlier, we can rewrite this inner product in the simpler form
\be
\begin{split}
&\<\ell,m_\perp;k|\ell',m_\perp';k'\> \\
& \qquad = \Lambda^2 \int dr^2 g_k^{(3)}(r) g_{k'}^{(3)}(r) \cdot \fr{3!}{32\pi^2} \int \fr{dz_1 \, dz_2 \, d\theta}{\sqrt{z_1 z_2 (1-z_2)}} F^{(3)}_{\ell,m_\perp}(z,r,\theta) F^{(3)}_{\ell',m_\perp'}(z,r,\theta).
\end{split}
\ee

The only $r$-dependence in the Casimir eigenfunctions $F_{\ell,m_\perp}$ is simply an overall factor of $r^{m_\perp}$, shown in eq.~(\ref{eq:Cas3PScheme}), which for notational simplicity we can instead choose to include in the weight functions $g_k(r)$. This removal of the $r$-dependence does not spoil the conformal structure, as the conformal Casimir is actually independent of $r$. The rescaled Casimir eigenfunctions are then automatically orthogonal with respect to this integration measure, such that we just need to properly normalize them. For the $m_\perp = 0$ case, we then obtain the final basis functions,
\be
F^{(3)}_{\ell,m_\perp=0}(z) = \fr{1}{\sqrt{\Ncal_{\ell,0}}} \, (1-z_1)^{\ell_2} \, P^{(2\ell_2,-\half)}_{\ell_1}(2z_1-1) \, P^{(-\half,-\half)}_{\ell_2}(2z_2-1),
\ee
where the overall coefficient is given by
\be
\Ncal_{\ell,0} \equiv \fr{3!}{16\pi} \cdot \fr{\G(\ell_1+2\ell_2+1)\G(\ell_1+\half)}{(2\ell_1 + 2\ell_2 + \half)\G(\ell_1+2\ell_2+\half)\G(\ell_1+\half)} \cdot \fr{\G^2(\ell_2+\half)(1+\de_{\ell_2,0})}{2\G^2(\ell_2+1)}.
\ee
We can then define the weight functions as the complete set of orthogonal polynomials for the resulting inner product,
\be
\Lambda^2 \int dr^2 g_k^{(3)}(r) g_{k'}^{(3)}(r) = \de_{kk'}.
\ee
Including the overall factor of $r^{m_\perp}$ from the original Casimir eigenfunctions, the resulting weight functions consist of the Zernike polynomials,
\be
g^{(3)}_k(r) = \fr{1}{\sqrt{\Ncal_k}} \, R^{m_\perp}_{2k+m_\perp}(r).
\ee
with the normalization coefficient
\be
\Ncal_k = \fr{\Lambda^2}{2k+m_\perp+1},
\ee
though in this work we only consider the $m_\perp=0$ sector.


\subsection{General Structure for Higher Particle Number}

This procedure can be continued to states with higher particle number, eventually generating the full UV basis for free scalar CFTs. For each sector, we just need to construct the all minus Casimir eigenfunctions, then act with the Pauli-Lubanski generator to obtain the remaining states. Rather than derive the general $n$-particle basis here in full detail, we will instead simply discuss the natural set of coordinates for the all minus states, as well as the basic structure of the resulting basis functions.

Following the approach for three particles, one should build the basis states recursively, constructing $n$-particle Casimir eigenfunctions from $(n-1)$-particle ones. We can make this structure manifest by defining the dimensionless variables
\be
z_i \equiv \fr{p_{i-}}{P_- - p_{1-} - \cdots - p_{(i-1)-}}.
\ee
Using these variables, the all minus Casimir greatly simplifies, and the resulting set of eigenfunctions consists of products of Jacobi polynomials in $z_i$ \cite{Xu}
\be
F^{(n)}_{\ell-}(z) = \prod_{i=1}^{n-1} (1-z_i)^{|\ell^{i+1}|} \, P^{(2|\ell^{i+1}| + \half(n-i)-1,-\half)}_{\ell_i}(2z_i - 1).
\ee
These functions are labeled by $n-1$ non-negative integers, $\ell_i$, and for simplicity, we've introduced the notation
\be
|\ell^i| \equiv \sum_{j=i}^{n-1} \ell_j.
\ee

We can then act on these Jacobi polynomials with the Pauli-Lubanski pseudoscalar to generate the remaining Casimir eigenfunctions. Just like in the three-particle case, it is simpler to express these functions in terms of invariant masses, rather than transverse momenta. We can then implicitly define the new variables $\mu_i$ via the relation
\be
\mu_i^2 + \cdots + \mu_{n-1}^2 = (p_i + \cdots + p_n)^2.
\ee
Our regulator then corresponds to the UV cutoff
\be
\mu^2 = \sum_{i=1}^{n-1} \mu_i^2 \leq \Lambda^2.
\ee
Given this cutoff, a natural set of integration variables is generalized spherical coordinates, which are defined in terms of the dimensionless ratios
\be
\fr{\mu_i}{\Lambda} = r \sin\theta_1 \cdots \sin\theta_{i-1} \cos \theta_i.
\ee
The resulting Casimir eigenfunctions can then be written in terms of generalized spherical harmonics.

Finally, we can use this basis to construct the appropriate inner product for the weight functions $g_k(r)$. The integration measure is always simply a monomial in the radial variable $r$, so the resulting orthogonal basis just consists of generalized Zernike polynomials.


\section{Imposing Symmetrization}
\label{app:Symmetrize}

While the basis derived in the previous section is complete, we actually need to impose an additional constraint on the resulting eigenfunctions of the conformal Casimir. Since these states are constructed from identical particles, their associated ``wavefunction'' $F_\Ocal(p)$ must be symmetric under the exchange of any two momenta. We therefore need to restrict our basis to only those functions which are invariant with respect to all such permutations.

Because the conformal Casimir is manifestly symmetric with respect to particle exchange, this symmetrization procedure only mixes states with the same Casimir eigenvalue. Restricting to symmetric functions therefore does not ruin the conformal structure of our basis, but instead just reduces our Hilbert space to Casimir eigenstates built from identical particles.

The Pauli-Lubanski generator is also manifestly invariant under permutations, such that acting with $W$ on a symmetric basis state yields another state which is automatically symmetric. In practice, we therefore only need to symmetrize a single spin component, which we can choose to be the all minus component for each operator with spin.

As a simple first example, let's consider the two-particle states. We need to reduce this basis to functions which are invariant under the exchange $p_1 \lra p_2$. Written in terms of the new variables $z$ and $r$, this corresponds to the simultaneous exchange
\be
z \ra 1-z, \qquad r \ra -r.
\ee
Fortunately, our basis functions transform very simply under this permutation,
\be
F^{(2)}_\ell(z) \ra (-1)^{\ell} F^{(2)}_\ell(z).
\ee
Our basis therefore reduces to those states which are even under this transformation, corresponding to $F^{(2)}_\ell$ with even spin $\ell$.

For our purposes, we can actually reduce our basis even further. We are specifically interested in those two-particle states that contribute to the spectral density of the operator $\phi^2$, which is even under the parity transformation $p_\perp \ra -p_\perp$. We can therefore restrict our basis to the symmetric, parity-even sector by only including the all minus states $F^{(2)}_{\ell-}$.

Just like the two-particle basis, our set of $n$-particle states must be symmetric under the exchange of any two momenta. Restricting this basis to symmetric states therefore corresponds to finding the set of functions invariant under the symmetric group $S_n$. For this work, we only need to consider up to three-particle states, but the overall symmetrization procedure can be generalized to arbitrary particle number.

The three-particle symmetric group $S_3$ can be generated using just two actions: a single permutation and a cyclic rotation. We therefore only need to reduce our basis to those functions which are invariant under these two transformations.

Let's start with the permutation, which we'll choose without loss of generality to be $p_2 \lra p_3$. Using our new integration variables, this action corresponds to the simultaneous exchange
\be
z_2 \ra 1-z_2, \quad \theta \ra -\theta,
\ee
with the remaining variables $z_1$ and $r$ unchanged. The states with $m_\perp=0$ are independent of $\theta$, and therefore must be symmetric under just $z_1 \ra 1-z_2$. Similar to the two-particle case, this reduces our basis to those wavefunctions $F^{(3)}_{\ell,0}$ with even $\ell_2$. Any other component we create by acting with $W$ on these even states will then automatically be invariant under $p_2 \lra p_3$.

To fully symmetrize our basis with respect to all permutations, we need to also consider the cyclic rotation
\be
p_1 \ra p_2, \quad p_2 \ra p_3, \quad p_3 \ra p_1.
\ee
The radial coordinate $r$ is manifestly invariant under any momentum exchange, such that the weight functions $g_k(r)$ are automatically symmetric. Turning to the angular variable $\theta$, we see that this transformation simply corresponds to the $z$-dependent rotation
\be
\theta \ra \theta + \alpha(z),
\ee
where $\alpha$ satisfies
\be
\cos\alpha = -\sqrt{\fr{z_1 z_2}{1-z_2(1-z_1)}}, \quad \sin\alpha = -\sqrt{\fr{1-z_2}{1-z_2(1-z_1)}}.
\ee
Under this rotation, our angular basis functions transform as
\be
\begin{split}
\cos m_\perp\theta &\ra \cos m_\perp\alpha \, \cos m_\perp\theta - \sin m_\perp\alpha \, \sin m_\perp\theta, \\
\sin m_\perp\theta &\ra \sin m_\perp\alpha \, \cos m_\perp\theta + \cos m_\perp\alpha \, \sin m_\perp\theta.
\end{split}
\ee
As we can see, this cyclic permutation \emph{preserves} the value of $m_\perp$. We therefore don't need to worry about different spin components mixing under symmetrization, and can just fully symmetrize the $m_\perp=0$ sector first.

These all minus basis functions only depend on $z_1,z_2$. Unfortunately, the transformations of the associated Jacobi polynomials under general permutations are very complicated. However, we can directly construct the full basis of symmetric states with $m_\perp=0$ by finding all linear combinations of Jacobi polynomials which are invariant under the simultaneous exchange
\be
z_1 \ra z_2(1-z_1), \quad z_2 \ra \fr{(1-z_1)(1-z_2)}{1-z_2(1-z_1)}.
\ee

However, this brute force symmetrization procedure is somewhat tedious, especially for states with large particle number. In future work, we plan to use an alternative strategy of constructing the basis functions directly in terms of polynomials which are \emph{manifestly} symmetric under particle exchange. Using this strategy to construct the symmetric all minus states, we can then act with the Pauli-Lubanski generator to obtain the remaining basis states, which are automatically symmetric. This new approach should allow us to more efficiently generate the full, symmetric basis of Casimir eigenstates.


\section{Matrix Elements for Casimir Basis}
\label{app:MasslessMatrix}

In this appendix, we use our basis of Casimir eigenstates to calculate the matrix elements for contributions to the operator $M^2$. We specifically consider matrix elements which preserve the conformal structure of our basis and do not mix distinct Casimir multiplets, in order to obtain the single multiplet results discussed in section~\ref{sec:MasslessResults}. These matrix elements take the generic form
\be
\<\Ccal,\ell;\vec{P},k| M^2 |\Ccal',\ell';\vec{P}',k'\> = 2 P_- (2\pi)^2 \de^2(P-P') \, \Mcal_{\Ccal \ell k,\Ccal' \ell' k'}.
\ee
In what follows, we focus only on the dynamical elements $\Mcal$, suppressing the overall kinematic normalization factor.

The invariant mass operator $M^2$ can be rewritten in terms of momentum generators as
\be
M^2 = 2P_+ P_- - P_\perp^2.
\ee
As discussed in appendix \ref{app:Interactions}, we can choose a particular reference frame for this Lorentz invariant inner product, fixing the overall $P_-$ and setting $P_\perp = 0$. These matrix elements then reduce to the simpler expression
\be
\Mcal_{\Ccal \ell k,\Ccal' \ell' k'} \equiv \<\Ccal,\ell;k| M^2 |\Ccal',\ell';k'\> = 2P_- \<\Ccal,\ell;k| P_+ |\Ccal',\ell';k'\>.
\ee


\subsection{Kinetic Terms}

Let's begin by computing the $M^2$ matrix elements for the original UV CFT. As shown in appendix~\ref{app:Interactions}, the lightcone Hamiltonian can be expanded in terms of raising/lowering operators as
\be
P_+^{(\textrm{CFT})} = \int \fr{d^2p}{(2\pi)^2} \, a^\dagger_p a_p \fr{p_\perp^2}{2p_-}.
\ee
This free Hamiltonian preserves particle number, such that we can consider each $n$-particle sector separately. First, we'll focus on the corresponding two-particle matrix element
\be
\<\ell;k|P_+^{(\textrm{CFT})}|\ell';k'\> = \int \fr{d^2p}{(2\pi)^2} \<\ell;k|a^\dagger_p a_p|\ell';k'\> \fr{p_\perp^2}{2p_-} = \<\ell;k|\left( \fr{p_{1\perp}^2}{2p_{1-}} + \fr{p_{2\perp}^2}{2p_{2-}} \right)|\ell';k'\>.
\ee

Unsurprisingly, the total lightcone energy $P_+$ just turns into a sum over the individual particle energies. We can then use our two-particle basis functions to rewrite the matrix element as the integral
\be
\Mcal_{\ell k,\ell'k'} = 2P_-\<\ell;k|P_+|\ell';k'\> = \int \fr{d\mu^2}{\mu} g^{(2)}_k(\mu) g^{(2)}_{k'}(\mu) \int \fr{dp_-}{\sqrt{p_-(P_- - p_-)}} F^{(2)}_{\ell}(p) F^{(2)}_{\ell'}(p) \, \mu^2.
\ee
As we can see, the resulting free Hamiltonian is completely independent of $p_-$, such that the associated matrix elements are diagonal in $\ell$,
\be
\Mcal_{\ell k,\ell'k'} = \de_{\ell\ell'} \int \fr{d\mu^2}{\mu} g^{(2)}_k(\mu) g^{(2)}_{k'}(\mu) \, \mu^2.
\ee
We now need to evaluate the $\mu^2$ integral, which we can rewrite in terms of the dimensionless variable $r$ as
\be
\int \fr{d\mu^2}{\mu} g^{(2)}_k(\mu) g^{(2)}_{k'}(\mu) \, \mu^2 = \Lambda^3 \int dr \, g_k(r) g_{k'}(r) \, r^2.
\ee
This integral can be evaluated analytically, obtaining the result
\be
\begin{split}
&\fr{\Lambda^3}{\sqrt{\Ncal_{k} \Ncal_{k'}}} \int dr \, P_{2k}(r) P_{2k'}(r) \, r^2 \\
& \quad = \Lambda^2 \left( \fr{8k^2+4k-1}{16k^2+8k-3} \de_{kk'} + \sqrt{\fr{2k_{\max}+\half}{2k_{\min}+\half}} \, \fr{2k_{\max}(2k_{\max}-1)}{(4k_{\max}-1)(4k_{\max}+1)} \, \de_{|k-k'|,1} \right).
\end{split}
\ee
We therefore find that these ``kinetic term'' matrix elements are quadratically sensitive to the invariant mass cutoff $\Lambda$. From a dimensional analysis perspective, this is unsurprising, as the original UV CFT possesses no other dimensionful parameters to set the overall energy scale for $M^2$. We can then combine these pieces together to construct the unperturbed two-to-two matrix elements $\Mcal^{(\textrm{CFT})}$, which can be diagonalized to obtain the $\phi^2$ spectral density discussed in section \ref{sec:MasslessResults}.

For the case of $N$ scalar fields, our two-particle sector is restricted to states which are $O(N)$ singlets. As discussed in appendix~\ref{app:CasimirBasis}, the associated basis is the same as the single field case, with an additional factor of $N$ in the overall normalization. However, when computing the associated kinetic term matrix elements, we obtain an overall multiplicity of $N$ which perfectly cancels the altered normalization. The two-particle matrix elements for $O(N)$ flavor singlets are therefore \emph{identical} to those for the single field case.

Next, we can turn to the independent three-particle sector. As before, the associated matrix elements turn into a sum over the individual particle energies,
\be
\<\ell,m_\perp;k|P_+^{(\textrm{CFT})}|\ell',m_\perp';k'\> = \<\ell,m_\perp;k|\left( \fr{p_{1\perp}^2}{2p_{1-}} + \fr{p_{2\perp}^2}{2p_{2-}} + \fr{p_{3\perp}^2}{2p_{3-}} \right)|\ell',m_\perp';k'\>,
\ee
which leads to the $M^2$ matrix element
\be
\begin{split}
&\Mcal_{\ell m_\perp k, \ell' m_\perp' k'} \\
& \quad = 3! \int d\mu^2 g_k^{(3)}(\mu) g_{k'}^{(3)}(\mu) \int \fr{d^2p_1 \, d^2p_2 \, d^2p_3}{32\pi^2 p_{1-} p_{2-} p_{3-}} \de^3\Big( \sum_i p_i - P \Big) F^{(3)}_{\ell,m_\perp}(p) F^{(3)}_{\ell',m_\perp'}(p) \, \mu^2.
\end{split}
\ee

Just like in the two-particle case, the invariant mass is simply $\mu^2$, such that the resulting matrix element takes the simple form
\be
\Mcal_{\ell m_\perp k, \ell' m_\perp' k'} = \de_{\ell_1 \ell_1'} \, \de_{\ell_2 \ell_2'} \, \de_{m_\perp m_\perp'} \int d\mu^2 g_k^{(3)}(\mu) g_{k'}^{(3)}(\mu) \, \mu^2.
\ee
We therefore see that the kinetic term is diagonal in both $\vec{\ell}$ and $m_\perp$. This structure is quite important, as we're specifically interested in studying the spectral density of the operator $\phi^3$, which only has support on states with $m_\perp=0$. As we will see, the other interactions we consider also have this structure, such that we can safely restrict our three-particle basis to the subspace of states with $m_\perp=0$.

Finally, we can evaluate the remaining integral over the weight functions, which is greatly simplified by the restriction to $m_\perp = 0$,
\be
\fr{2\Lambda^4}{\sqrt{\Ncal_{k}\Ncal_{k'}}} \int dr \, r \, R^{0}_{2k}(r) R^{0}_{2k'}(r) \, r^2 = \Lambda^2 \left( \half \de_{kk'} + \fr{k_{\max}}{2\sqrt{(2k+1)(2k'+1)}} \de_{|k-k'|,1} \right).
\ee
These matrix elements are therefore quadratically sensitive to the UV cutoff, just like in the two-particle case. We can then combine the individual terms together to construct the three-to-three component of $\Mcal^{(\textrm{CFT})}$, which we can use to obtain the $\phi^3$ spectral density.


\subsection{Large-$N$ Interaction}

Next, we can consider the quartic interaction matrix elements in the $O(N)$ model. In the large-$N$ limit, matrix elements which change particle number are suppressed, such that we can just focus on the two-to-two processes
\be
\de P^{(\lambda)}_+ = \fr{\lambda}{2} \int \fr{d^2p \, d^2q \, d^2k}{(2\pi)^6 \sqrt{8p_- q_- k_-}} \, \fr{a^\dagger_{p,i} a^\dagger_{q,i} a_{k,j} a_{p+q-k,j} + 2a^\dagger_{p,i} a^\dagger_{k,j} a_{q,i} a_{p+q-k,j}}{\sqrt{2(p_- + q_- - k_-)}}.
\ee
The resulting two-particle matrix elements can then be written as
\benn
\begin{split}
&\de \Mcal_{\ell k,\ell' k'}^{(\lambda)} = 2P_- \<\ell;k| \de P_+^{(\lambda)} |\ell';k'\> \\
&= \fr{\lambda}{8\pi^2} \int \fr{d\mu^2}{\mu} \fr{d\mu^{\prime2}}{\mu'} g^{(2)}_k(\mu) g^{(2)}_{k'}(\mu') \int \fr{dp_- \, dp_-'}{\sqrt{p_-(P_- - p_-)p_-'(P_- - p_-')}} F^{(2)}_{\ell}(p) F^{(2)}_{\ell'}(p') \Big( N + 2 \Big).
\end{split}
\eenn
Note that we've included the modified normalization for two-particle states in the $O(N)$ model. The first term clearly dominates at large-$N$, such that we can safely ignore the second contribution. We can then rewrite the simplified matrix element as
\be
\de \Mcal_{\ell k,\ell' k'}^{(\lambda)} = \fr{\lambda N}{2} \<\ell;k|\phivec(0)\>\<\phivec(0)|\ell';k'\>,
\ee
where we've explicitly factorized this expression into two copies of the same integral,
\be
\<\phivec(0)|\ell;k\> = \int \fr{d\mu^2}{\mu} g^{(2)}_k(\mu) \int \fr{dp_-}{2\pi\sqrt{p_-(P_- - p_-)}} F^{(2)}_{\ell}(p).
\ee
We then just need to evaluate the two independent integrals. Starting with the $p_-$ integration, we obtain
\be
\fr{1}{2\pi} \int \fr{dz}{\sqrt{z (1 - z)}} F^{(2)}_\ell(z) = \fr{1}{2\pi\sqrt{\Ncal_{\ell}}} \int \fr{dz}{\sqrt{z(1-z)}} P^{(-\half,-\half)}_{\ell}(2z-1) = \fr{1}{2\sqrt{\pi}} \, \de_{\ell,0}.
\ee
This inner product therefore projects onto the Casimir eigenstate for $\phivec$, with $\ell=0$. Similarly, the integral over $\mu^2$ results in
\be
\Lambda \int dr \, g^{(2)}_k(r) = \fr{\Lambda}{\sqrt{\Ncal_k}} \int dr \, P_{2k}(r) = \sqrt{2\Lambda} \, \de_{k,0}.
\ee
Combining these results together, we then obtain the final expression
\be
\de \Mcal_{\ell k,\ell' k'}^{(\lambda)} = \fr{\lambda N \Lambda}{4\pi} \, \de_{\ell,0} \, \de_{k,0} \, \de_{\ell',0} \, \de_{k',0}.
\label{eq:LargeNElement}
\ee
The single nonzero matrix element is therefore proportional to the number of fields $N$, such that the true interaction scale is $\kappa \equiv \lambda N$.


\section{Modified Basis with Dirichlet Boundary Conditions}
\label{app:MassiveBasis}

In this appendix, we impose vanishing Dirichlet boundary conditions on the conformal Casimir eigenfunctions to obtain a new basis of states. As we shall see, this modified basis arises naturally from divergences in the mass term
\be
\de P_+^{(m)} = \int \fr{d^2p}{(2\pi)^2} \, a^\dagger_p a_p \, \fr{m^2}{2p_-}.
\ee
These divergences reorganize the Casimir basis derived in appendix~\ref{app:CasimirBasis} into new linear combinations which manifestly vanish when any individual lightcone momentum goes to zero,
\be
F^{(n)}_\Ocal(p) \ra \Ft^{(n)}_\Ocal(p) \sim p_{1-} p_{2-} \cdots p_{n-} F^{(n)}_\Ocal(p).
\ee
This new ``Dirichlet basis'' eliminates the divergences in the mass term, such that the entire resulting mass spectrum is finite. In constructing these new linear combinations, one only needs to impose Dirichlet boundary conditions on one component in each multiplet, then act with the Pauli-Lubanski pseudoscalar on this modified state to generate the remaining spin components.

We first present a simple two-particle example, in order to explicitly demonstrate the reorganization of our basis states by the mass term. We then show that the resulting basis can easily be obtained by finding the complete basis of polynomials which are orthogonal with respect to a modified inner product. Finally, we discuss the resulting basis of two- and three-particle states, which can then be generalized to arbitrary particle number.


\subsection{New Boundary Conditions from Mass Term}

In order to study theories with a massive scalar field, we need to consider the matrix elements associated with the mass term
\be
\de \Lcal = -\half m^2 \phi^2.
\ee
For two-particle states, this relevant perturbation leads to the $M^2$ matrix correction,
\be
\begin{split}
\de\Mcal^{(m)}_{\ell k,\ell'k'} &= 2P_-\<\ell;k|\de P_+^{(m)}|\ell';k'\> \\
&= \int \fr{d\mu^2}{\mu} g^{(2)}_k(\mu) g^{(2)}_{k'}(\mu) \int \fr{dp_-}{\sqrt{p_-(P_- - p_-)}} F^{(2)}_{\ell}(p) F^{(2)}_{\ell'}(p) \, \fr{m^2 P_-^2}{p_-(P_- - p_-)}.
\end{split}
\ee
As we can see, this integrand only depends on $p_-$, such that the resulting matrix elements are diagonal in $k$,
\be
\de\Mcal_{\ell k,\ell'k'} = \de_{kk'} \int \fr{dp_-}{\sqrt{p_-(P_- - p_-)}} F^{(2)}_{\ell}(p) F^{(2)}_{\ell'}(p) \, \fr{m^2 P_-^2}{p_-(P_- - p_-)}.
\ee
The mass term therefore has \emph{no effect} on the weight functions $g_k(\mu)$. We can thus ignore them for the rest of this discussion and focus solely on the basis functions $F_\Ocal(p)$.

Looking more carefully at the structure of the integrand, we see that it diverges when $p_- \ra 0,P_-$, which corresponds to the lightcone momentum of either particle vanishing. We can see this divergence explicitly by switching to the dimensionless variable $z$ and imposing a small cutoff $\epsilon$ on the range of integration,
\be
\de\Mcal_{\ell,\ell'} = m^2 \int_\epsilon^{1-\epsilon} \fr{dz}{\sqrt{z(1-z)}} F^{(2)}_{\ell}(z) F^{(2)}_{\ell'}(z) \, \fr{1}{z(1-z)}.
\ee
Focusing specifically on the all minus basis functions, we can then evaluate this integral and take the limit $\epsilon \ra 0$, obtaining
\be
\begin{split}
\de\Mcal_{\ell,\ell'} &= \fr{m^2}{\sqrt{\Ncal_\ell \Ncal_{\ell'}}} \int_\epsilon^{1-\epsilon} \fr{dz}{\sqrt{z(1-z)}} P^{(-\half,-\half)}_\ell(2z-1) P^{(-\half,-\half)}_{\ell'}(2z-1) \fr{1}{z(1-z)} \\
&= \fr{2m^2}{\sqrt{(1+\de_{\ell,0})(1+\de_{\ell',0})}} \left(-4 \Lmax + \fr{4}{\pi\sqrt{\epsilon}} \right).
\end{split}
\ee
Each of these matrix elements therefore diverges as $\epsilon \ra 0$.

Let's now isolate this divergent piece in order to understand its effects on the resulting spectrum of mass eigenstates. The associated matrix elements take the simple form
\be
\de\Mcal^{(\epsilon)}_{\ell,\ell'} = \fr{8m^2}{\pi\sqrt{\epsilon(1+\de_{\ell,0})(1+\de_{\ell',0})}} = \fr{8m^2}{\pi\sqrt{\epsilon}} \begin{pmatrix} \half & \fr{1}{\sqrt{2}} & \fr{1}{\sqrt{2}} & \cdots \\ \fr{1}{\sqrt{2}} & 1 & 1 & \cdots \\ \fr{1}{\sqrt{2}} & 1 & 1 & \cdots \\ \vdots & \vdots & \vdots & \ddots \end{pmatrix}.
\ee
This matrix can be rewritten as simply an outer product of the vector
\be
\begin{pmatrix} \fr{1}{\sqrt{2}} & 1 & 1 & \cdots \end{pmatrix} = \fr{1}{\sqrt{2}} F^{(2)}_{0}(z) + F^{(2)}_{2}(z) + F^{(2)}_{4}(z) + \cdots
\label{eq:DivergentEigen}
\ee
The divergent piece in $\de\Mcal$ is therefore a \emph{projection operator}, such that it effectively removes this single vector from our Hilbert space in the limit $\epsilon \ra 0$. The reduced space then consists of all states which are orthogonal to this one, which we can easily construct out of all linear combinations of the form,
\be
F^{(2)}_{\ell}(z) - \sqrt{2} F^{(2)}_{0}(z).
\label{eq:ZeroEigen}
\ee
However, this particular combination perfectly cancels the constant term in each basis function $F^{(2)}_{\ell}(z)$. The divergent mass term therefore just reshuffles our basis to eliminate the constant term in each basis function, which is equivalent to imposing vanishing Dirichlet boundary conditions in $z$.

To see this explicitly, let's consider a simple example. Truncating our basis to $\ell \leq 2$, we obtain the divergent term
\be
\de\Mcal^{(\epsilon)}_{\ell,\ell'} = \fr{8m^2}{\pi\sqrt{\epsilon}} \begin{pmatrix} \half & \fr{1}{\sqrt{2}} \\ \fr{1}{\sqrt{2}} & 1 \end{pmatrix}.
\ee
We can then find the two orthonormal eigenvectors for this matrix,
\be
\begin{split}
&\begin{pmatrix} \fr{1}{\sqrt{3}} & \sqrt{\fr{2}{3}} \end{pmatrix} = \sqrt{\fr{2}{3}} \left( \fr{1}{\sqrt{2}} F^{(2)}_{0}(z) +  F^{(2)}_{2}(z) \right), \\
&\begin{pmatrix} -\fr{2}{\sqrt{3}} & \fr{1}{\sqrt{3}} \end{pmatrix} = \fr{1}{\sqrt{3}} \left( F^{(2)}_{2}(z) -\sqrt{2} F^{(2)}_{0}(z) \right),
\end{split}
\ee
which match the form of eq.~(\ref{eq:DivergentEigen}) and (\ref{eq:ZeroEigen}), respectively. The first state has the divergent eigenvalue $\fr{12m^2}{\pi\sqrt{\epsilon}}$, while the second state has eigenvalue $0$. If we diagonalize the full matrix $M^2$ and then take the limit $\epsilon\ra0$, we'll therefore find one unphysical eigenvalue which diverges as $O(1/\sqrt{\epsilon})$, and a second physical eigenvalue which remains finite.

Looking at the full expression for the remaining physical eigenstate, we see that it takes the simple form
\be
\Ft^{(2)}_2(z) = F^{(2)}_{2}(z) -\sqrt{2} F^{(2)}_{0}(z) = -8 \sqrt{\fr{2}{\pi}} \, z(1-z).
\ee
Comparing this expression to the Casimir eigenstate,
\be
F^{(2)}_{2}(z) = \sqrt{\fr{2}{\pi}} \Big(1 - 8 z(1-z)\Big),
\ee
we see that the divergent mass term simply removes the constant piece, leaving a function which manifestly vanishes when $z \ra 0,1$.

This general structure continues as we include basis states with larger $\ell$. The divergence in the mass term simply removes a single linear combination from our basis in the limit $\epsilon\ra0$, leaving only states with vanishing Dirichlet boundary conditions. While these new combinations $F_{\ell} - \sqrt{2} F_{0}$ are no longer orthogonal, one can simply use Gram-Schmidt to re-orthogonalize this shifted basis.

In the following two subsections, we explicitly construct the all minus Dirichlet basis functions for states with two and three particles. Rather than use Gram-Schmidt, however, we note that the resulting basis must be orthogonal with respect to a modified integration measure. We then simply obtain the complete set of orthgonal polynomials for this new inner product, which is \emph{identical} to the basis one obtains through directly applying Gram-Schmidt to the states with Dirichlet boundary conditions. The advantage of this approach is that it naturally generalizes to arbitrary particle number, simplifying the construction of the new Dirichlet basis.


\subsection{Two-Particle States}

Starting with the two-particle sector, we can define the new all minus Dirichlet basis states,
\be
|\Lt;\vec{P},k\> \equiv \int d\mu^2 g^{(2)}_k(\mu) \int \fr{dp_- \, dp_\perp}{(2\pi)^2 4p_-(P_- - p_-)} (2\pi) \de\bigg(P_+ - \fr{\mu^2}{2P_-}\bigg) \Ft^{(2)}_{\ell-}(p)|p,P-p\>.
\ee
Because the weight functions are unchanged by the new boundary conditions in $p_-$, the resulting inner product is the same as for the Casimir basis,
\be
\<\Lt;k|\Lt';k'\> = \de_{kk'} \int \fr{dz}{\sqrt{z(1 - z)}} \Ft^{(2)}_{\ell-}(z) \Ft^{(2)}_{\ell'-}(z).
\ee
However, the basis functions $\Ft_{\ell}$ are polynomials which must vanish as $z\ra0,1$, so they must be proportional to an overall factor of $z(1-z)$,
\be
\Ft^{(2)}_{\ell-}(z) = z(1-z) f_\ell(z).
\ee
We can then derive the form of the basis functions by finding the complete basis of polynomials which are orthogonal with respect to the modified integration measure,
\be
\int \fr{dz}{\sqrt{z(1 - z)}} \Ft^{(2)}_{\ell-}(z) \Ft^{(2)}_{\ell'-}(z) = \int dz \, z^{\fr{3}{2}} (1 - z)^{\fr{3}{2}} f_{\ell}(z) f_{\ell'}(z).
\ee
The resulting functions are simply Jacobi polynomials for a new measure, leading to the Dirichlet basis functions
\be
F^{(2)}_{\ell-}(z) = \fr{1}{\sqrt{\Ncal_\ell}} \, z(1-z) P^{(\fr{3}{2},\fr{3}{2})}_{\ell-2}(2z-1),
\ee
with the overall normalization factor
\be
\Ncal_\ell = \fr{\G^2(\ell+\half)}{2\ell\G(\ell-1)\G(\ell+2)}.
\ee

These basis functions are polynomials in $z$ of degree $\ell$, which means they can be written solely in terms of Casimir basis functions with $\ell' \leq \ell$,
\be
\Ft^{(2)}_{\ell-}(z) = \fr{1}{\sqrt{(\ell-1)(\ell+1)}} \sum_{\ell'=0}^{\ell-2} \sqrt{1+\de_{\ell',0}} \, F^{(2)}_{\ell'-}(z) - \sqrt{\fr{\ell-1}{\ell+1}} \, F^{(2)}_{\ell-}(z),
\ee
which are also the linear combinations we would obtain by directly applying Gram-Schmidt to the set of functions $F^{(2)}_{\ell-} - F^{(2)}_0$. We can therefore still restrict the basis to $\Ccal \leq \Cmax$ by truncating in $\ell$.

Intriguingly, these Dirichlet boundary conditions appear to reorganize our Casimir eigenstates into operators of the schematic form
\be
\widetilde{\Ocal}_\ell \sim \Big( \p_- \phi(x) \Big) \lrpar_{\mu_1} \cdots \lrpar_{\mu_\ell} \Big( \p_- \phi(x) \Big),
\ee
analogous to primary operators built from scalar fields in 2D. While these Dirichlet basis states are \emph{not} eigenstates of the conformal Casimir, this structure suggests they may be eigenstates of some other differential operator with a sensible physical interpretation.


\subsection{Three-Particle States}

Now that we have a Dirichlet basis for the two-particle subspace, we can move on to the three-particle states. The three-particle mass term diverges when any of the three lightcone momenta vanish,
\be
\de M^2 = m^2 P_- \left( \fr{1}{p_{1-}} + \fr{1}{p_{2-}} + \fr{1}{p_{3-}} \right).
\ee
Following the same procedure as for the two-particle states, we see that that these divergences rearrange the Casimir basis into linear combinations which vanish when any $p_{i-}\ra0$. The resulting basis states take the general form,
\be
|\Lt,m_\perp;\vec{P},k\> \equiv \int d\mu^2 g_k^{(3)}(\mu) \int \fr{d^2p_1 \, d^2p_2 \, d^2p_3}{(2\pi)^6 2p_{1-} 2p_{2-} 2p_{3-}} (2\pi)^3 \de^3\Big( \sum_i p_i - P \Big) \Ft^{(3)}_{\ell,m_\perp}(p) |p_1,p_2,p_3\>.
\ee
Focusing specifically on the $m_\perp = 0$ components, we can then consider the inner product
\be
\<\Lt,m_\perp=0;k|\Lt',m_\perp'=0;k'\> = \de_{kk'} \, \fr{3!}{16\pi} \int \fr{dz_1 \, dz_2}{\sqrt{z_1 z_2 (1-z_2)}} \Ft^{(3)}_{\ell,0}(z) \Ft^{(3)}_{\ell',0}(z).
\ee

In order to satisfy all three boundary conditions, the basis functions must take the general form,
\be
\Ft^{(3)}_{\ell,0}(z) = p_{1-} p_{2-} p_{3-} f_\ell(z) = z_1 (1-z_1)^2 z_2 (1-z_2) f_\ell(z).
\ee
We then need to find a complete basis of orthogonal polynomials for the new measure,
\be
\int \fr{dz_1 \, dz_2}{\sqrt{z_1 z_2 (1-z_2)}} \Ft^{(3)}_{\ell,0}(z) \Ft^{(3)}_{\ell',0}(z) = \int dz_1 \, dz_2 \, z_1^{\fr{3}{2}} (1-z_2)^4 z_2^{\fr{3}{2}} (1-z_2)^{\fr{3}{2}} f_\ell(z) f_{\ell'}(z).
\ee
The resulting Dirichlet basis functions are
\be
\Ft_{\ell,0}(z) = \fr{1}{\sqrt{\Ncal_{\ell}}} \, z_1(1-z_1)^{\ell_2} z_2(1-z_2) \, P^{(2\ell_2,\fr{3}{2})}_{\ell_1-1}(2z_1-1) \, P^{(\fr{3}{2},\fr{3}{2})}_{\ell_2-2}(2z_2-1),
\ee
with the overall normalization factor
\be
\Ncal_{\ell} = \fr{3!}{16\pi} \cdot \fr{\G(\ell_1+2\ell_2)\G(\ell_1+\fr{3}{2})}{(2\ell_1+2\ell_2+\half)\G(\ell_1)\G(\ell_1+2\ell_2+\fr{3}{2})} \cdot \fr{\G^2(\ell_2+\half)}{2\ell_2\G(\ell_2-1)\G(\ell_2+3)}.
\ee


\subsection{General Multi-Particle States}

Now that we understand the general procedure, we can provide the basic structure for the all minus Dirichlet basis functions. Consider an arbitrary $n$-particle state, whose form is analogous to that of the two- and three-particle basis states,
\be
|\widetilde{\Ccal},\Lt;\vec{P},k\> = \int d\mu^2 g_k(\mu) \int \fr{d^2p_1 \cdots d^2p_n}{(2\pi)^{2n} 2p_{1-} \cdots 2p_{n-}} (2\pi)^3 \de^3\Big(\sum_i p_i - P\Big) \Ft^{(n)}_\Ocal(p)|p_1,\cdots,p_n\>.
\ee
The Dirichlet boundary conditions resulting from the mass term restrict these new basis states to take the general form
\be
\Ft^{(n)}_{\ell-}(z) = p_{1-} \cdots p_{n-} f_\ell(z) = z_1 (1-z_1)^{n-1} z_2 (1-z_2)^{n-2} \cdots z_{n-1} (1-z_{n-1}) f_\ell(z),
\ee
where we've used the general $z$ variables introduced in appendix~\ref{app:CasimirBasis},
\be
z_i \equiv \fr{p_{i-}}{P_- - p_{1-} - \cdots - p_{(i-1)-}}.
\ee

This overall prefactor modifies the inner product for the general polynomial $f_\ell(z)$. The resulting all minus basis functions are then built from a product of Jacobi polynomials, just like the original Casimir eigenstates, but with a shifted integration measure \cite{Xu},
\be
\Ft^{(n)}_{\ell-}(z) = \prod_{i=1}^{n-2} \left( z_i(1-z_i)^{|\ell^{i+1}|} P^{(2|\ell^{i+1}| + \half(n-i)-1,\fr{3}{2})}_{\ell_i-1}(2z_i - 1) \right) z_{n-1}(1-z_{n-1}) P^{(\fr{3}{2},\fr{3}{2})}_{\ell_{n-1}-2}(2z_{n-1} - 1).
\ee
We can then act with the Pauli-Lubanski generator $W$ to obtain the Dirichlet basis functions for the other components.


\section{Matrix Elements for Dirichlet Basis}
\label{app:MassiveMatrix}

In this appendix, we use the new Dirichlet basis to calculate matrix elements for the invariant mass $M^2$. We specifically compute matrix elements associated with the orthogonal two- and three-particle polynomials derived in appendix \ref{app:MassiveBasis}, which can then be combined together into symmetric combinations following the procedure discussed in appendix \ref{app:Symmetrize}.

The matrix elements for the unperturbed UV CFT only depend on the weight functions $g_k(\mu)$, which are unaffected by our new Dirichlet boundary conditions for $F_\Ocal(p)$. We can therefore reuse the results in appendix~\ref{app:MasslessMatrix} to obtain spectral densities for a free, massless scalar field. Here, we consider the corrections arising from the relevant deformations discussed in appendix \ref{app:Interactions}, which can then be used to calculate the results presented in section~\ref{sec:MassiveResults}.


\subsection{Mass Terms}

The simplest deformation of the UV theory is the addition of a mass term, which shifts the Hamiltonian by
\be
\de P_+^{(m)} = \int \fr{d^2p}{(2\pi)^2} \, a^\dagger_p a_p \fr{m^2}{2p_-}.
\ee
Just like the original kinetic term, this new ``interaction'' doesn't mix states with different particle number, so we can consider the two- and three-particle sectors independently.

Starting with the two-particle case, we see that this mass term leads to a matrix element correction of the form
\be
\de\Mcal^{(m)}_{\Lt k,\Lt' k'} = 2P_-\<\Lt;k|\de P_+|\Lt';k'\> = 2P_-\<\Lt;k|\left( \fr{m^2}{2p_{1-}} + \fr{m^2}{2p_{2-}} \right)|\Lt';k'\>.
\ee
We can then use our new Dirichlet basis functions to rewrite this correction as the integral
\be
\de\Mcal_{\Lt k,\Lt' k'} = m^2 \int \fr{d\mu^2}{\mu} g^{(2)}_k(\mu) g^{(2)}_{k'}(\mu) \int \fr{dp_-}{\sqrt{p_-(P_- - p_-)}} \Ft^{(2)}_{\ell}(p) \Ft^{(2)}_{\ell'}(p) \left(\fr{P_-}{p_-} + \fr{P_-}{P_- - p_-}\right).
\ee
Because of the eventual permutation symmetry of our basis states, the two contributions to this integral must be identical. We can therefore simplify the computation of these matrix elements by rewriting them as two copies of the first term,
\be
\de\Mcal_{\Lt k,\Lt' k'} = 2m^2 \int \fr{d\mu^2}{\mu} g^{(2)}_k(\mu) g^{(2)}_{k'}(\mu) \int \fr{dp_-}{\sqrt{p_-(P_- - p_-)}} \Ft^{(2)}_{\ell}(p) \Ft^{(2)}_{\ell'}(p) \, \fr{P_-}{p_-}.
\ee

As we can see, this operator has no $\mu$-dependence, which means the associated matrix elements are diagonal with respect to $k$,
\be
\de\Mcal_{\Lt k,\Lt' k'} = 2 m^2 \, \de_{k k'} \int \fr{dp_-}{\sqrt{p_-(P_- - p_-)}} \Ft^{(2)}_{\ell}(p) \Ft^{(2)}_{\ell'}(p) \, \fr{P_-}{p_-}.
\ee
We can now evaluate the remaining $p_-$ integral, which can be rewritten in terms of the variable $z$ as
\be
\int \fr{dp_-}{\sqrt{p_-(P_- - p_-)}} \Ft^{(2)}_{\ell}(p) \Ft^{(2)}_{\ell'}(p) \, \fr{P_-}{p_-} = \int \fr{dz}{\sqrt{z(1-z)}} \Ft^{(2)}_{\ell}(z) \Ft^{(2)}_{\ell'}(z) \, \fr{1}{z}.
\ee
Given our two-particle basis functions, this integral can then be evaluated analytically,
\be
\begin{split}
\fr{1}{\sqrt{\Ncal_{\ell}\Ncal_{\ell'}}} \int dz \, z^{\fr{3}{2}} (1-z)^{\fr{3}{2}} \, P^{(\fr{3}{2},\fr{3}{2})}_{\ell-2}(2z-1) \, P^{(\fr{3}{2},\fr{3}{2})}_{\ell'-2}(2z-1) \, \fr{1}{z} = (-1)^{\ell+\ell'} \fr{4}{3} \sqrt{\fr{\ell \ell' (\ell_{\min}-1)_3}{(\ell_{\max}-1)_3}},
\end{split}
\label{eq:MassTermMixing}
\ee
where for notational simplicity we've used the Pochhammer symbol $(q)_n \equiv \fr{\G(q+n)}{\G(q)}$.

We can then use these results to obtain the full two-particle matrix correction $\de \Mcal^{(m)}$. Just like with the kinetic term, the mass term matrix elements are unchanged in the $O(N)$ model.

Next, we can turn to the three-particle subspace, which receives a very similar matrix correction
\be
\de\Mcal^{(m)}_{\Lt m_\perp k, \Lt' m_\perp' k'} = 2P_-\<\Lt,m_\perp;k|\left( \fr{m^2}{2p_{1-}} + \fr{m^2}{2p_{2-}} + \fr{m^2}{2p_{3-}} \right)|\Lt',m_\perp';k'\>.
\ee
Just like in the two-particle case, we can use the permutation symmetry of our basis states to rewrite the contributions of each individual particle as three copies of the same integral,
\be
\begin{split}
&\de\Mcal_{\Lt m_\perp k, \Lt' m_\perp' k'} \\
& \quad = 3m^2 \cdot 3! \int d\mu^2 g_k^{(3)}(\mu) g_{k'}^{(3)}(\mu) \int \fr{d^2p_1 \, d^2p_2 \, d^2p_3}{32\pi^2 p_{1-} p_{2-} p_{3-}} \de^3\Big( \sum_i p_i - P \Big) \Ft^{(3)}_{\ell,m_\perp}(p) \Ft^{(3)}_{\ell',m_\perp'}(p) \, \fr{P_-}{p_{1-}}.
\end{split}
\ee

Note that this operator again only depends on $p_-$, which means these matrix elements are diagonal with respect to $k$,
\be
\de\Mcal_{\Lt m_\perp k, \Lt' m_\perp' k'} = 3m^2 \, \de_{kk'} \, \fr{3!}{32\pi^2} \int \fr{dp_{1-} \, dp_{2-}}{\sqrt{p_{1-} p_{2-} P_- (P_- - p_{1-} - p_{2-})}} \, \Ft^{(3)}_{\ell,m_\perp}(p) \Ft^{(3)}_{\ell',m_\perp'}(p) \, \fr{P_-}{p_{1-}}.
\ee
These matrix elements are also diagonal with respect to $m_\perp$, which means we can safely restrict our basis to $m_\perp=0$. We then need to evaluate the $p_-$ integrals, which can be rewritten using a change of variables as
\be
\begin{split}
&\fr{3!}{32\pi^2} \int \fr{dp_{1-} \, dp_{2-}}{\sqrt{p_{1-} p_{2-} P_- (P_- - p_{1-} - p_{2-})}} \, \Ft^{(3)}_{\ell,0}(p) \Ft^{(3)}_{\ell',0}(p) \, \fr{P_-}{p_{1-}} \\
& \qquad = \fr{3!}{32\pi^2} \int \fr{dz_1 \, dz_2 \, d\theta}{\sqrt{z_1 z_2 (1-z_2)}} \, \Ft^{(3)}_{\ell,0}(p) \Ft^{(z)}_{\ell',0}(z) \, \fr{1}{z_1}.
\end{split}
\ee
As we can see, this operator now has no $z_2$-dependence, such that the matrix elements are also diagonal in $\ell_2$. The remaining $z_1$ integral is then greatly simplified by the restriction $\ell_2=\ell_2'$, and can be evaluated analytically to obtain
\be
\begin{split}
&\fr{1}{\sqrt{\Ncal_{\ell_1}\Ncal_{\ell_1'}}} \int dz_1 \, z_1^{\fr{3}{2}} (1-z_1)^{2\ell_2} P^{(2\ell_2,\fr{3}{2})}_{\ell_1-1}(2z_1-1) \, P^{(2\ell_2,\fr{3}{2})}_{\ell_1'-1}(2z_1-1) \, \fr{1}{z_1} \\
&=(-1)^{\ell_1+\ell_1'} \fr{2}{3} \sqrt{\fr{(2\ell_1+2\ell_2+\half)(2\ell_1'+2\ell_2+\half)(\ell_{1\min})_{3/2}(\ell_{1\min}+2\ell_2)_{3/2}}{(\ell_{1\max})_{3/2} (\ell_{1\max}+2\ell_2)_{3/2}}}.
\end{split}
\ee
We can then combine together all of these results to calculate the full three-particle mass correction $\de \Mcal^{(m)}$.


\subsection{Interaction Terms}

We can now consider the addition of interactions which mix states with different particle number. The simplest such correction to the Hamiltonian is the cubic interaction
\be
\de P^{(g)}_+ = \fr{g}{2} \int \fr{d^2p \, d^2q}{(2\pi)^4 \sqrt{8 p_- q_- (p_- + q_-)}} \Big( a^\dagger_p a^\dagger_q a_{p+q} + a^\dagger_{p+q} a_p a_q \Big).
\ee
This operator clearly mixes states whose particle numbers differ by one. We are specficially interested in studying the one-particle mass shift in the perturbative regime $g/m^{3/2} \ll 1$, which means we can focus on the one-to-two matrix elements
\be
\begin{split}
\de\Mcal^{(g)}_{\phi,\Lt k} &\equiv 2P_- \<\phi|\de P^{(g)}_+|\Lt;k\> = 2P_- \cdot \fr{g}{2} \int \fr{d^2p \, d^2q}{(2\pi)^4 \sqrt{8 p_- q_- (p_- + q_-)}} \<\phi| a^\dagger_{p+q} a_p a_q |\Lt;k\> \\
&= \fr{g}{2} \<\phi^2(0)|\Lt;k\> = \fr{g}{4\pi} \int \fr{d\mu^2}{\mu} g^{(2)}_k(\mu) \int \fr{dp_-}{\sqrt{p_-(P_- - p_-)}} \Ft^{(2)}_{\ell}(p).
\end{split}
\label{eq:PhiCubedIntegral}
\ee
Note that this interaction matrix element is actually proportional to the overlap of our basis states with $\phi^2$, which means we can also reuse this calculation in determining the $\phi^2$ spectral density. Evaluating the $p_-$ integral first, we obtain
\be
\begin{split}
\fr{1}{4\pi} \int \fr{dz}{\sqrt{z (1 - z)}} \Ft^{(2)}_{\ell}(z) &= \fr{1}{4\pi\sqrt{\Ncal_{\ell}}} \int dz \sqrt{z(1-z)} \, P^{(\fr{3}{2},\fr{3}{2})}_{\ell-2}(2z-1) \\
&= \fr{1}{\sqrt{8\pi(\ell^2-1)} } \qquad (\ell \textrm{ even}).
\end{split}
\ee
This inner product therefore vanishes unless $\ell$ is even, though this is already required for all of our basis states by permutation symmetry. Turning to the $\mu^2$ integral, we find
\be
\Lambda \int dr \, g^{(2)}_k(r) = \fr{\Lambda}{\sqrt{\Ncal_k}} \int dr \, P_{2k}(r) = \sqrt{2\Lambda} \, \de_{k,0}.
\ee
As we can see, these matrix elements are sensitive to the UV cutoff $\Lambda$ and only mix the single-particle state with basis states that have $k = 0$.

Next, we can turn to the quartic interaction, whose Hamiltonian correction contains two distinct terms, one which preserves particle number and one which mixes states whose particle numbers differ by two,
\be
\de P^{(\lambda)}_+ = \fr{\lambda}{24} \int \fr{d^2p \, d^2q \, d^2k}{(2\pi)^6\sqrt{8p_- q_- k_-}} \left( \fr{4 a^\dagger_p a^\dagger_q a^\dagger_k a_{p+q+k}}{\sqrt{2(p_- + q_- + k_-)}} + h.c. + \fr{6 a^\dagger_p a^\dagger_q a_k a_{p+q-k}}{\sqrt{2(p_- + q_- - k_-)}} \right).
\ee
We're again specifically interested in the perturbative one-particle mass shift, which means we can focus on the first term, which leads to the one-to-three matrix elements
\be
\begin{split}
\de\Mcal^{(\lambda)}_{\phi,\Lt m_\perp k} &\equiv 2P_- \<\phi|\de P^{(\lambda)}_+|\Lt,m_\perp;k\> \\
&= 2P_-  \fr{\lambda}{6} \int \fr{d^2p \, d^2q \, d^2k}{(2\pi)^6\sqrt{8p_- q_- k_-}} \fr{\<\phi|a^\dagger_{p+q+k} a_p a_q a_k|\Lt,m_\perp;k\>}{\sqrt{2(p_- + q_- + k_-)}} = \fr{\lambda}{6} \<\phi^3(0)|\Lt,m_\perp;k\>.
\end{split}
\ee
Similar to before, these matrix elements are proportional to the overlap of our basis with $\phi^3$, such that we can use these results to also calculate the $\phi^3$ spectral density. We can then write this matrix element as the integral
\be
\de\Mcal_{\phi,\Lt m_\perp k} = \fr{\lambda}{64\pi^3} \int d\mu^2 g_k^{(3)}(\mu) \int \fr{d^2p_1 \, d^2p_2 \, d^2p_3}{p_{1-} p_{2-} p_{3-}} \de^3\Big( \sum_i p_i - P \Big) \Ft^{(3)}_{\ell,m_\perp}(p).
\ee
Using the dimensionless integration variables for $p$, we can rewrite this expression as
\be
\de\Mcal_{\phi,\Lt m_\perp k} = \fr{\lambda}{64\pi^3} \int d\mu^2 g_k^{(3)}(\mu) \int \fr{dz_1 \, dz_2 \, d\theta}{\sqrt{z_1 z_2 (1-z_2)}} \, \Ft^{(3)}_{\ell,m_\perp}(z,\theta).
\ee

As we can see, the integrand is independent of $\theta$, such that the one-particle state only interacts with basis states with $m_\perp = 0$,
\be
\fr{1}{64\pi^3} \int \fr{dz_1 \, dz_2 \, d\theta}{\sqrt{z_1 z_2 (1-z_2)}} \, \Ft^{(3)}_{\ell,m_\perp}(z,\theta) = \de_{m_\perp,0} \, \fr{1}{32\pi^2} \int \fr{dz_1 \, dz_2}{\sqrt{z_1 z_2 (1-z_2)}} \, \Ft^{(3)}_{\ell,0}(z).
\ee
We can then evaluate the independent $z$ integrals to obtain
\be
\begin{split}
&\fr{1}{32\pi^2 \sqrt{2\pi\Ncal_{\ell}}} \int dz_1 \sqrt{z_1} \, (1-z_1)^{\ell_2} P^{(2\ell_2,\fr{3}{2})}_{\ell_1-1}(2z_1-1) \int dz_2 \sqrt{z_2 (1-z_2)} \, P^{(\fr{3}{2},\fr{3}{2})}_{\ell_2-2}(2z_2-1) \\
&\qquad = \sqrt{\fr{(2\ell_1+2\ell_2+\half)\G(\ell_1+2\ell_2)\G(\ell_1+2\ell_2+\fr{3}{2})}{768\pi(\ell_2^2-1)\G(\ell_1)\G(\ell_1+\fr{3}{2})}} \, \fr{\G(\ell_2+1)}{\G(\ell_2+\fr{5}{2})\G(2\ell_2+1)} \\
& \qquad \qquad \otimes \phantom{}_3F_2\left(1-\ell_1,\ell_1+2\ell_2+\fr{3}{2},\ell_2+1;2\ell_2+1,\ell_2+\fr{5}{2};1\right) \qquad (\ell_2 \textrm{ even}).
\end{split}
\ee
Similar to before, this inner product vanishes unless $\ell_2$ is even, which is already required by permutation symmetry.

We can then evaluate the remaining $\mu^2$ integral, which simply projects onto those basis states with $k = 0$,
\be
\fr{2\Lambda^2}{\sqrt{\Ncal_{k}}} \int dr \, r \, R^{0}_{2k} (r) = \Lambda \, \de_{k,0}.
\ee
These matrix elements are therefore also sensitive to the UV cutoff $\Lambda$.

Finally, let's consider the $O(N)$ version of the quartic interaction. As discussed in appendix~\ref{app:MasslessMatrix}, in the large-$N$ limit the Hamiltonian simplifies such that we only need to consider the single interaction term
\be
\de P^{(\lambda)}_+ = \fr{\lambda}{2} \int \fr{d^2p \, d^2q \, d^2k}{(2\pi)^6 \sqrt{8p_- q_- k_-}} \, \fr{a^\dagger_{p,i} a^\dagger_{q,i} a_{k,j} a_{p+q-k,j}}{\sqrt{2(p_- + q_- - k_-)}},
\ee
with the resulting two-particle matrix elements
\be
\de \Mcal_{\Lt k,\Lt' k'}^{(\lambda)} = \fr{\lambda N}{8\pi^2} \int \fr{d\mu^2}{\mu} \fr{d\mu^{\prime2}}{\mu'} g^{(2)}_k(\mu) g^{(2)}_{k'}(\mu') \int \fr{dp_- \, dp_-'}{\sqrt{p_-(P_- - p_-)p_-'(P_- - p_-')}} \Ft^{(2)}_{\ell}(p) \Ft^{(2)}_{\ell'}(p').
\ee
As we can see, this expression factorizes into two copies of the same integral,
\be
\de \Mcal_{\Lt k,\Lt' k'}^{(\lambda)} = \fr{\lambda N}{2} \<\Lt;k|\phivec(0)\>\<\phivec(0)|\Lt';k'\>,
\ee
where the overlap with $\phivec$ is given by
\be
\<\phivec(0)|\Lt;k\> = \fr{1}{2\pi} \int \fr{d\mu^2}{\mu} g^{(2)}_k(\mu) \int \fr{dp_-}{\sqrt{p_-(P_- - p_-)}} \Ft^{(2)}_{\ell}(p).
\ee
Comparing this integral to eq.~(\ref{eq:PhiCubedIntegral}), we see that it has the same form as the one-to-two matrix elements associated with the $\phi^3$ interaction. We can then use our previous results to obtain the final expression
\be
\de \Mcal_{\Lt k,\Lt' k'}^{(\lambda)} = \fr{\kappa \Lambda}{2\pi\sqrt{(\ell^2-1)(\ell^{\prime 2}-1)}} \, \de_{k,0} \, \de_{k',0}.
\ee
where we've again defined the interaction scale $\kappa \equiv \lambda N$.


\section{Conjectured Formalism for General CFTs}
\label{app:GeneralCFT}

For any CFT, our prescription for defining a discretized Hilbert space of states is to use the local operators, $\mathcal{O}(x)$, to define conformal Casimir and momentum eigenstates in the following way,
\be
|\Ccal,\ell;\vec{P},k\> \equiv \int d\mu^2 g_k(\mu) \int d^3x \, e^{-iP\cdot x} \Ocal(x)|0\>,
\label{app:states}
\ee
where $\mu^2 \equiv P^2$. The $g_{k}(\mu)$ are a set of ``weight functions'' that discretize the continuous invariant mass parameter $\mu^{2}$. 

This definition of Hilbert space states works for any CFT, and an important feature of defining states in this way is that: 
\begin{enumerate}
\itemsep0em
\item[(i)] Inner products between states are defined in terms of CFT two-point functions.
\item[(ii)] Hamiltonian matrix elements between states are defined in terms of CFT three-point functions. 
\end{enumerate}
Thus, the ingredients needed to implement conformal truncation are expressible solely in terms of the CFT data. In this work, we have specifically focused on the case of initiating the truncation from a free UV CFT. However, if one is given CFT data, it should be possible to formulate conformal truncation around any interacting CFT. In this appendix, we discuss how such a formulation should proceed. 

The case of a free massless scalar CFT is still a useful illustrative example. For concreteness, consider the states associated with the operator $\phivec$ in the large-$N$ setup of section~\ref{sec:MasslessResults},
\be
|\phivec;\vec{P},k\> \equiv \int d\mu^2 g_k(\mu) \int d^3x \, e^{-iP\cdot x} \phivec(x)|0\>.
\label{eq:PhiVecStates}
\ee
In the main text, we expanded the operator $\phivec(x)$ in terms of Fock space modes in order to determine the weight functions $g_k(\mu)$ and the corresponding Hamiltonian matrix elements. In reality, though, all we were doing was making use of facts (i) and (ii) above.    

To see this, we start with the simple observation that eq.~(\ref{app:states}) fixes the inner product in terms of the two-point function,
\be
\<\phivec;\vec{P},k|\phivec;\vec{P}',k'\> = \int d\mu^2 d^3x \, g_k(\mu) \, e^{iP\cdot x} \int d\mu^{\prime \, 2} d^3x' \, g_{k'}(\mu') \, e^{-iP'\cdot x'} \<\phivec(x)\phivec(x')\>,
\ee
The key point is that the two-point function sets the integration measure for the weight functions. Indeed, we can use the \KL spectral representation to rewrite the inner product as
\be
\<\phivec;\vec{P},k|\phivec;\vec{P}',k'\> = 2P_-(2\pi)^2 \de^2(P-P') \, (2\pi)^2 \int d\mu^2 \, \rho_{\phivec}(\mu) \, g_k(\mu) g_{k'}(\mu).
\ee
Using the free $\phivec$ spectral density from eq.~(\ref{eq:PhiSquareDensity}), the inner product for the weight functions then takes the form
\be
(2\pi)^2 \int_0^{\Lambda^2} \fr{d\mu^2}{4\pi\mu} \, g_k(\mu) g_{k'}(\mu) = \delta_{kk'}.
\ee
This integral precisely matches the inner product derived from the Fock space modes in eq.~(\ref{eq:2pWeightIntegral}), up to an overall normalization factor. This slight difference in the overall coefficient simply arises from the normalization convention for the operator $\phivec$ in eq.~(\ref{app:states}) and cancels in any final matrix element. We then obtain the familiar weight functions
\be
g_k(\mu) = P_{2k} \Big( \fr{\mu}{\Lambda} \Big).
\ee

Given this set of weight functions, we can then consider the matrix elements arising from the relevant deformation,
\be
\de \Lcal = - \fr{\lambda N}{4} \big( \phivec \big)^2,
\ee
with the associated Hamiltonian correction
\be
\de P_+^{(\lambda)} = \fr{\lambda N}{4} \int d^2x \, \big(\phivec\big)^2(x).
\ee
In the main text, we computed the matrix elements for this operator via Fock space modes. However, we could have instead recast this computation as taking the Fourier transform of a three-point function. Indeed, from the definition of our $\phivec$ states in eq.~(\ref{eq:PhiVecStates}), it is clear that
\be
\begin{split}
&\<\phivec;\vec{P},k|\de P_+|\phivec;\vec{P}',k'\> \\
& \quad = \fr{\lambda N}{4} \int d\mu^2 d^3x \, g_k(\mu) \, e^{iP\cdot x} \int d\mu^{\prime \, 2} d^3x' \, g_{k'}(\mu') \, e^{-iP'\cdot x'} \int d^2y \, \< \phivec(x) \big(\phivec\big)^2(y) \phivec(x') \>.
\label{app:ME2}
\end{split}
\ee
We thus see explicitly that the Hamiltonian matrix element is fixed by the relevant three-point function.

In the large-$N$ limit, this particular three-point function can be written in the simple form
\be
\< \phivec(x) (\phivec)^2(y) \phivec(x') \> = 2 \, \<\phivec(x)\phivec(y)\> \, \<\phivec(y)\phivec(x')\>.
\ee 
Inserting the spectral representation of these two-point functions into eq.~(\ref{app:ME2}), we can then obtain the simplified $M^2$ matrix elements
\be
\<\phivec;k|\de M^2 |\phivec;k'\> = 2\pi^2 \lambda N \int d\mu^2 \, \rho_{\phivec}(\mu) \, g_k(\mu) \int d\mu^{\prime \, 2} \, \rho_{\phivec}(\mu') \, g_{k'}(\mu').
\ee
This form for the matrix element makes several features manifest. First, all other three-point functions involving $\big(\phivec\big)^2$ are suppressed by $1/N$, making it clear that this interaction only affects the $\phivec$ Casimir multiplet in the large-$N$ limit. In addition, the kinematic structure of this particular three-point function leads to the factorized behavior for the matrix elements, with the two weight functions each being integrated against unity, such that this expression vanishes unless $k=k'=0$. Evaluating the trivial integrals yields the final matrix element
\be
\<\phivec;k|\de M^2|\phivec;k'\> = \fr{\kappa\Lambda}{4\pi} \, \delta_{k,0} \,\delta_{k',0},
\ee
which reproduces the Fock space result in eq.~(\ref{eq:LargeNElement}).

The procedure outlined here should be applicable to any interacting CFT. Given a CFT operator, one can then use its two-point function to find the corresponding measure for the weight functions $g_k(\mu)$ and obtain an orthonormal basis of states. For example, the states built from any scalar operator $\Ocal$ would have the resulting inner product,
\be
\<\De;\vec{P},k|\De;\vec{P}',k'\> = 2P_-(2\pi)^2 \de^2(P-P') \, (2\pi)^2 \int d\mu^2 \, \rho_{\Ocal}(\mu) \, g_k(\mu) g_{k'}(\mu).
\ee 
With basis states thus defined, one can then use the CFT three-point functions to determine the Hamiltonian matrix elements resulting from any relevant deformation,
\be
\de \Lcal = - \lambda \Ocal_R,
\ee
leading to the general expression
\be
\begin{split}
&\<\Ccal,\ell;\vec{P},k|\de P_+^{(\Ocal_R)}|\Ccal',\ell';\vec{P}',k'\> \\
& \qquad = \lambda \int d\mu^2 d^3x \, g_k(\mu) \, e^{iP\cdot x} \int d\mu^{\prime \, 2} d^3x' \, g_{k'}(\mu') \, e^{-iP'\cdot x'} \int d^2y \, \< \Ocal(x) \Ocal_R(y) \Ocal'(x') \> \\
& \qquad = (2\pi)^2 \de^2(P-P') \int d\mu^2 g_k(\mu) \int d\mu^{\prime \, 2} g_{k'}(\mu') \, \Mcal^{(\Ocal_R)}_{\Ccal\ell,\Ccal'\ell'}(\mu,\mu').
\end{split}
\ee
It would be very interesting to test this procedure in systems where the full set of CFT data is known, such as the 2D Ising model, to determine dynamical correlation functions in the presence of relevant deformations.

\bibliographystyle{utphys}
\bibliography{BibScalar3d}

\end{document}